\documentclass[prb,amsmath,amssymb,superscriptaddress,twocolumn]{revtex4}
\usepackage{braket}
\usepackage{graphicx}
\usepackage{multirow}
\usepackage{bbm,color,ulem}

\usepackage{bm}
\usepackage{extarrows}
\usepackage{array}
\newcolumntype{M}[1]{>{\centering\arraybackslash}m{#1}}
\newcolumntype{N}{@{}m{0pt}@{}}

\DeclareMathAlphabet{\mathpzc}{OT1}{pzc}{m}{it}
\begin{document}
\title{Fast pulse sequences for dynamically corrected gates in single-triplet qubits}
\author{Robert E.\ Throckmorton}
\email{rthrockm@umd.edu}
\affiliation{Condensed Matter Theory Center and Joint Quantum Institute, Department of Physics, University of Maryland, College Park, Maryland 20742-4111 USA}
\author{Chengxian Zhang}
\affiliation{Department of Physics, City University of Hong Kong, Tat Chee Avenue, Kowloon, Hong Kong SAR, China}
\affiliation{City University of Hong Kong Shenzhen Research Institute, Shenzhen, Guangdong 518057, China}
\author{Xu-Chen Yang}
\affiliation{Department of Physics, City University of Hong Kong, Tat Chee Avenue, Kowloon, Hong Kong SAR, China}
\affiliation{City University of Hong Kong Shenzhen Research Institute, Shenzhen, Guangdong 518057, China}
\author{Xin Wang}
\email{x.wang@cityu.edu.hk}
\affiliation{Department of Physics, City University of Hong Kong, Tat Chee Avenue, Kowloon, Hong Kong SAR, China}
\affiliation{City University of Hong Kong Shenzhen Research Institute, Shenzhen, Guangdong 518057, China}
\author{Edwin Barnes}
\affiliation{Department of Physics, Virginia Tech, Blacksburg, Virginia 24061, USA}
\author{S.\ Das Sarma}
\affiliation{Condensed Matter Theory Center and Joint Quantum Institute, Department of Physics, University of Maryland, College Park, Maryland 20742-4111 USA}
\date{\today}
\begin{abstract}
We present a set of experimentally feasible pulse sequences that implement any single-qubit gate on a singlet-triplet
spin qubit and demonstrate that these new sequences are up to three times faster than existing sequences in the literature.
We show that these sequences can be extended to incorporate built-in dynamical error correction, yielding gates that
are robust to both charge and magnetic field noise and up to twice as fast as previous dynamically corrected gate schemes.
We present a thorough comparison of the performance of our new sequences with that of several existing ones using randomized
benchmarking, considering both quasistatic and $1/f^\alpha$ noise models.  We provide our results both as a function of
evolution time and as a function of the number of gates, which respectively yield both an effective coherence time and
an estimate of the number of gates that can be performed within this coherence time.  We determine which set of pulse
sequences gives the best results for a wide range of noise strengths and power spectra.  Overall, we find that the traditional,
slower sequences perform best when there is no field noise or when the noise contains significant high-frequency components;
otherwise, our new, fast sequences exhibit the best performance.
\end{abstract}
\maketitle

\section{Introduction} \label{Sec:Introduction}
The problem of how to efficiently and precisely control qubits is of fundamental interest for quantum computing
and quantum information.  This includes determining how to perform gates within the limitations of a given platform
and how to combat errors while the gates are being implemented.  One goal is to achieve at least $99\%$ average
fidelity for all gates, which is the threshold above which error correction based on surface code architectures
may be implemented \cite{FowlerPRA2012}.  This $>99\%$ error constraint for surface codes is still much lower than
the very stringent quantum error correction threshold of $99.99\%$ required for general circuit level quantum
computations, and the current experimental activity in quantum computing hardwares is aimed at achieving $>99\%$
error threshold to enable practical surface code architectures.  One platform that is particularly promising is that
of electron spins in semiconductor quantum dots, due to their compatibility with the existing semiconductor industry,
as well as to the fast electrical control and long coherence times available in these systems.  Several different
implementations of qubits using such electronic spins include the single-spin Loss-DiVincenzo qubit\cite{LossPRA1998,NowackScience2011,PlaNature2012,PlaNature2013,VeldhorstNatNano2014,BraakmanNatNano2013,OtsukaSciRep2016,ItoSciRep2016},
the singlet-triplet qubit\cite{LevyPRL2002,PettaScience2005,FolettiNatPhys2009,VanWeperenPRL2011,MauneNature2012,ShulmanScience2012,DialPRL2013,ShulmanNatCommun2014,ReedPRL2016,MartinsPRL2016},
the triple-dot exchange-only qubit\cite{DiVincenzoNature2000,MedfordNatNano2013,MedfordPRL2013,EngSciAdv2015,ShimPRB2016},
and ``hybrid'' qubits with three electrons in two dots\cite{ShiPRL2012,KimNature2014,KimNPJQI2015}.  While there has been
much experimental progress on improving the fidelity of gate operations\cite{NowackScience2011,VanWeperenPRL2011,ShulmanScience2012,VeldhorstNature2016},
with fidelities as high as about $99\%$ for single-qubit gates and about $90\%$ for two-qubit gates having been
demonstrated\cite{NicholnpjQI2017}, more work must still be done to comfortably exceed the 99\% surface-code threshold
for all gate operations.

One of the main problems that must be tackled in order to achieve such efficient and precise control of qubits
is noise-induced gate error.  In semiconductor spin qubits, the two dominant types of noise are (magnetic) field
noise due to fluctuations both in nuclear spins and in applied magnetic fields, and charge noise due to voltage
fluctuations in the gates and in the semiconductor environment used to define the quantum dots or to nearby
charged impurities.  In fact, charge noise in various forms is present in all of the currently pursued experimental
qubits, including superconducting and ion trap qubits, in addition to semiconductor spin qubits of interest in the
current paper.  While noise may be reduced to an extent at the material platform level through techniques such as
isotopic purification for Si, dynamical nuclear spin polarization in GaAs, and improved materials and device designs,
it has proven challenging to suppress it to a sufficiently low level.  Therefore, various dynamical decoupling
techniques have been developed to further mitigate the effects of noise at the qubit control level. For example,
by applying appropriate pulse sequences or by implementing carefully designed feedback control, it is possible to
dramatically extend the coherence time of spin qubits\cite{BluhmPRL2010,BluhmNatPhys2011,SergeevichPRA2011,ShulmanNatCommun2014,MuhonenNatNanotechnol2014,MalinowskiNatNanotechnol2017}
with generalized spin echo type (e.g. CPMG) pulse sequences being well-known examples of extending spin coherence
in semiconductors\cite{WitzelPRL2007,WitzelPRB2007,LeePRL2008}.  Several control techniques have also been proposed
that are capable of not only enhancing the lifetime of qubits, but also of performing gate operations while noise
errors are simultaneously suppressed \cite{WangNatCommun2011,KestnerPRL2013,WangPRA2014,GoelmanJMagnReson1989,KhodjastehPRA2009,KhodjastehPRL2010,BenskyEPL2010,GracePRA2012,GreenPRL2012,KosutPRA2013}. However, such techniques generally need to be optimized according to the specific noise profiles and physical constraints of
a given type of qubit.

In this work, we focus on singlet-triplet qubits, which have been used to experimentally demonstrate some of the
highest single- and two-qubit gate fidelities achieved so far in a spin-qubit system \cite{NicholnpjQI2017}. Such
a qubit consists of two electrons occupying a double quantum dot, with an electrostatically controlled Heisenberg
exchange interaction between them and a magnetic field gradient across the two dots, the latter typically produced
with either a micromagnet or by polarizing the nuclear spins \cite{FolettiNatPhys2009,BluhmPRL2010,BrunnerPRL2011,PetersenPRL2013,WatsonArXiv2017}.
The (effective) Hamiltonian of this qubit within the logical subspace (total $S_z=0$) has the form,
\begin{equation}
\hat{H}=J\hat{\sigma}_z+h\hat{\sigma}_x, \label{Eq:Hamiltonian}
\end{equation}
where $J$ is the exchange coupling and $h$ is the magnetic field difference between the two dots.  In practice, it is
very difficult to change $h$ quickly, and thus all operations are performed by tuning $J$ using external electrical gates
that can rapidly modify the electron wavefunctions in the quantum dot.  If $J$ is switched on and held at a fixed value,
then the associated time evolution operator is
\begin{equation}
\hat{U}(t,0)=\exp\left [-i(J\hat{\sigma}_z+h\hat{\sigma}_x)t\right ]. \label{Eq:TimeEvolution}
\end{equation}
This operator rotates the qubit about an axis in the $xz$ plane which is determined by the magnitude of $J$ relative to
$h$. The value of $J$ is restricted in actual experiments to positive (and, strictly speaking, non-zero) values due to
the nature of the exchange coupling. (Note that recent experimental findings\cite{MartinsArXiv2017} indicate that it may
be possible to relax these restrictions in multi-electron systems.) As a result, with a single square $J$ pulse we can
only perform rotations about axes in the $xz$ plane at an angle $\theta\in\left (\theta_{min},\frac{\pi}{2}\right )$
with respect to the $z$ axis; all other rotations require multiple pulses to perform. The minimal value $\theta_{min}$
reflects the fact that in real systems, it is not possible to make $J$ arbitrarily large, which would be necessary to
implement a pure $z$ rotation with a single pulse when $h\ne0$. Instead, it is restricted to $J\le J_{max}$, where
$J_{max}$ is determined by the geometry of the quantum dots. In terms of the effective Hamiltonian, Eq. (1), charge
noise induces fluctuations in the exchange coupling $J$, while field noise produces fluctuations in the gradient $h$.
The task of designing robust gate operations for singlet-triplet qubits thus requires that we somehow mitigate (at
least) the leading-order errors in $\hat U(t,0)$ caused by these fluctuations by modulating $J$ in a prescribed manner
while respecting the $0<J\le J_{max}$ constraint.

We outline here a two-pronged approach to combatting the effects of noise during gate operations: faster pulse sequences
and dynamical error correction.  We begin by reviewing existing pulse sequences in the literature and proposing new sequences
capable of implementing faster gates.  We will consider five sets of pulse sequences in this work, three of which are not
corrected for noise-induced error, and the other two of which are.  The first uncorrected set, which we will call the
``uncorrected unoptimized'' (UCUO) sequences, is based on the composite $z$ rotation sequence proposed by Guy Ramon\cite{RamonPRB2011},
and $x$ rotations are performed by switching off $J$ for an interval of time. Other rotations can then be performed using
the standard $z$-$x$-$z$ decomposition\cite{NielsenBook}.  The second is the ``uncorrected optimized, type I'' (UCO-I)
set, for which $z$ rotations are performed using the $\theta$-$2\theta$-$\theta$ sequence proposed in Ref.\ \onlinecite{ZhangPRL2017},
and $x$ rotations are implemented using a Hadamard-$z$-Hadamard sequence.  The third, a new set that we introduce in the
present work, is the ``uncorrected optimized, type II'' (UCO-II) set; this set still uses the $\theta$-$2\theta$-$\theta$
sequence from the UCO-I set to perform $z$ rotations, but also includes new sequences for performing other rotations, which
we will describe in detail.  These new pulse sequences, and the comparison of their efficacies with the existing pulse sequences
for singlet-triplet gate operations, are one of the main results of this work.

The two sets of error-corrected pulse sequences are derived using the \textsc{supcode} method\cite{WangNatCommun2011,KestnerPRL2013,WangPRA2014},
which modifies a given (na\"ive) pulse sequence that implements a desired gate operation by combining it with a noisy identity
operation designed in such a way as to cancel the effects of noise-induced error to leading order. \textsc{supcode} is specifically
designed for singlet-triplet qubits, where the positivity of the exchange coupling between the electron spins precludes the
applicability of most dynamical gate correction methods, although the basic strategy of combining naive sequences with noisy
identity operations to cancel errors can be applied to a broad range of systems. This strategy will be exploited in the present
work to design new sequences that implement dynamically corrected gates. These sequences are derived under the assumption of
quasistatic noise (i.e., fluctuations in $J$ and $h$ are treated as constant over the duration of the gate), but still work
even when time dependence of the noise is included, assuming that the high-frequency components are sufficiently small.  The
first set of corrected sequences we consider, the ``corrected unoptimized'' (CUO) set, is based on traditional pulse sequences,
and is the set outlined in Ref.\ \onlinecite{WangPRA2014}.  The uncorrected pulse sequences that this set is derived from assume
that $x$ rotations can be performed with a single pulse; $z$ rotations are then performed using the Hadamard-$x$-Hadamard sequence,
and all others with an $x$-$z$-$x$ decomposition.  The other corrected set is the ``corrected optimized, type II'' (CO-II) set.
These are derived from our UCO-II pulse sequences and offer two major advantages.  First of all, the sequences that the CUO set
is based on can be up to seven pulses long, while none of the UCO-II sequences are longer than five, and thus the UCO-II sequences
are already faster.  Second of all, the uncorrected identities are faster (and, in some cases, consist of fewer pulses) than those
for the corresponding CUO sequence. The different sets of sequences are summarized in Table \ref{Tab:pulses}.
\begin{table*}
	\centering
		\begin{tabular}{| m{20mm} | m{40mm} | m{75mm} |}
		\hline
			\centering Acronym & \centering Full name & {\centering Meaning} \\
		\hline
		\hline
			\centering UCUO & \centering UnCorrected UnOptimized & Standard decomposition of an arbitrary rotation into $z$ rotations (done by the Ramon sequence\cite{RamonPRB2011}) and $x$ rotations (constructed using the Hadamard-$z$-Hadamard sequence). \\
		\hline
			\centering UCO-I & \centering UnCorrected Optimized, type I & Similar to the UCUO sequences, except that $z$ rotations are performed using the
$\theta$-$2\theta$-$\theta$ sequence as proposed in Ref.\ \onlinecite{ZhangPRL2017}. \\
		\hline
			\centering UCO-II & \centering UnCorrected Optimized, type II & Still uses the $\theta$-$2\theta$-$\theta$
sequence to perform $z$ rotations, but includes new sequences to be detailed in Sec.~\ref{sec:UCOII}. \\
		\hline
			\centering CUO & \centering Corrected UnOptimized & Traditional \textsc{supcode} sequences correcting noise errors, as outlined in Ref.\ \onlinecite{WangPRA2014}. \\
		\hline
			\centering CO-II & \centering Corrected Optimized, \newline type II & Sequences derived from UCO-II and capable of correcting noise errors. \\
		\hline
		\end{tabular}
	\caption{Summary of different sets of sequences considered in this paper.}
	\label{Tab:pulses}
\end{table*}

Having proposed these five sets of pulse sequences, we now wish to evaluate their relative merits.  The main part
of this analysis is a determination of how much time is required to execute the sequences.  We will show that the
UCO-II sequences can be made faster than any of the other sets of uncorrected sequences, due in large part to
the fact that many of the UCO-II sequences consist of fewer pulses than their UCUO and UCO-I counterparts and to the
fact that these sequences can be made faster through adjustment of parameters.  We show that the UCO-I sequences
can be up to three times faster than the UCUO sequences, while the UCO-II sequences are up to twice as fast as the
UCO-I sequences, all depending on the gate in question.  We also compare our two sets of corrected sequences, and
find that the CO-II sequences are up to twice as fast as their CUO counterparts.  There are some advantages, however,
that the UCUO and CUO sequences have over their faster counterparts.  Due to the fact that they contain pure $x$
rotations (during which $J$ is set to zero), they turn out to be resistant to the effects of charge noise, and thus
work very well in systems with low magnetic noise, such as isotopically-purified Si.  Unfortunately, these sequences
are incompatible with existing experimental systems for singlet-triplet qubits (at least those based on two electrons
in a double quantum dot), as it is very challenging to completely turn off the exchange coupling.

We then proceed with a randomized benchmarking\cite{MagesanPRL2011,MagesanPRA2012} analysis of these pulse sequences.
In such an analysis, one generates random sequences of Clifford gates of a given length and determines the average
fidelity of these sequences.  The single-qubit Clifford gates are the set of rotations that map the three Cartesian
coordinate axes, $x$, $y$, and $z$, onto themselves; there are $24$ such gates in all.  Performing these simulations
for varying numbers of gates allows one to determine valuable information about the qubit, such as the effective
decoherence time $T_2$ and the number of gates that one can perform before the loss of information in the qubit reaches
an unacceptable level.  In this analysis, we will consider both (Gaussian) quasistatic and $1/f^\alpha$ noise.  We
also compare barrier control and tilt control \cite{ReedPRL2016,MartinsPRL2016,ZhangPRL2017}, which correspond to
different amounts of charge noise, with the amount present when using barrier control an order of magnitude smaller
than when using tilt control\cite{MartinsPRL2016}.

In the case of quasistatic noise, we find that, among the uncorrected sequences, the UCUO sequences result in the longest
$T_2$ by an order of magnitude in the absence of field noise.  This results from the fact that there are segments of these
sequences during which the exchange coupling is completely turned off.  The amount of charge noise in the system, quantified
by the standard deviation of the distribution that we draw the exchange coupling from, is proportional to the mean exchange
coupling \cite{ReedPRL2016,MartinsPRL2016,BarnesPRB2016}, so that setting the exchange coupling to zero results in zero noise.
As a result, the UCUO sequences are resistant to the effects of charge noise.  An immediate corollary is that, in systems
with no field noise, the UCUO sequence set is the optimal dynamical decoupling technique for singlet-triplet qubits under
quasistatic charge noise---in fact, this remains true even for dynamic charge noise as discussed in the next paragraph.  On
the other hand, the exchange coupling is never turned off in the UCO-I and UCO-II sequences, and thus they are subject to
the effects of charge noise for their entire durations. When we introduce field noise at levels typical of GaAs experiments,
on the other hand, the UCUO sequences have the shortest $T_2$ of all the uncorrected sequences, due to the fact that they
take the longest to perform and to the fact that they are now subject to noise-induced error for their entire durations.  We
observe similar results with the dynamically-corrected sequence sets: in the absence of field noise, the CUO sequences result
in a longer $T_2$ than the CO-II sequences.  This is again due to the fact that the CUO sequences have segments during which
the exchange coupling is completely turned off.  However, when sufficiently strong field noise is present, the CO-II sequences
have longer $T_2$.  We also consider the fidelities of these sets as a function of the number of gates, and find similar results.

For the case of $1/f^\alpha$ noise, we first consider the case of $1/f^{2.6}$ field noise and $1/f^{0.7}$ charge noise,
roughly corresponding to the experimentally measured power spectra \cite{DialPRL2013,MedfordPRL2012}.  Among the uncorrected
pulse sequences, we find, once again, that the UCUO sequences result in the best $T_2$ values when there is no field
noise present, while the UCO-II sequences are best when field noise is present.  We also find that, regardless of whether
or not field noise is present, the CUO sequences give the better $T_2$ values, compared with the CO-II sequences.  This
latter result is likely due to the fact that the corrected pulse sequences are longer than the uncorrected sequences and
that the charge noise possesses significant high-frequency components, which are more detrimental to CO-II.  When we look at the fidelities as a function of
the number of gates, however, we find that, in the case of barrier control and in the presence of field noise, we find
that we can perform about the same number of gates using either of the corrected sequences.

We then vary the exponents characterizing the two types of noise, $\alpha_h$ and $\alpha_J$ for field and charge noise,
respectively.  We consider two cases here---one in which we fix $\alpha_h=2.6$ and vary $\alpha_J$, and one in which
we set $\alpha_h=\alpha_J=\alpha$ and vary $\alpha$.  We again consider both the uncorrected and corrected sequences,
as well as cases in which field noise is absent and those in which it is present.  In the case of $\alpha_h=\alpha_J=\alpha$,
we find that, as before, the UCUO sequences give the best $T_2$ of the uncorrected sequences and the CUO sequences perform
the best among the corrected sequences if field noise is turned off completely.  When field noise is present, however,
the situation becomes more complex---which set of pulse sequences is best depends on the value of $\alpha$ for both the
uncorrected and corrected sequences.  The general tendency is for the ``optimized'' sequences to have longer $T_2$ when
$\alpha_J$ is large, while the ``unoptimized'' sequences do better when $\alpha_J$ is small.  We also see similar complex
behavior when we fix $\alpha_h=2.6$ and vary $\alpha_J$.

We also provide similar plots of the fidelity as a function of the number of gates, in order to help quantify how many
gates one can reliably perform for different amounts of noise, using each set of pulse sequences.  The results are
not qualitatively different from what we see from the plots as a function of time, though some of the points at which
a different sequence set becomes more advantageous are different; the values of $\alpha$ at which one begins to be able
to perform more gates using the ``optimized'' sequences are smaller than those at which they begin to have better $T_2$.
This is due to the fact that the ``optimized'' sequences take less time to perform, so that, even if $T_2$ is shorter
when using them, the shorter durations of the pulse sequences allow one to perform more gates within this shorter time
span.  Thus, a long $T_2$ is neither necessary nor sufficient; what matters is the precise gating pulse sequence determining
the dimensionless coherence time measured in units of the gating time.

Overall, our results show that ``optimized'' sequences without any segments during which the exchange coupling is
turned off completely do best when there is field noise present, which would be the situation in GaAs and natural
Si, while the ``unoptimized'' sequences tend to do better in the absence of field noise, which is the case in isotopically-purified
Si.  This suggests that, if it were possible to completely turn off the exchange coupling, then the ``unoptimized''
pulse sequences will be more reliable in isotopically-purified Si.  Furthermore, we see that ``optimized'' sequences
do better when high-frequency components in the noise are small, such as in the quasistatic limit and when the exponents
$\alpha_h$ and $\alpha_J$ are large (i.e., $\alpha_{h,J}\gtrsim 1$) in the $1/f^\alpha$ model.

The rest of the paper is organized as follows.  Sec.\ \ref{Sec:PulseSequences} is dedicated to reviewing existing
pulse sequences in the literature, introducing our new pulse sequences, reviewing the \textsc{supcode} method, and
introducing dynamically-corrected versions of our (na\"ive) pulse sequences.  We discuss the relative merits of
these pulse sequences in Sec.\ \ref{Sec:Evaluation}.  In Sec.\ \ref{Sec:RBResults}, we evaluate our na\"ive and
corrected sequences using randomized benchmarking.  Finally, we summarize our conclusions in Sec.\ \ref{Sec:Conclusion}.

\section{Pulse sequences} \label{Sec:PulseSequences}
As pointed out in the introduction, it is only possible to implement rotations about axes in the $xz$ plane at angles
$\theta_{min}<\theta<\frac{\pi}{2}$ with respect to the $z$ axis with single pulses, due to the form of the Hamiltonian, Eq.\
\eqref{Eq:Hamiltonian}, and experimental constraints on the exchange coupling.  Therefore, we need to implement all
other rotations using sequences of multiple pulses.  We now review existing pulse sequences found in the literature
and introduce new sequences.  Let $R(\vec{n},\phi)$ denote the matrix describing a rotation by an angle $\phi$ about
an axis given by the normal vector $\vec{n}$, i.e.,
\begin{equation}
R(\vec{n},\phi)=e^{-i\phi\vec{n}\cdot\vec{\sigma}/2}, \label{Eq:RotFormula}
\end{equation}
where $\vec{\sigma}$ is the vector of Pauli matrices, $\vec{\sigma}=\sigma_x\hat{\vec{x}}+\sigma_y\hat{\vec{y}}+\sigma_z\hat{\vec{z}}$.
We then see that the time evolution operator, Eq.\ \eqref{Eq:TimeEvolution}, performs a rotation
about the axis given by
\begin{equation}
\vec{n}=\frac{h}{\sqrt{h^2+J^2}}\hat{\vec{x}}+\frac{J}{\sqrt{h^2+J^2}}\hat{\vec{z}}
\end{equation}
and by an angle
\begin{equation}
\phi=2t\sqrt{h^2+J^2}.
\end{equation}
We will also make use of the shorthand, $R(\theta,\phi)$, to denote rotations about the axis given
by $\vec{n}=\sin{\theta}\hat{\vec{x}}+\cos{\theta}\hat{\vec{z}}$.  In this case, the time required
to perform the rotation is given by
\begin{equation}
ht=\tfrac{1}{2}\phi\sin{\theta}. \label{Eq:PulseTiming}
\end{equation}
Experimental restrictions require that $\theta_{min}<\theta<\frac{\pi}{2}$.

\subsection{Uncorrected unoptimized (UCUO) pulse sequences}
In this work, we will be comparing the performance of five different sets of pulse sequences, which we
will now detail.  We begin with what will henceforth be referred to as the ``uncorrected unoptimized''
(UCUO) sequences.  These are the ``unoptimized'' sequences employed in Ref.\ \onlinecite{ZhangPRL2017},
which we will review here. Since $J\le J_{max}$, it is not possible to implement a pure $z$ rotation with a single pulse, and composite pulses must be used instead. The UCUO sequences are based on the following $z$ rotation sequence proposed by Guy
Ramon\cite{RamonPRB2011} (note that our notation differs from that of the original paper):
\begin{equation}
R(\hat{\vec{z}},\phi)=R(\theta,\chi)R(\hat{\vec{x}},\alpha)R(\theta,\chi), \label{Eq:ZRotGR}
\end{equation}
and may be summarized as follows.
\begin{enumerate}
	\item {Rotate by angle $\chi$ about an axis at an angle $\theta$ with respect to the $z$ axis.}
	\item {Rotate by angle $\alpha$ about the $x$ axis.}
	\item {Repeat the first rotation.}
\end{enumerate}
If we now multiply out the right-hand side and set both sides equal, we find two solutions for
$\alpha$; one is
\begin{equation}
\alpha=-2\arcsin\left [\tan{\theta}\sin\left (\frac{\phi}{2}\right )\right ] \label{Eq:AlphaSoln1}
\end{equation}
and the other is
\begin{equation}
\alpha=2\pi+2\arcsin\left [\tan{\theta}\sin\left (\frac{\phi}{2}\right )\right ]. \label{Eq:AlphaSoln2}
\end{equation}
The solutions for $\chi$ are
\begin{widetext}
\begin{equation}
\chi=\arccos\left\{\frac{\pm\cos\left (\frac{\phi}{2}\right )\sqrt{1-\sin^2\left (\frac{\phi}{2}\right )\tan^2{\theta}}-\sin^2\left (\frac{\phi}{2}\right )\sin^2{\theta}}{\cos^2\left (\frac{\phi}{2}\right )+\sin^2\left (\frac{\phi}{2}\right )\cos^2{\theta}}\right\}, \label{Eq:ChiSoln}
\end{equation}
\end{widetext}
where we use the plus sign for $\pm$ if we use Eq.\ \eqref{Eq:AlphaSoln1} for $\alpha$ and the minus
sign if we use Eq.\ \eqref{Eq:AlphaSoln2}.  Which solution we choose depends on the sign of
$\phi$.  If $\phi$ is positive, then we use Eq.\ \eqref{Eq:AlphaSoln1} for $\alpha$, and we use
Eq.\ \eqref{Eq:AlphaSoln2} if $\phi$ is negative.  We can see from Eqs.\ \eqref{Eq:AlphaSoln1} and
\eqref{Eq:AlphaSoln2} that we can only obtain real values of $\alpha$ for any value of $\phi$ if
we restrict $\theta$ to the interval, $\left (0,\frac{\pi}{4}\right )$.

We note, however, that the value of $\alpha$ that we obtain from Eq.\ \eqref{Eq:AlphaSoln1} is negative.
Because $J$ and $h$ are restricted to be positive, we cannot perform a rotation with negative $\alpha$.
To fix this problem, we can add $2\pi$ to the obtained value of $\alpha$, at the expense of introducing
an overall minus sign on top of peforming the $z$ rotation by $\phi$:
\begin{equation}
R(\hat{\vec{z}},\phi)=-R(\theta,\chi)R(\hat{\vec{x}},\alpha+2\pi)R(\theta,\chi).
\end{equation}

\begin{figure}[ht]
\includegraphics[width=0.49\columnwidth]{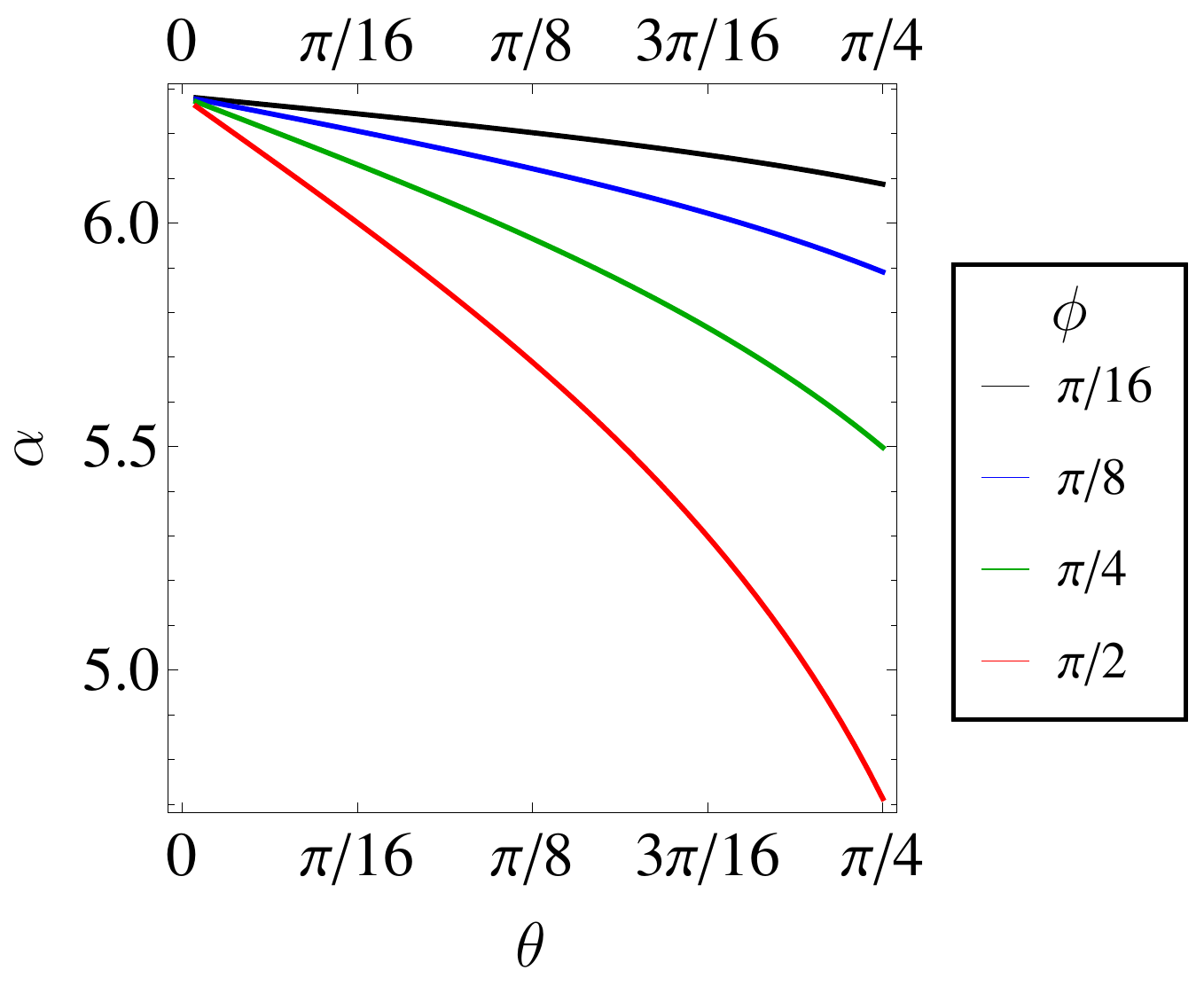}
\includegraphics[width=0.49\columnwidth]{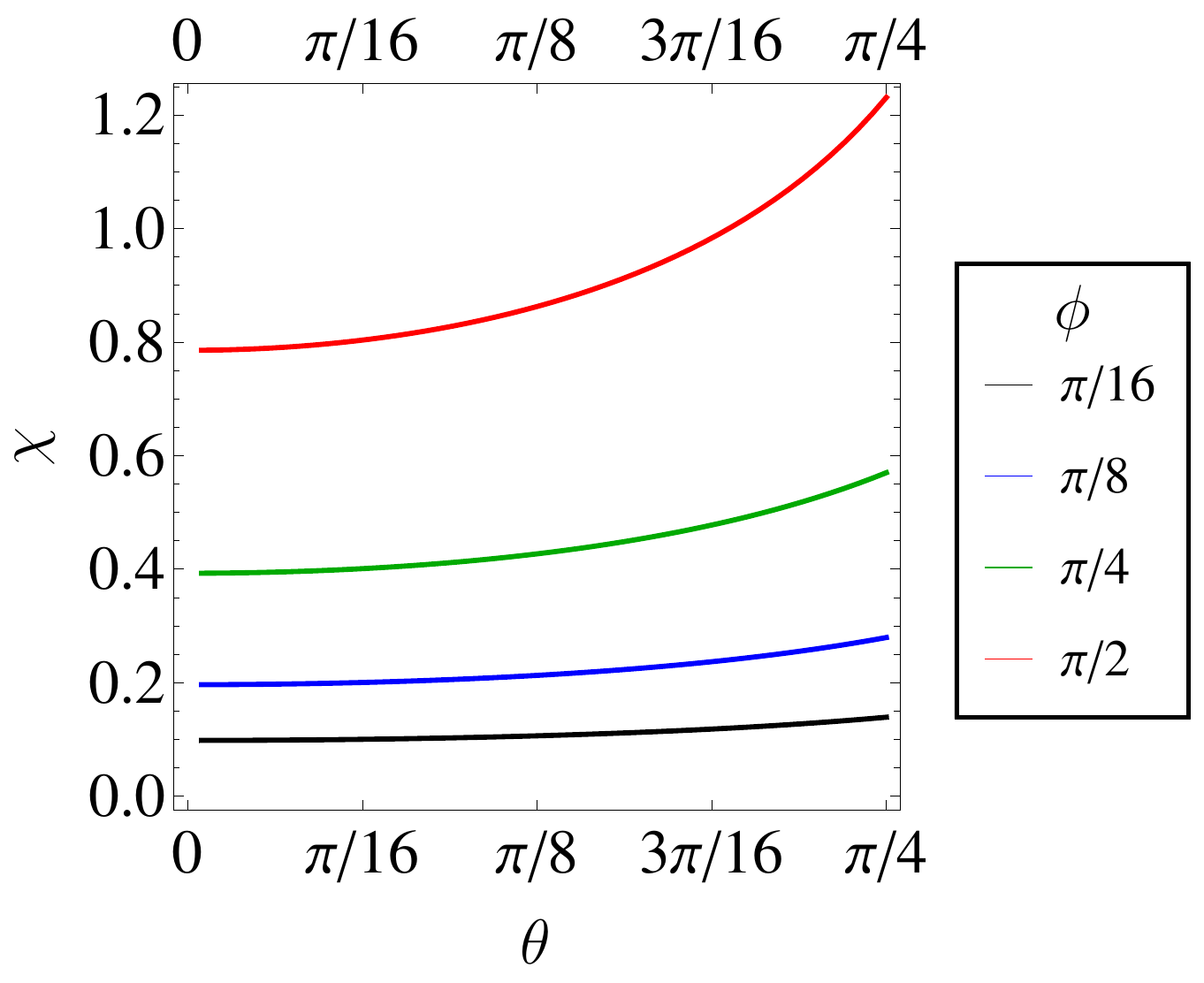}
\caption{Plots of $\alpha$ (left) and $\chi$ (right) as functions of $\theta$ for the Ramon sequence,
Eq.\ \eqref{Eq:ZRotGR}, for positive values of $\phi$.  Note that we add $2\pi$ to the (negative) value
of $\alpha$ obtained from Eq.\ \eqref{Eq:AlphaSoln1}; this results in an overall minus sign on top of
the rotation.}
\label{Fig:AlphaAndChi_PiOver2}
\end{figure}

Let us consider four values of $\phi$, namely, $\pi/2$, $\pi/4$, $\pi/8$, and $\pi/16$, as examples, and
determine $\chi$ and $\alpha$ as functions of $\theta$.  As noted earlier, the value of $\alpha$ that
we obtain from Eq.\ \eqref{Eq:AlphaSoln1} is negative, and thus we must add $2\pi$ to the obtained value.
This is exactly what we do, and we plot the resulting values of $\alpha$ and $\chi$ in Fig.\ \ref{Fig:AlphaAndChi_PiOver2}.

\begin{figure}[ht]
\includegraphics[width=0.49\columnwidth]{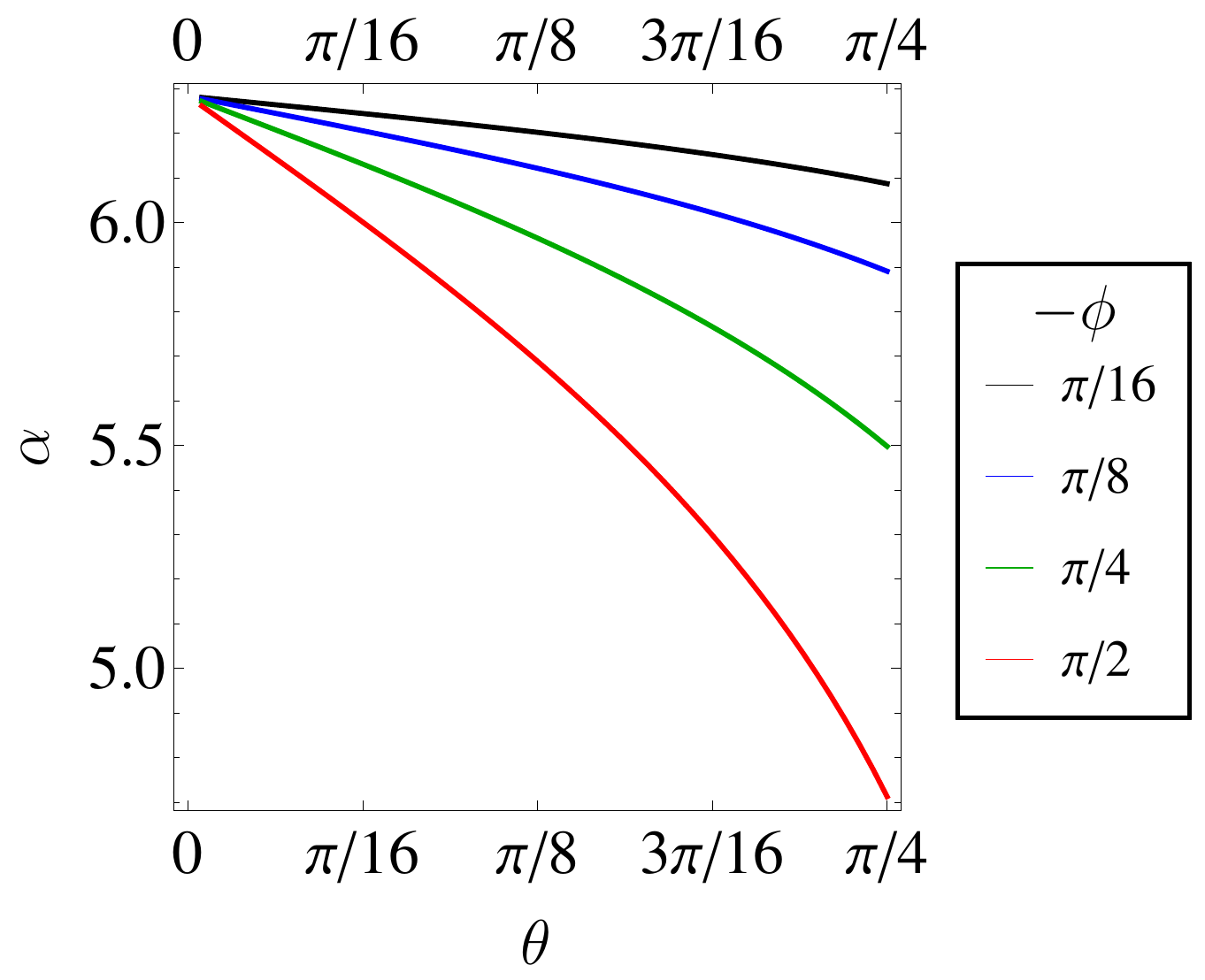}
\includegraphics[width=0.49\columnwidth]{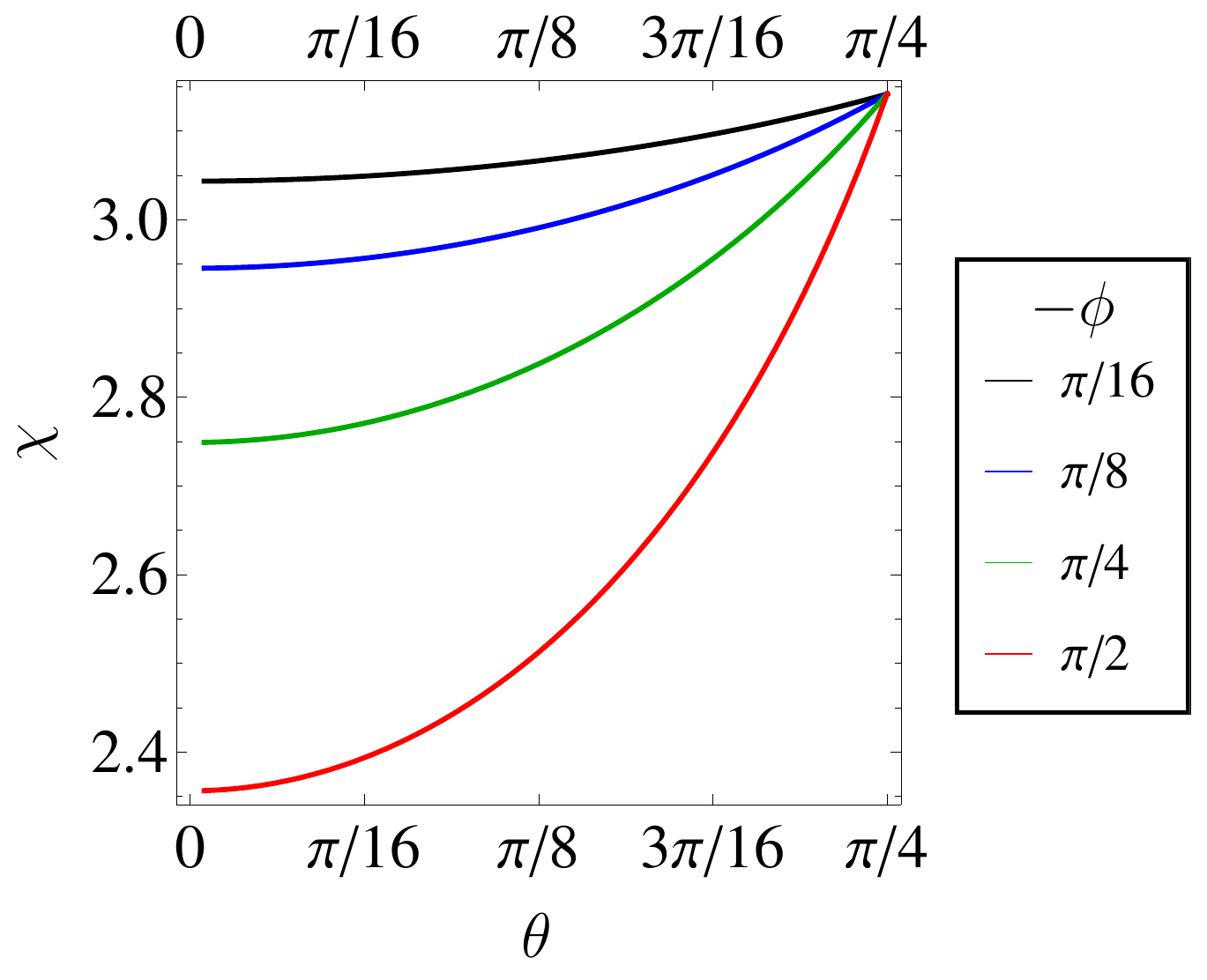}
\caption{Plots of $\alpha$ (left) and $\chi$ (right) as functions of $\theta$ for the Ramon sequence,
Eq.\ \eqref{Eq:ZRotGR}, for negative values of $\phi$.  Unlike the case of positive values of $\phi$,
we do not need to add $2\pi$ to the value of $\alpha$ obtained from Eq.\ \eqref{Eq:AlphaSoln2}, and thus
we perform a proper $z$ rotation.}
\label{Fig:AlphaAndChi_PiOver2_NegPhi}
\end{figure}

We also consider four negative values of $\phi$: $-\pi/2$, $-\pi/4$, $-\pi/8$, and $-\pi/16$.  In
this case, as stated earlier, $\alpha$ is now given by Eq.\ \eqref{Eq:AlphaSoln2} and we choose the minus
sign in the formula for $\chi$.  This time, we obtain a positive value for $\alpha$, and thus we do not
have to add $2\pi$ to it.  Therefore, the Ramon sequence performs a proper $z$ rotation; we do not acquire
an overall minus sign.  We plot our results in Fig.\ \ref{Fig:AlphaAndChi_PiOver2_NegPhi}.

To implement other rotations, one approach is to use the standard $z$-$x$-$z$ decomposition for arbitrary
rotations\cite{NielsenBook}:
\begin{equation}
R(\vec{n},\phi)=R(\hat{\vec{z}},\phi_3)R(\hat{\vec{x}},\phi_2)R(\hat{\vec{z}},\phi_1),
\end{equation}
where the angles $\phi_1$, $\phi_2$, and $\phi_3$ depend on the rotation that one wishes to perform.

We note a problem with this set of pulse sequences---they require that one be able to completely
turn off the exchange coupling, which is challenging in existing experimental setups.  However, if it
were possible to do so (for example, with singlet-triplet qubits defined in multi-electron quantum dots \cite{MartinsArXiv2017}), then this fact will actually prove advantageous when dealing with noise, as
we will see later.  The strength of the charge noise in our system, quantified as noise in the exchange
coupling, is roughly proportional to the exchange coupling itself, and thus the UCUO pulse sequences
will be largely immune to charge noise.

\subsection{Uncorrected optimized pulse sequences, type I (UCO-I)}
We will now give the ``uncorrected optimized'' (UCO) pulse sequences.  The first such set,
``type I'' (UCO-I), is the same as the UCUO set, except that we perform $z$ rotations using
a generalization of the Hadamard-$x$-Hadamard sequence that we will call the ``$\theta$-$2\theta$-$\theta$''
sequence\cite{ZhangPRL2017}:
\begin{equation}
R(\hat{\vec{z}},\phi)=-R(\theta,\pi)R(2\theta,\phi)R(\theta,\pi). \label{Eq:ZRotRT}
\end{equation}
One may verify this by using the identity,
\begin{equation}
R(\vec{n},\phi)=\cos\left (\frac{\phi}{2}\right )-i\sin\left (\frac{\phi}{2}\right )\vec{n}\cdot\vec{\sigma},
\end{equation}
and the fact that the Pauli matrices are mutually anticommuting.  We thus realize a $z$ rotation
(with an overall minus sign) by an angle $\phi$ via the following sequence:

\begin{enumerate}
	\item {Rotate by angle $\pi$ about an axis at an angle $\theta$ with respect to the $z$ axis.}
	\item {Rotate by angle $\phi$ about an axis at an angle $2\theta$ with respect to the $z$ axis.}
	\item {Repeat the first rotation.}
\end{enumerate}

The restriction on $\theta$ imposed by the restrictions on the Hamiltonian parameters requires that
$\theta_{min}<\theta<\frac{\pi}{4}$.

We also present here the values of $J$ needed for each type of rotation.  If we take $J_1$ to
be the value of $J$ used for the rotations by $\pi$, then the value of $J$ needed for the
rotation by $\phi$, which we denote by $J_2$, is
\begin{equation}
J_2=\frac{J_1^2-h^2}{2J_1},
\end{equation}
which may be verified via the trigonometric identity,
\begin{equation}
\cot{2\theta}=\frac{\cot^2{\theta}-1}{2\cot{\theta}}.
\end{equation}
As in Ref.\ \onlinecite{ZhangPRL2017}, $x$ rotations are performed using the Hadamard-$z$-Hadamard sequence in order to avoid switching off $J$,
and all other rotations are performed via the $z$-$x$-$z$ decomposition, with each piece implemented by composite pulses

\subsection{Uncorrected optimized pulse sequences, type II (UCO-II)}\label{sec:UCOII}
We now introduce the second set of UCO pulse sequences, or ``type II'' (UCO-II).  Unlike the previous
two sets of pulse sequences, we do not simply introduce a new sequence for performing $z$ rotations and
then build the other sequences from it using standard decompositions.  We instead introduce new sequences
for performing these other rotations that are shorter and, as we will show in the next section, faster
than those obtained from said standard decompositions.  We still use the $\theta$-$2\theta$-$\theta$
sequence detailed above for $z$ rotations, but the other rotations are performed using the sequences
that we describe below.

\subsubsection{$x$ rotations}
First, we introduce a new sequence for performing $x$ rotations.  This sequence is based on a generalization
of the ``Ramon'' sequence for performing $z$ rotations:
\begin{equation}
R(\hat{\vec{x}},\phi)=R(\theta,\chi)R(\theta',\alpha)R(\theta,\chi). \label{Eq:XRotGGR}
\end{equation}
It may be summarized as follows:
\begin{enumerate}
	\item {Rotate by angle $\chi$ about an axis at an angle $\theta$ with respect to the $z$ axis.}
	\item {Rotate by angle $\alpha$ about an axis at an angle $\theta'$ with respect to the $z$ axis.}
	\item {Repeat the first rotation.}
\end{enumerate}
If we expand the right-hand side and solve for $\alpha$ and $\chi$, we obtain two possible values
of $\alpha$,
\begin{equation}
\alpha=-2\arcsin\left [\frac{\cos{\theta}}{\sin(\theta-\theta')}\sin\left (\frac{\phi}{2}\right )\right ] \label{Eq:AlphaSolnX1}
\end{equation}
and
\begin{equation}
\alpha=2\pi+2\arcsin\left [\frac{\cos{\theta}}{\sin(\theta-\theta')}\sin\left (\frac{\phi}{2}\right )\right ], \label{Eq:AlphaSolnX2}
\end{equation}
and two possible values of $\chi$,
\begin{widetext}
\begin{equation}
\chi=\arccos\left\{\frac{\pm\cos\left (\frac{\phi}{2}\right )\sqrt{\sin^2(\theta-\theta')-\cos^2{\theta}\sin^2\left (\frac{\phi}{2}\right )}-\sin^2\left (\frac{\phi}{2}\right )\sin{\theta}\cos{\theta}\cos(\theta-\theta')}{\left [\cos^2\left (\frac{\phi}{2}\right )+\sin^2\left (\frac{\phi}{2}\right )\sin^2{\theta}\right ]\sin(\theta-\theta')}\right\}. \label{Eq:ChiSolnX}
\end{equation}
\end{widetext}
We can deduce from Eqs.\ \eqref{Eq:AlphaSolnX1} and \eqref{Eq:AlphaSolnX2} that we can only obtain
real-valued solutions for $\alpha$ for arbitrary $\phi$ if $\theta\geq\frac{\theta'}{2}+\frac{\pi}{4}$.
We find by direct substitution back into these equations that, for positive values of $\phi$, we must
use Eq.\ \eqref{Eq:AlphaSolnX1} to obtain $\alpha$ and choose the plus sign in Eq.\ \eqref{Eq:ChiSolnX}.
On the other hand, we must use Eq.\ \eqref{Eq:AlphaSolnX2} for $\alpha$ and choose the minus sign in Eq.\
\eqref{Eq:ChiSolnX} if $\phi$ is negative.  Note that, under the stated constraint on $\theta$, we will
always obtain a negative value for $\alpha$ from Eq.\ \eqref{Eq:AlphaSolnX1}; this means that we must
add $2\pi$ to the value so obtained, at the expense of introducing an overall minus sign to the $x$
rotation that this sequence performs.

\begin{figure}[ht]
\includegraphics[width=0.49\columnwidth]{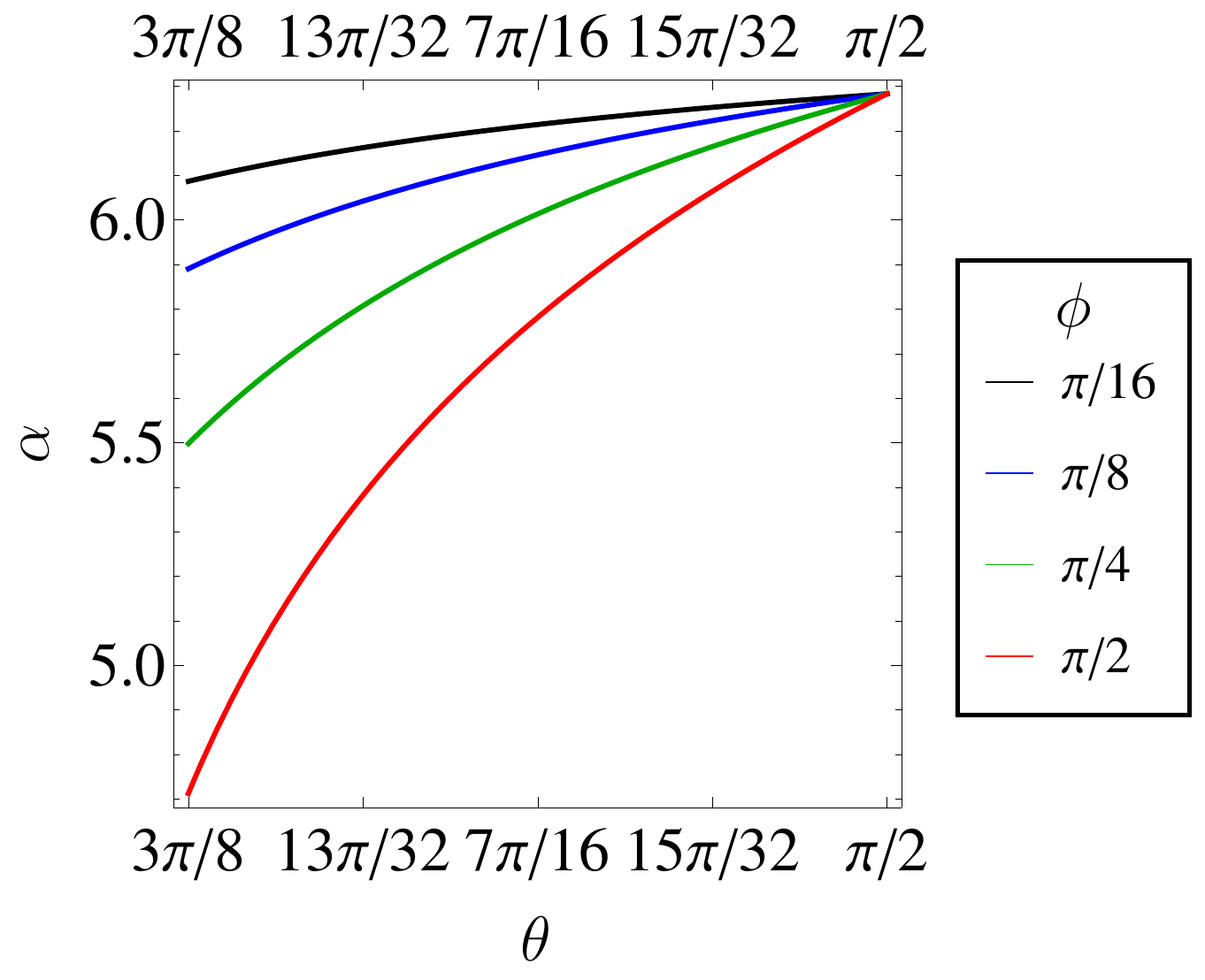}
\includegraphics[width=0.49\columnwidth]{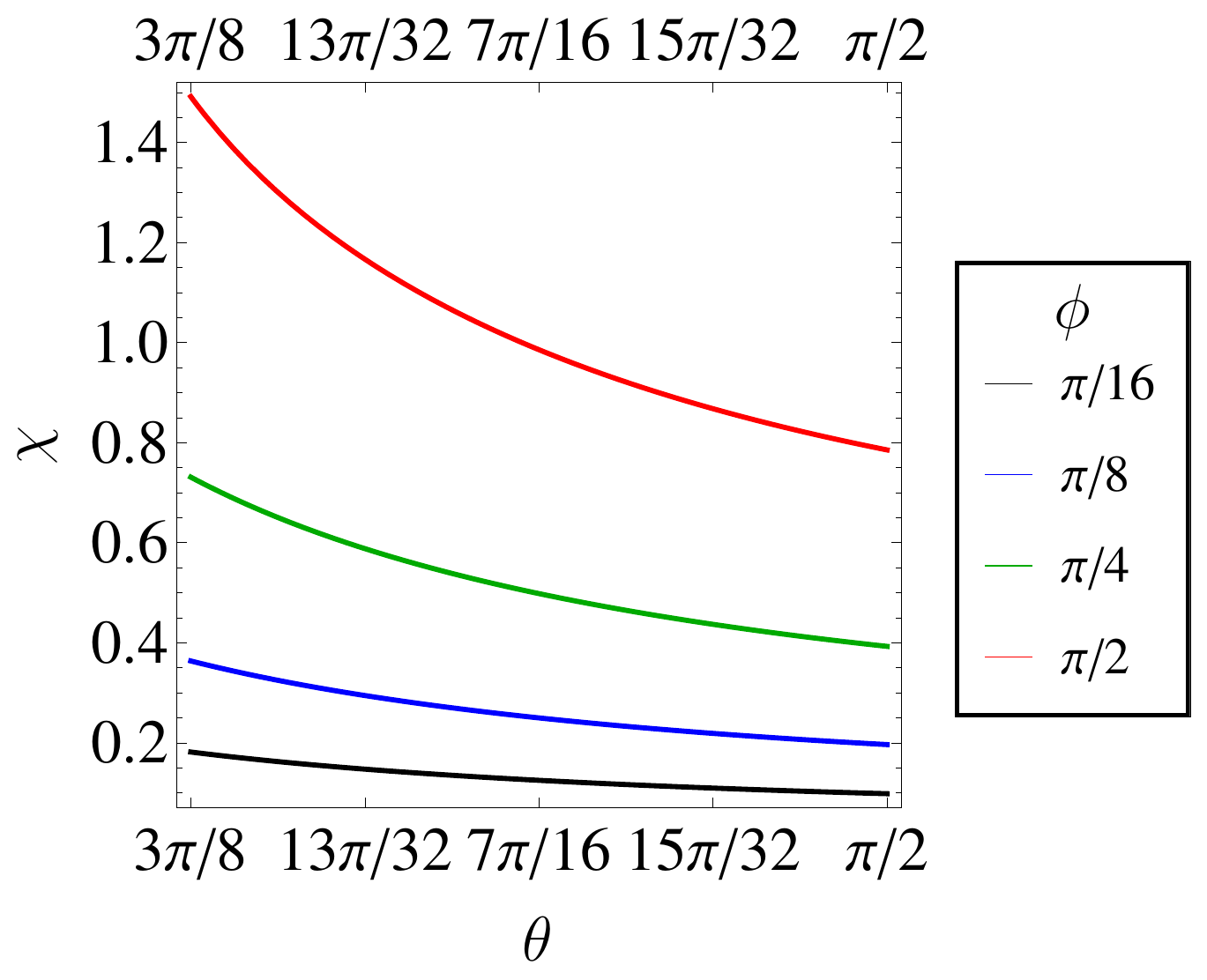}
\includegraphics[width=0.49\columnwidth]{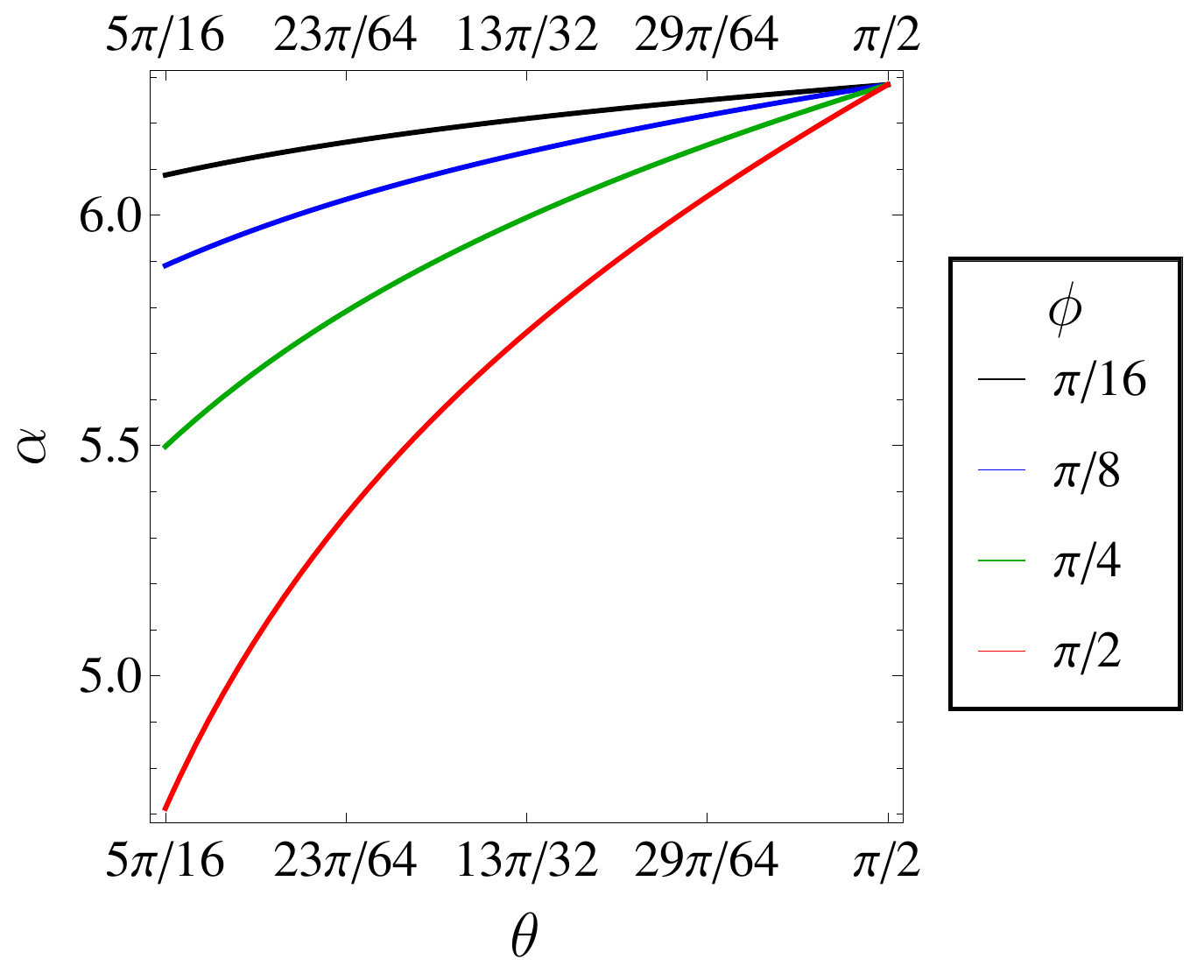}
\includegraphics[width=0.49\columnwidth]{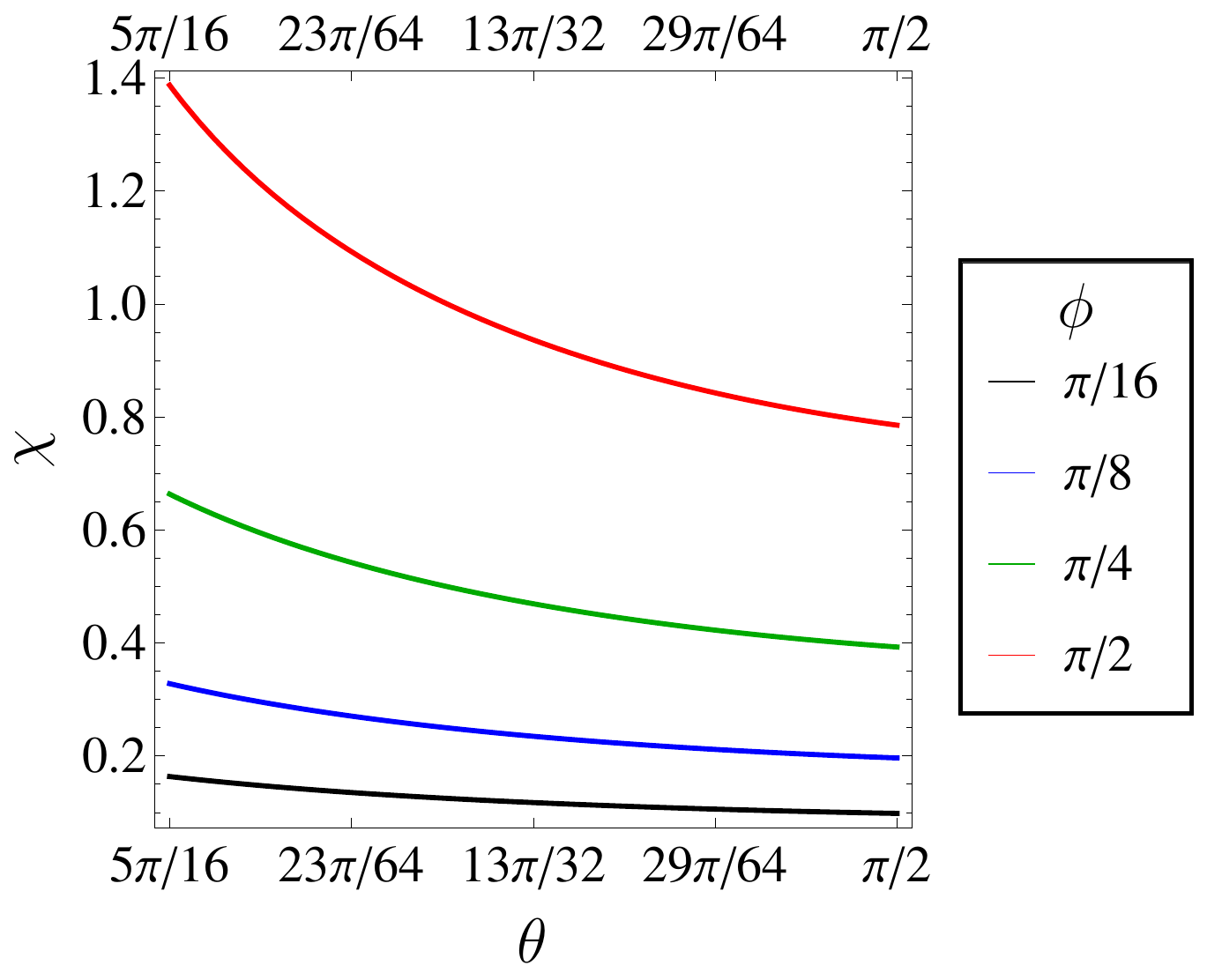}
\includegraphics[width=0.49\columnwidth]{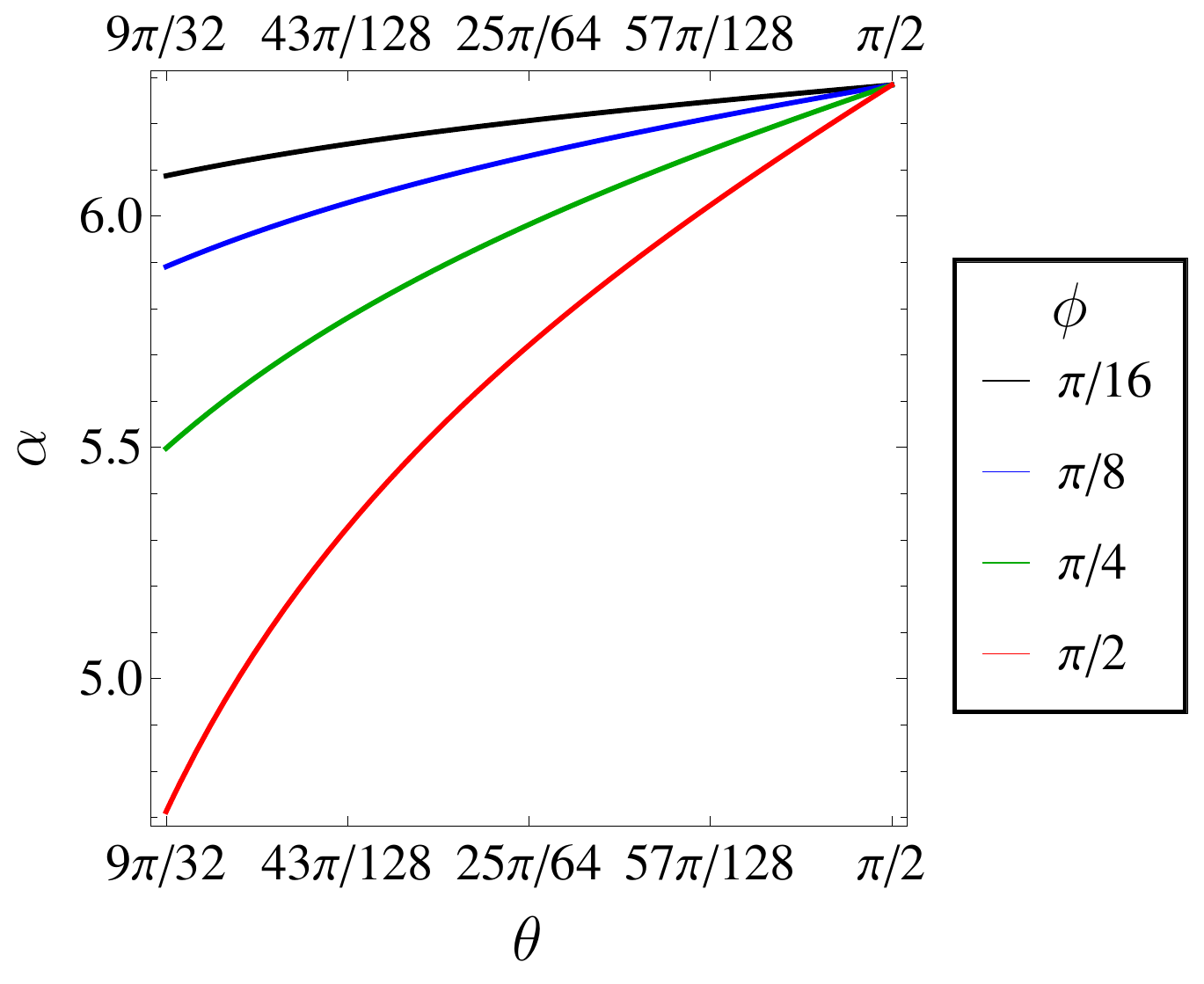}
\includegraphics[width=0.49\columnwidth]{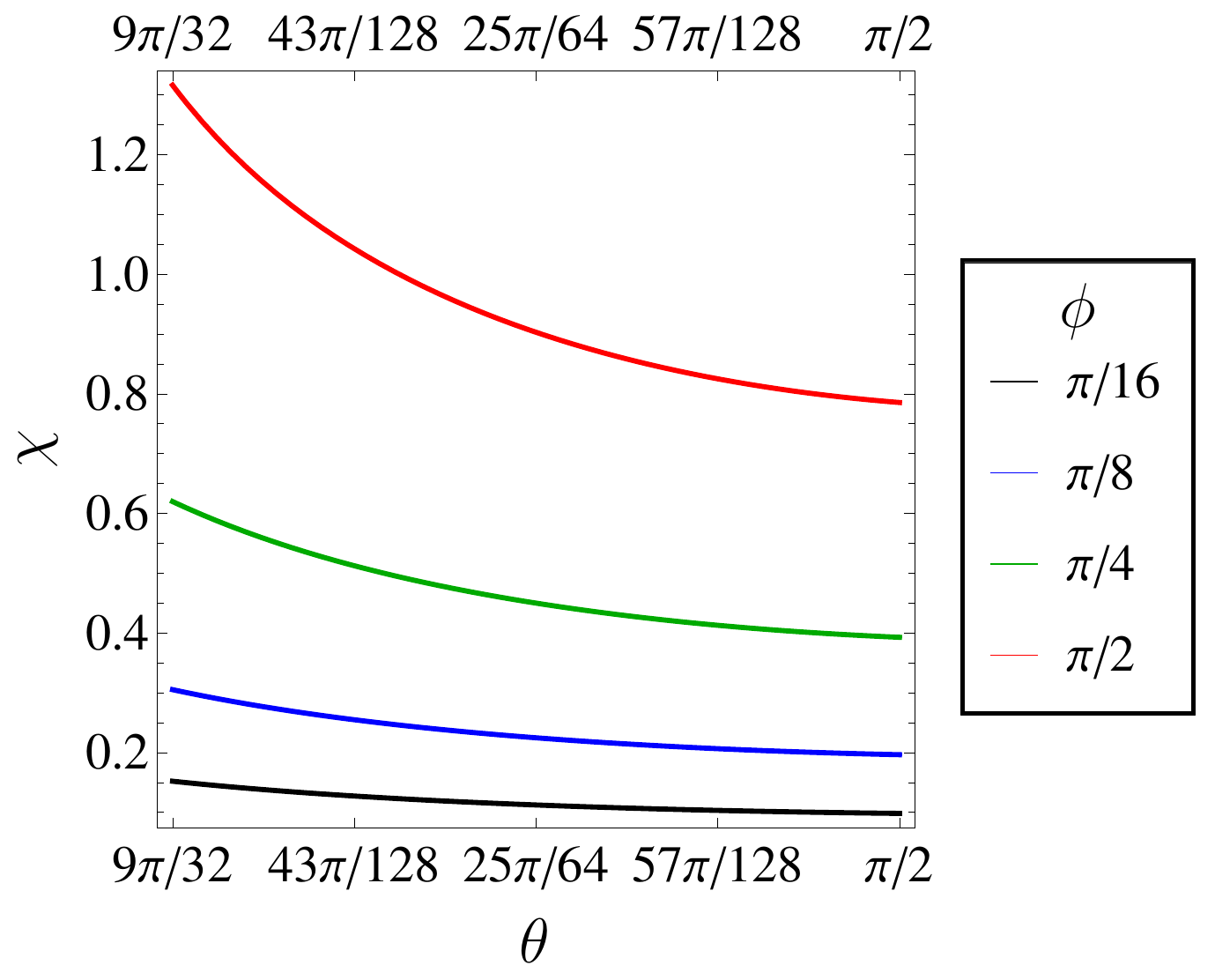}
\caption{Plots of $\alpha$ (left) and $\chi$ (right) for the modified Ramon sequence, Eq.\ \eqref{Eq:XRotGGR}
with $\theta'=\pi/4$ (top row), $\pi/8$ (middle row), and $\pi/16$ (bottom row) and for positive values of $\phi$.}
\label{Fig:AlphaAndChi_XRot}
\end{figure}

We will now consider four positive values of $\phi$, namely, $\frac{\pi}{2}$, $\frac{\pi}{4}$, $\frac{\pi}{8}$,
and $\frac{\pi}{16}$, as well as the corresponding negative values, and $\theta'=\frac{\pi}{4}$, $\frac{\pi}{8}$,
and $\frac{\pi}{16}$.  As pointed out before, we must add $2\pi$ to the values of $\alpha$ that we obtain from
Eq.\ \eqref{Eq:AlphaSolnX1}.  We plot our results for $\alpha$ and $\chi$ in Fig.\ \ref{Fig:AlphaAndChi_XRot}.

\begin{figure}[ht]
\includegraphics[width=0.49\columnwidth]{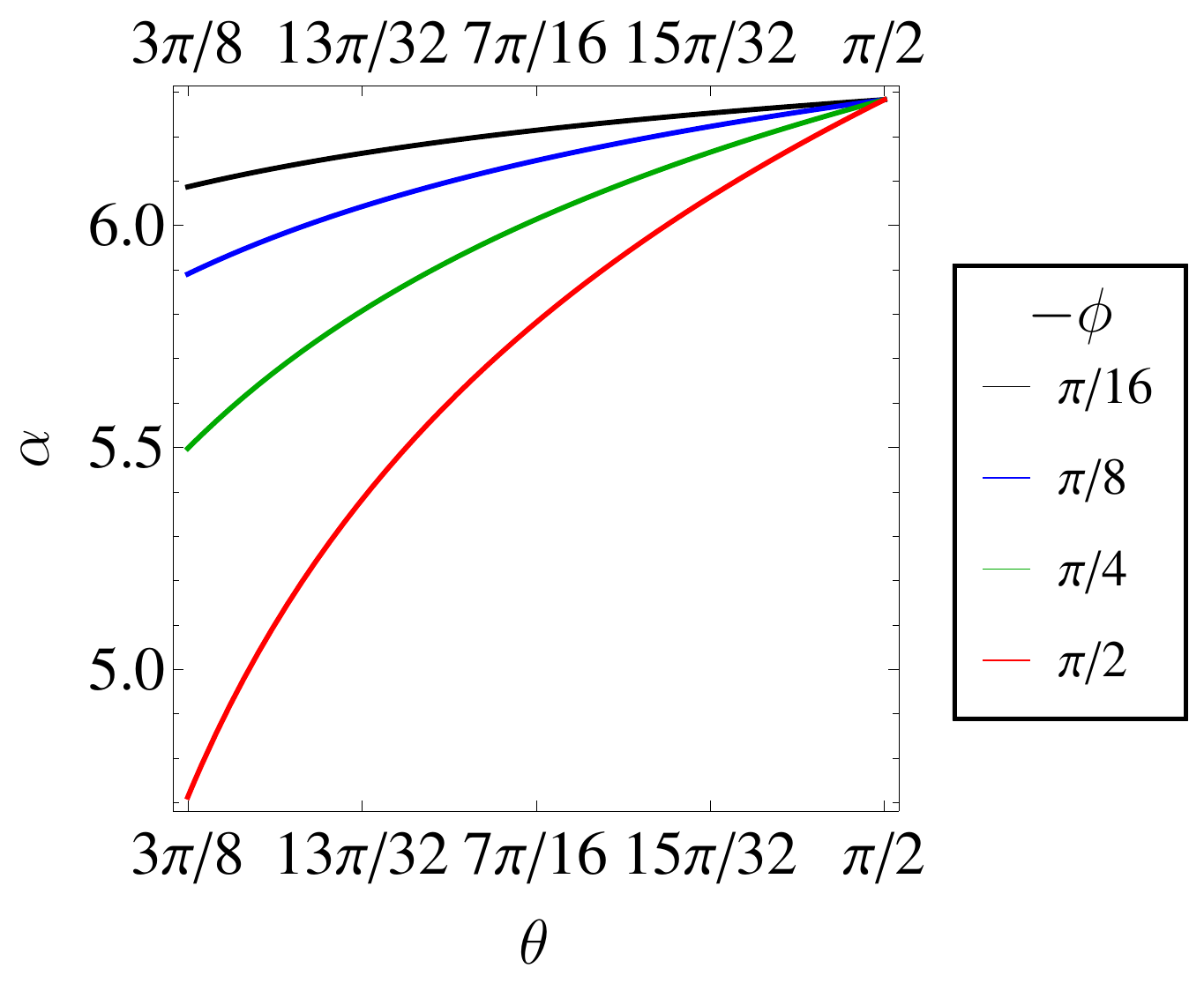}
\includegraphics[width=0.49\columnwidth]{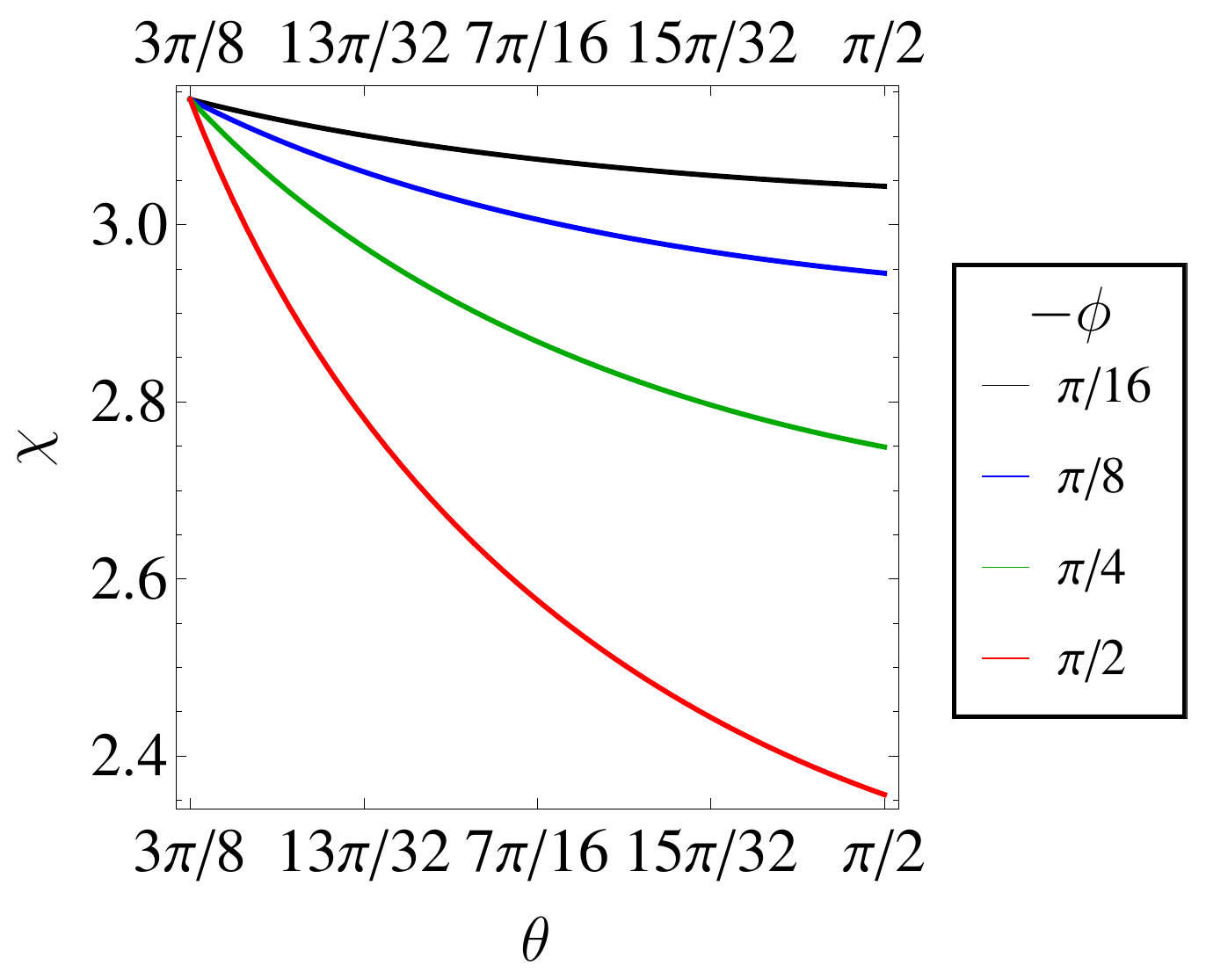}
\includegraphics[width=0.49\columnwidth]{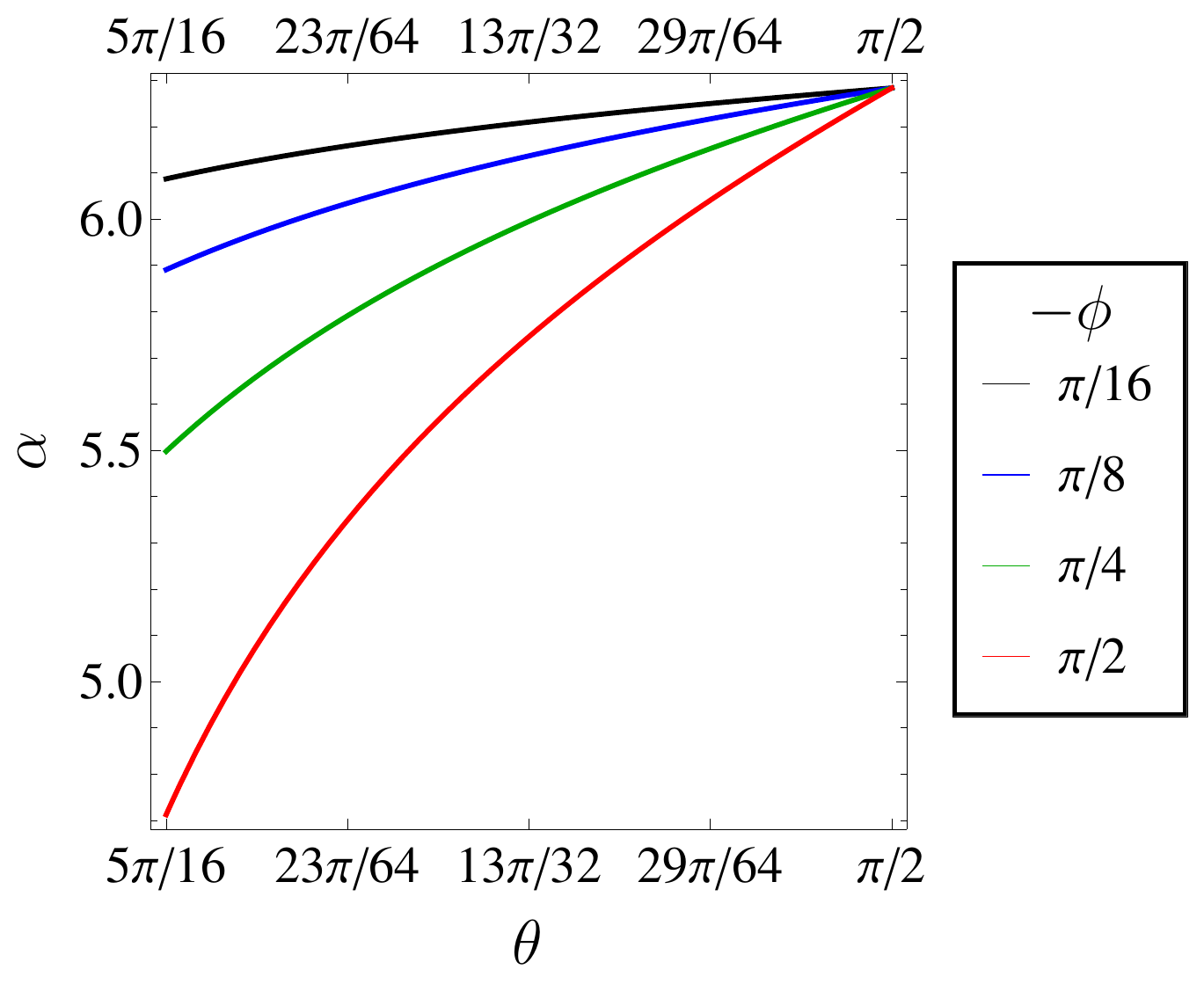}
\includegraphics[width=0.49\columnwidth]{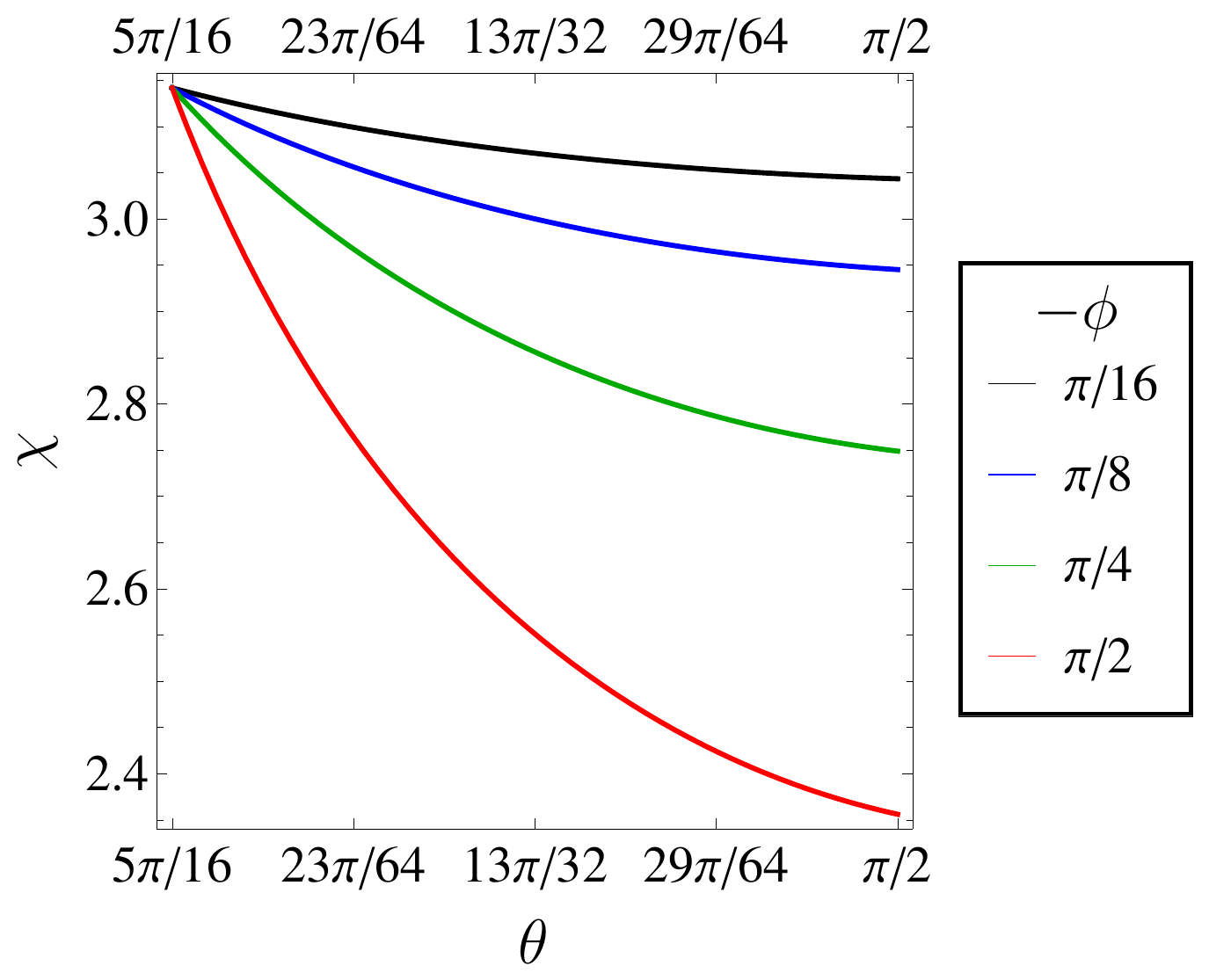}
\includegraphics[width=0.49\columnwidth]{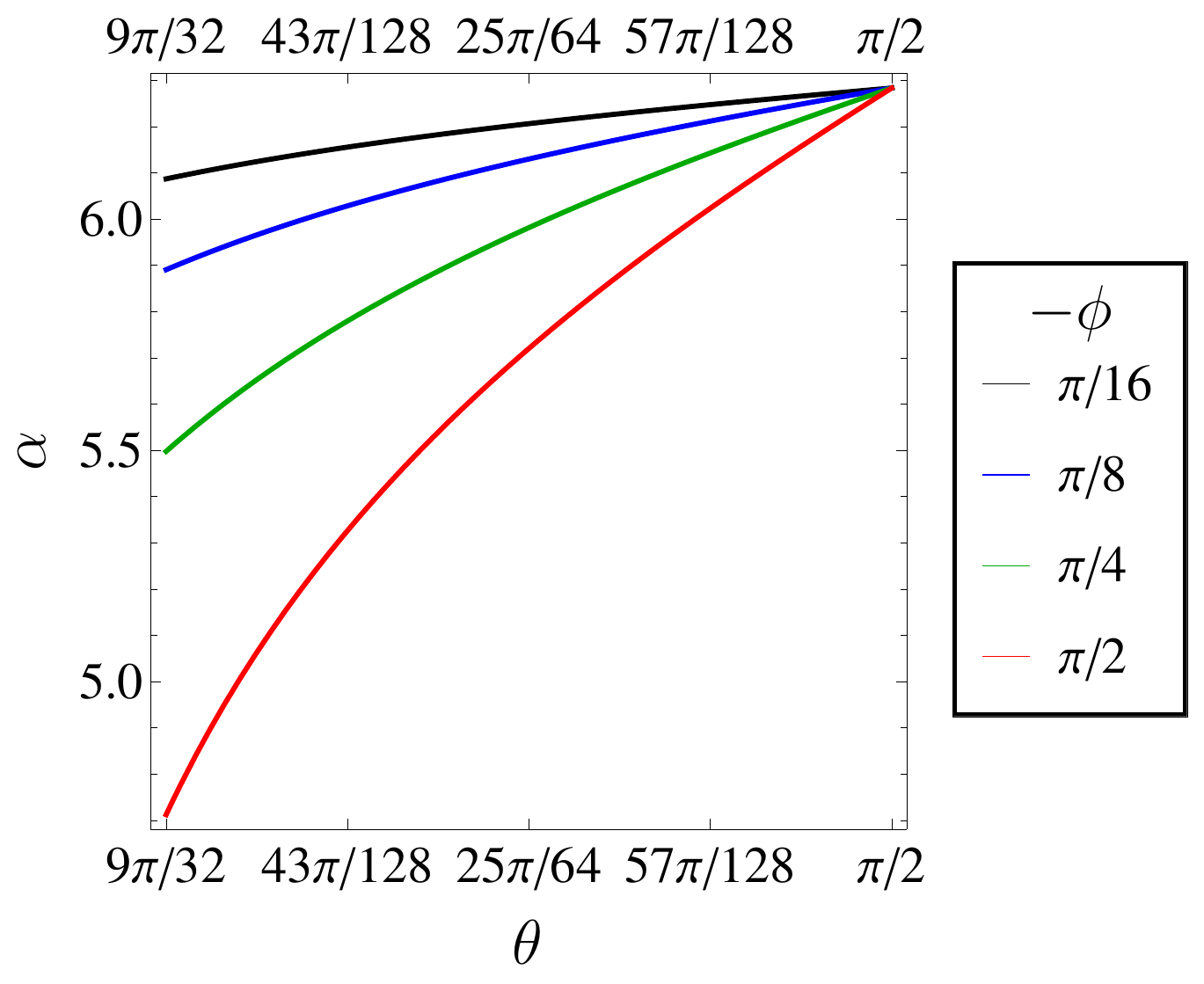}
\includegraphics[width=0.49\columnwidth]{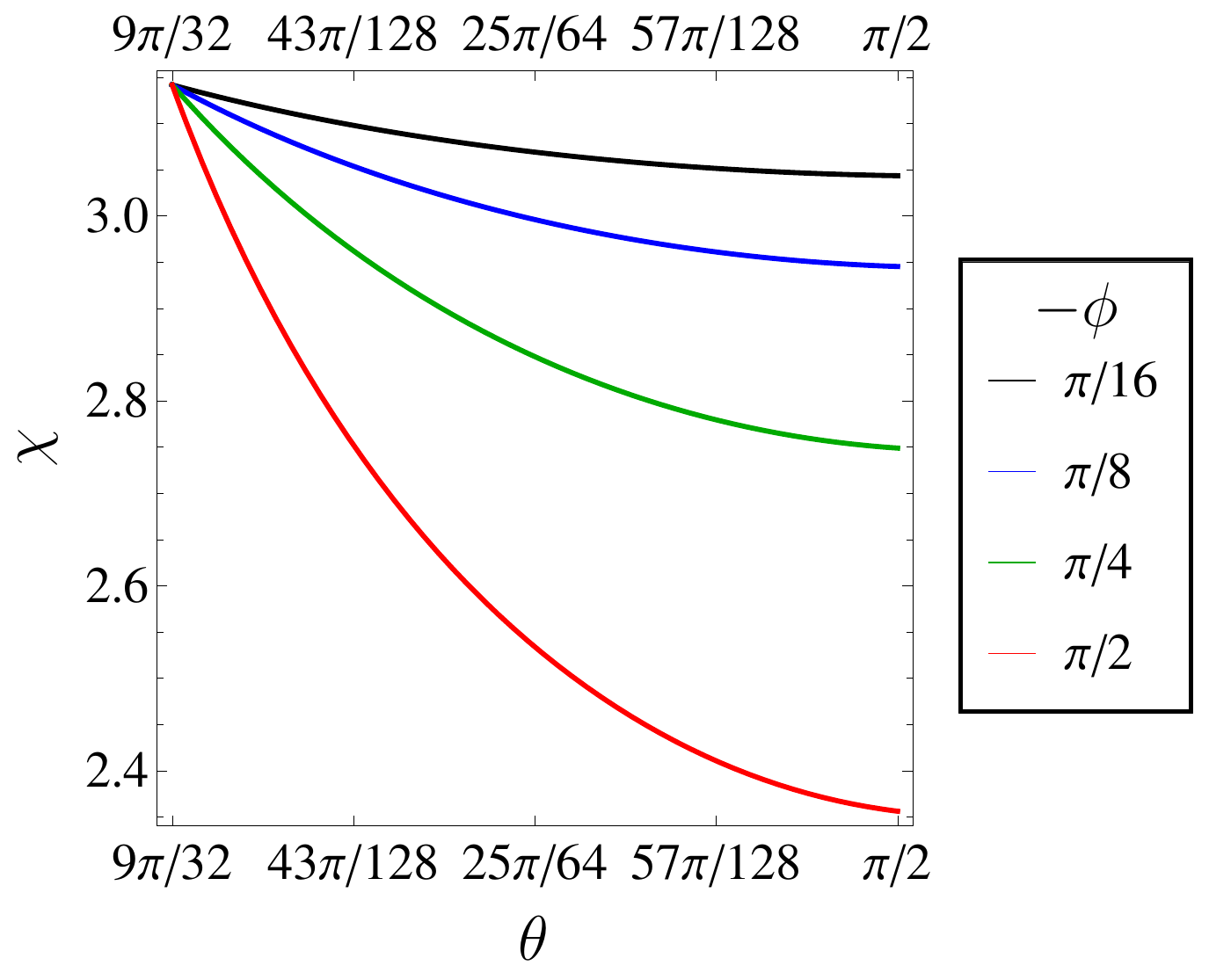}
\caption{Similar to Fig.\ \ref{Fig:AlphaAndChi_XRot}, but for negative values of $\phi$.}
\label{Fig:AlphaAndChi_XRot_NegPhi}
\end{figure}

Now let us consider rotations by the corresponding negative angles.  We plot the values of $\alpha$
and $\chi$ that we obtain in Fig.\ \ref{Fig:AlphaAndChi_XRot_NegPhi}.

\subsubsection{$y$ rotations}
We now give a sequence that performs a $y$ rotation.  We could, in principle, perform a $y$ rotation by
simply decomposing it into a $z$-$x$-$z$ sequence and using the above sequences for performing $z$ and
$x$ rotations, yielding a nine-pulse sequence.  However, it turns out that there is a shorter, five-pulse
sequence for performing $y$ rotations:
\begin{equation}
R(\hat{\vec{y}},\phi)=R\left (\theta_1,\frac{\pi}{2}\right )R(\theta_2,\pi)R(\theta_3,\phi)R(\theta_2,\pi)R\left (\theta_1,\frac{3\pi}{2}\right ).
\end{equation}
The angles $\theta_k$ are the angles with respect to the $z$ axis of the axes of rotation for
each rotation.  These angles are not independent---if we multiply out the rotation operators on
the right-hand side, we find that, in order to obtain a $y$ rotation, we must take
\begin{equation}
\theta_2=\frac{\theta_1+\theta_3}{2}+\frac{\pi}{4}.
\end{equation}
Since we require that $0<\theta_2<\frac{\pi}{2}$, we must necessarily take $\theta_1+\theta_3<\frac{\pi}{2}$.
We will show in the next section that this sequence takes less time to execute than the nine-pulse sequence.

\subsubsection{$x-z$ rotations}
We now consider a sequence for performing rotations about $\hat{\vec{x}}-\hat{\vec{z}}$.  It turns out that
this can be done, up to an extra minus sign, as a symmetric three-part sequence:
\begin{equation}
R(\theta_1,\pi)R(\theta_2,\phi)R(\theta_1,\pi)=-R(2\theta_1-\theta_2,-\phi).
\end{equation}
In order to perform a rotation about $\hat{\vec{x}}-\hat{\vec{z}}$, we set
\begin{equation}
2\theta_1-\theta_2=-\frac{\pi}{4},
\end{equation}
or
\begin{equation}
\theta_2=2\theta_1+\frac{\pi}{4}.
\end{equation}
We require that $0<\theta_2<\frac{\pi}{2}$, which implies that $0<\theta_1<\frac{\pi}{8}$.  The following sequence
thus performs a rotation by $-\phi$ about $\hat{\vec{x}}-\hat{\vec{z}}$:
\begin{equation}
R(\hat{\vec{x}}-\hat{\vec{z}},-\phi)=-R(\theta,\pi)R\left (2\theta+\frac{\pi}{4},\phi\right )R(\theta,\pi).
\end{equation}

In words, the sequence just described is as follows:

\begin{enumerate}
	\item Perform a rotation by angle $\pi$ about an axis at an angle $\theta$ with respect to the $z$ axis.
	\item Perform a rotation by angle $\phi$ about an axis at an angle $2\theta+\frac{\pi}{4}$ with respect to the $z$ axis.
	\item Repeat the first rotation.
\end{enumerate}

The fact that this sequence is symmetric will prove beneficial in when we derive a dynamically-corrected version of
it using \textsc{supcode}---we will be able to use fewer pulses to correct this sequence than an asymmetric sequence,
as is shown in Ref.\ \onlinecite{WangPRA2014}.

\subsubsection{Other rotations}
We now consider the remaining rotations.  We find that all of these rotations may be expressed in the
form,
\begin{equation}
R(\theta_1,2\pi-\psi)R(\theta_2,\phi)R(\theta_1,\psi)=-R(\vec{n},\phi),
\end{equation}
where the unit vector $\hat{\vec{n}}$ is given by
\begin{eqnarray}
\vec{n}&=&[\sin{\theta_1}\cos(\theta_1-\theta_2)-\cos{\theta_1}\sin(\theta_1-\theta_2)\cos{\psi}]\hat{\vec{x}} \cr
&+&\sin(\theta_1-\theta_2)\sin{\psi}\hat{\vec{y}} \cr
&+&[\cos{\theta_1}\cos(\theta_1-\theta_2)+\sin{\theta_1}\sin(\theta_1-\theta_2)\cos{\psi}]\hat{\vec{z}}. \nonumber \\
\end{eqnarray}

In order to obtain the desired sequence, we simply set the components of $\vec{n}$ to the appropriate
values and solve for $\theta_1$ and $\theta_2$ for a given value of $\psi$.  While these equations could,
in principle, be solved analytically, the resulting expressions are incredibly unwieldy, and thus we elect
to instead solve them numerically.

In words, the sequence just described is as follows:

\begin{enumerate}
	\item Perform a rotation by angle $\psi$ about an axis at an angle $\theta_1$ with respect to the $z$ axis.
	\item Perform a rotation by angle $\phi$ about an axis at an angle $\theta_2$ with respect to the $z$ axis.
	\item Perform a rotation by angle $2\pi-\psi$ about an axis at an angle $\theta_1$ with respect to the $z$ axis.
\end{enumerate}

\subsection{Corrected unoptimized pulse sequences (CUO)} \label{Sec:CUOSeq}
So far, we have detailed sets of pulse sequences that, under ideal conditions, can be used to perform the
$24$ Clifford gates required for randomized benchmarking simulations.  However, we will still encounter error
in performing these sequences due to charge and nuclear spin noise in our qubit.  Therefore, we require
a means of combatting the effects of this noise.  The method that we use to do this is \textsc{supcode},
which is detailed in Ref.\ \onlinecite{WangPRA2014}.  We will begin with a review of the method and the basic
results that we employ throughout.  The basic idea is to combine a pulse sequence that implements a noisy gate with another sequence that implements a noisy identity operation. This identity operation is designed to have errors which precisely cancel those of the noisy gate. If we let $U(\theta,\phi)$ represent an uncorrected rotation (i.e., with noise),
then we may write our ``identity'' operation as a sequence of interrupted identities:
\begin{eqnarray}
\tilde{I}^{(n)}&=&U(\theta_n,m_n\pi-\phi_n)\cdots U(\theta_2,m_2\pi-\phi_2)U(\theta_1,2m_1\pi) \cr
&\times& U(\theta_2,m_2\pi+\phi_2)\cdots U(\theta_n,m_n\pi+\phi_n).
\end{eqnarray}
One of the major strengths of this form of the uncorrected identity is that we may derive higher ``levels''
of this uncorrected identity recursively:
\begin{eqnarray}
&&\tilde{I}^{(n+1)}= \cr && U(\theta_{n+1},m_{n+1}\pi-\phi_{n+1})\tilde{I}^{(n)}U(\theta_{n+1},m_{n+1}\pi+\phi_{n+1}). \cr &&
\end{eqnarray}
Unless specified otherwise, we will be assuming that all of the $\phi_n=0$, $m_1=2$, and the remaining $m_n=1$,
resulting in a symmetric form for $\tilde{I}^{(n)}$.

To first order in the errors in the magnetic field gradient $\delta h$
and in the exchange $\delta J$, the uncorrected single-pulse rotations are given
by
\begin{equation}
U(\theta,\phi)=R(\theta,\phi)\left (I-i\sum_{i=x,y,z}\Delta_i\sigma_i\right ),
\end{equation}
where the $\sigma_i$ are the Pauli matrices, $R(\theta,\phi)$ is the ideal rotation,
and the $\Delta_i$ give the first-order error in the rotation.  The expressions for
the $\Delta_i$ are
\begin{eqnarray}
\Delta_x&=&\frac{h^2\phi+J^2\sin{\phi}}{2(h^2+J^2)^{3/2}}\,\delta h+\frac{Jh(\phi-\sin{\phi})}{2(h^2+J^2)^{3/2}}\,\delta J, \nonumber \\ \\
\Delta_y&=&-\frac{J(1-\cos{\phi})}{2(h^2+J^2)}\,\delta h+\frac{h(1-\cos{\phi})}{2(h^2+J^2)}\,\delta J, \\
\Delta_z&=&\frac{Jh(\phi-\sin{\phi})}{2(h^2+J^2)^{3/2}}\,\delta h+\frac{J^2\phi+h^2\sin{\phi}}{2(h^2+J^2)^{3/2}}\,\delta J, \nonumber \\
\end{eqnarray}
where $J=h\cot{\theta}$ is the ideal exchange coupling.

The parameter, $\epsilon$, which may be the detuning between the two quantum dots or a
small voltage shift that changes the height of the potential barrier between them, sets the exchange coupling,
so that $\delta J$ takes the form,
\begin{equation}
\delta J=g(J)\,\delta\epsilon.
\end{equation}
Here, we will assume that $g(J)=\frac{J}{\epsilon_0}$, corresponding to the approximately exponential dependence of $J$ on detuning seen in experiments \cite{DialPRL2013}.

We let the uncorrected identity take the form,
\begin{equation}
\tilde{I}^{(n)}=I-i\sum_{i=x,y,z}(\alpha_i\,\delta h+\beta_i\,\delta\epsilon)\sigma_i,
\end{equation}
where $\delta h$ and $\delta\epsilon$ are the errors in the magnetic field
gradient and the detuning, respectively.  To find the corrected sequence, we simply
derive equations for $\alpha_i$ and $\beta_i$ such that the total first-order error
in the sequence under consideration is zero.

The sequences we refer to as "corrected unoptimized" (CUO) were first developed in Ref.\ \onlinecite{WangPRA2014}.  In order to provide a self-contained
presentation of our results, we will review the sequences and list the appropriate parameters
here.

\subsubsection{Identity, $x$, and $x+z$ rotations}
We start with the identity, the rotations by $\pm\frac{\pi}{2}$ and by $\pi$ about the $x$
axis, and the rotation by $\pi$ about $\hat{\vec{x}}+\hat{\vec{z}}$. The na\"ive versions
of these rotations are implemented with a single pulse, and thus are the simplest to treat.
We implement them as interrupted rotations by $2\pi+\phi$, with the identity placed between
the two ``halves'' of the pulse:
\begin{equation}
U_C(\theta,\phi)=
U(\theta,\pi+\tfrac{1}{2}\phi)\tilde{I}^{(5)}U(\theta,\pi+\tfrac{1}{2}\phi).
\end{equation}
We use a level $5$ identity; i.e., it is specified with $5$ exchange couplings.  It is shown
in Ref.\ \onlinecite{WangPRA2014} that, in the case of a symmetric sequence such as that given
above, a level $4$ identity suffices to cancel all the errors to leading order.  However, a level $4$ identity requires
that the exchange coupling become negative at some point in the sequence, and thus we will require
a level $5$ identity.  As the equations that we obtain from multiplying out the above identity
are very complicated, we must solve them numerically.  We now present the parameters required to
perform these rotations in Table \ref{Tab:IdentityAndPiXPlusZ_CUO}.
\begin{table*}
	\centering
		\begin{tabular}{| c | c | c | c | c | c | c | c |}
		\hline
			Operation & $J/h$ & $\phi$ & $J_1/h$ & $J_2/h$ & $J_3/h$ & $J_4/h$ & $J_5/h$ \\
		\hline
			$I$ & $1$ & $0$ & $0.64714$ & $3.7138$ & $0$ & $2.2988$ & $0.54893$ \\
		\hline
			$R(\hat{\vec{x}},-\pi/2)$ & $0$ & $-\pi/2$ & $0.52870$ & $4.1944$ & $0$ & $4.5149$ & $0.79467$ \\
		\hline
			$R(\hat{\vec{x}},\pi)$ & $0$ & $-\pi$ & $0.52902$ & $7.2860$ & $0$ & $3.0639$ & $0.86059$ \\
		\hline
			$R(\hat{\vec{x}}+\hat{\vec{z}},\pi)$ & $1$ & $-\pi$ & $0.49263$ & $6.3648$ & $0$ & $2.0008$ & $0.67803$ \\
		\hline
		\end{tabular}
	\caption{Parameters for the CUO identity, rotations by $-\frac{\pi}{2}$ and $\pi$ about the	$x$ axis,
	and the rotation by $\pi$ about $\hat{\vec{x}}+\hat{\vec{z}}$.  Here, $J=h\cot{\theta}$.}
	\label{Tab:IdentityAndPiXPlusZ_CUO}
\end{table*}

The corrected sequence for an $x$ rotation about $\frac{\pi}{2}$ uses a level $6$ identity, rather than level
$5$; we give the parameters in Table \ref{Tab:XRotPi_CUO}.
\begin{table*}
	\centering
		\begin{tabular}{| c | c | c | c | c | c | c | c | c |}
		\hline
			Operation & $J/h$ & $\phi$ & $J_1/h$ & $J_2/h$ & $J_3/h$ & $J_4/h$ & $J_5/h$ & $J_6/h$ \\
		\hline
			$R(\hat{\vec{x}},\pi/2)$ & $0$ & $\pi/2$ & $0.83930$ & $0$ & $1.1402$ & $0.0025406$ & $2.7063$ & $0.46095$ \\
		\hline
		\end{tabular}
	\caption{Parameters for the CUO rotation by $\frac{\pi}{2}$ about the $x$ axis.}
	\label{Tab:XRotPi_CUO}
\end{table*}

\subsubsection{$z$ rotations}
We now consider $z$ rotations; here, they are implemented using the Hadamard-$x$-Hadamard sequence.
The $x$ rotation is split into two ``halves'' and a level $5$ identity is inserted between the two:
\begin{eqnarray}
&&U_C(\hat{\vec{z}},\phi)= \cr
&&U(J=1,\pi)U(\hat{\vec{x}},\pi+\tfrac{1}{2}\phi)\tilde{I}^{(5)}U(\hat{\vec{x}},\pi+\tfrac{1}{2}\phi)U(J=1,\pi). \nonumber \\
\end{eqnarray}
We list the parameters specifying the uncorrected identity for each rotation in Table \ref{Tab:ZRots_CUO}.
\begin{table*}
	\centering
		\begin{tabular}{| c | c | c | c | c | c | c | c |}
		\hline
			Operation & $\phi$ & $J_1/h$ & $J_2/h$ & $J_3/h$ & $J_4/h$ & $J_5/h$ \\
		\hline
			$R(\hat{\vec{z}},-\pi/2)$ & $-\pi/2$ & $2.1165$ & $0$ & $0.91080$ & $0.35565$ & $5.5498$ \\
		\hline
			$R(\hat{\vec{z}},\pi/2)$ & $\pi/2$ & $0.95366$ & $0$ & $0.70853$ & $0.021024$ & $2.5518$ \\
		\hline
			$R(\hat{\vec{z}},\pi)$ & $\pi$ & $0.66942$ & $0$ & $0.76034$ & $0.0079157$ & $2.0111$ \\
		\hline
		\end{tabular}
	\caption{Parameters for CUO rotations by $-\frac{\pi}{2}$, $\frac{\pi}{2}$, and $\pi$ about the
	$z$ axis.}
	\label{Tab:ZRots_CUO}
\end{table*}

\subsubsection{Other rotations}
Finally, we consider the remaining rotations.  These are accomplished via an $x$-$z$-$x$ decomposition,
\begin{eqnarray}
R(\vec{n},\phi)&=&R(\hat{\vec{x}},\phi_3)R(J=1,\pi)R(\hat{\vec{x}},\phi_2) \cr
&\times&R(J=1,\pi)R(\hat{\vec{x}},\phi_1).
\end{eqnarray}
We obtain the corrected versions of these gates by inserting a level $7$ identity between the second
Hadamard gate and the preceding $z$ rotation:
\begin{eqnarray}
U_C(\vec{n},\phi)&=&U(\hat{\vec{x}},\phi_3)U(J=1,\pi)U(J_7,\pi-\theta_7)\tilde{I}^{(6)}U(\hat{\vec{x}},\phi_2) \cr
&\times&U(J_7,\pi+\theta_7)U(J=1,\pi)U(\hat{\vec{x}},\phi_1).
\end{eqnarray}
We list the parameters for the rest of the Clifford gates in Table \ref{Tab:OtherRots_CUO}.
\begin{table*}
	\centering
		\begin{tabular}{| c | c | c | c | c | c | c | c | c | c | c | c |}
		\hline
			Operation & $\phi_1$ & $\phi_2$ & $\phi_3$ & $J_1/h$ & $J_2/h$ & $J_3/h$ & $J_4/h$ & $J_5/h$ & $J_6/h$ & $J_7/h$ & $\theta_7$ \\
		\hline
			$R(\hat{\vec{y}},-\pi/2)$ & $\pi/2$ & $3\pi/2$ & $3\pi/2$ & $0.75330$ & $0.56113$ & $0$ & $1.6884$ & $0$ & $1.0914$ & $0.60835$ & $1.2726$ \\
		\hline
			$R(\hat{\vec{y}},\pi/2)$ & $3\pi/2$ & $3\pi/2$ & $5\pi/2$ & $0.81782$ & $0$ & $1.3113$ & $0.55040$ & $1.0366$ & $0$ & $1.6911$ & $-1.1929$ \\
		\hline
			$R(\hat{\vec{y}},\pi)$ & $\pi/2$ & $\pi$ & $3\pi/2$ & $0.46134$ & $0.68677$ & $0$ & $1.7332$ & $0$ & $0.90639$ & $0.41421$ & $1.9727$ \\
		\hline
			$R(\hat{\vec{x}}-\hat{\vec{z}},\pi)$ & $\pi/2$ & $3\pi/2$ & $\pi/2$ & $0.71967$ & $1.3078$ & $0$ & $0.81623$ & $0$ & $1.5118$ & $0$ & $-3\pi/4$ \\
		\hline
			$R(\hat{\vec{x}}+\hat{\vec{y}},\pi)$ & $3\pi$ & $\pi/2$ & $0$ & $0.54448$ & $0.63330$ & $0$ & $1.4188$ & $0$ & $1.7652$ & $0.041400$ & $1.7384$ \\
		\hline
			$R(\hat{\vec{x}}-\hat{\vec{y}},\pi)$ & $2\pi$ & $5\pi/2$ & $\pi$ & $0.60618$ & $0.71995$ & $0$ & $0.88507$ & $0$ & $2.2037$ & $0.019841$ & $2.1125$ \\
		\hline
			$R(\hat{\vec{y}}+\hat{\vec{z}},\pi)$ & $2\pi$ & $\pi$ & $7\pi/2$ & $1.1424$ & $0$ & $0.59501$ & $0.0042383$ & $1.4268$ & $0$ & $0.62132$ & $2.1010$ \\
		\hline
			$R(\hat{\vec{y}}-\hat{\vec{z}},\pi)$ & $0$ & $\pi$ & $\pi/2$ & $0.31843$ & $0.84663$ & $0$ & $1.2694$ & $0$ & $0.92116$ & $0.20711$ & $1.7924$ \\
		\hline
			$R(\hat{\vec{x}}+\hat{\vec{y}}+\hat{\vec{z}},2\pi/3)$ & $\pi/2$ & $\pi/2$ & $0$ & $0.40554$ & $1.1271$ & $0$ & $1.0423$ & $0$ & $1.1682$ & $0.022417$ & $2.0737$ \\
		\hline
			$R(\hat{\vec{x}}+\hat{\vec{y}}+\hat{\vec{z}},4\pi/3)$ & $2\pi$ & $7\pi/2$ & $7\pi/2$ & $1.1099$ & $0.67185$ & $0$ & $0.58455$ & $0$ & $3.5271$ & $0.72636$ & $1.3825$ \\
		\hline
			$R(\hat{\vec{x}}+\hat{\vec{y}}-\hat{\vec{z}},2\pi/3)$ & $4\pi$ & $3\pi/2$ & $5\pi/2$ & $0.81495$ & $0$ & $0.53383$ & $0.16963$ & $1.0824$ & $0$ & $0.73536$ & $-1.7509$ \\
		\hline
			$R(\hat{\vec{x}}+\hat{\vec{y}}-\hat{\vec{z}},4\pi/3)$ & $3\pi/2$ & $\pi/2$ & $0$ & $0.46515$ & $0.90353$ & $0$ & $1.2451$ & $0$ & $1.2943$ & $0.035404$ & $1.8526$ \\
		\hline
			$R(\hat{\vec{x}}-\hat{\vec{y}}+\hat{\vec{z}},2\pi/3)$ & $2\pi$ & $5\pi/2$ & $\pi/2$ & $0.59703$ & $0.74094$ & $0$ & $0.88895$ & $0$ & $2.0930$ & $0.029762$ & $1.9536$ \\
		\hline
			$R(\hat{\vec{x}}-\hat{\vec{y}}+\hat{\vec{z}},4\pi/3)$ & $3\pi/2$ & $7\pi/2$ & $2\pi$ & $0.96348$ & $1.0402$ & $0$ & $0.47533$ & $0$ & $5.0237$ & $0.19295$ & $-2.2348$ \\
		\hline
			$R(-\hat{\vec{x}}+\hat{\vec{y}}+\hat{\vec{z}},2\pi/3)$ & $2\pi$ & $\pi/2$ & $3\pi/2$ & $0.52445$ & $0.69563$ & $0$ & $1.36738$ & $0$ & $1.6155$ & $0.0095420$ & $2.2507$ \\
		\hline
			$R(-\hat{\vec{x}}+\hat{\vec{y}}+\hat{\vec{z}},4\pi/3)$ & $5\pi/2$ & $7\pi/2$ & $4\pi$ & $1.3517$ & $0.79872$ & $0$ & $0.40171$ & $0$ & $8.0500$ & $0.97474$ & $1.5893$ \\
		\hline
		\end{tabular}
	\caption{Parameters for CUO versions of the other Clifford gates.}
	\label{Tab:OtherRots_CUO}
\end{table*}

\subsection{Corrected optimized, type II, pulse sequences (CO-II)} \label{Sec:CO-IISeq}
We now detail what we will call the ``corrected optimized, type II'' (CO-II) pulse sequences, which we present here for the first time.  These
sequences are dynamically-corrected versions of the UCO-II pulse sequences detailed earlier.  One
problem that arises with the CUO sequences is that they require the exchange coupling to go to zero,
which is difficult to achieve in actual experiments. 
We will once again employ \textsc{supcode} to derive these dynamically-corrected sequences.

\subsubsection{Identity and $x+z$ rotations}
We start with the identity and the rotation by $\pi$ about $\hat{\vec{x}}+\hat{\vec{z}}$.
Unlike the CUO case, these are the only two rotations that are implemented with a single
pulse.  As with their CUO counterparts, the corrected sequences are implemented as
interrupted rotations by $2\pi+\phi$, with the identity placed between the two ``halves''
of the pulse:
\begin{equation}
U_C(\theta,\phi)=U(\theta,\pi+\tfrac{1}{2}\phi)\tilde{I}^{(5)}U(\theta,\pi+\tfrac{1}{2}\phi).
\end{equation}
As before, we use a level $5$ identity.  We present the parameters required to perform
these rotations in Table \ref{Tab:IdentityAndPiXPlusZ_COII}.
\begin{table*}
	\centering
		\begin{tabular}{| c | c | c | c | c | c | c | c |}
		\hline
			Operation & $J/h$ & $\phi$ & $J_1/h$ & $J_2/h$ & $J_3/h$ & $J_4/h$ & $J_5/h$ \\
		\hline
			$I$ & $0.93248$ & $0$ & $0.93248$ & $30$ & $0.32914$ & $30$ & $0.93248$ \\
		\hline
			$R(\hat{\vec{x}}+\hat{\vec{z}},\pi)$ & $1$ & $-\pi$ & $0.60016$ & $30$ & $0.15535$ & $2.8328$ & $0.74943$ \\
		\hline
		\end{tabular}
	\caption{Parameters for the CO-II identity operation and rotation by $\pi$ about $\hat{\vec{x}}+\hat{\vec{z}}$.}
	\label{Tab:IdentityAndPiXPlusZ_COII}
\end{table*}

\subsubsection{$z$ rotations}
We now consider how to perform corrected $z$ rotations.  These corrected sequences are based on our
$\theta$-$2\theta$-$\theta$ sequence.  We create the corrected version of this sequence by splitting
the rotation by $\phi$ into two rotations by $\pi+\tfrac{1}{2}\phi$ and inserting a symmetric level
$5$ uncorrected identity between the two halves:
\begin{eqnarray}
&&U_C(\hat{\vec{z}},\phi)= \cr
&&-U(\theta,\pi)U(2\theta,\pi+\tfrac{1}{2}\phi)\tilde{I}^{(5)}U(2\theta,\pi+\tfrac{1}{2}\phi)U(\theta,\pi). \nonumber \\
\end{eqnarray}
We present the parameters for rotations by $\phi=-\frac{\pi}{2}$, $\frac{\pi}{2}$, and $\pi$ in
Table \ref{Tab:ZRots_COII}.
\begin{table*}
	\centering
		\begin{tabular}{| c | c | c | c | c | c | c | c |}
		\hline
			Operation & $\theta/\pi$ & $\phi$ & $J_1/h$ & $J_2/h$ & $J_3/h$ & $J_4/h$ & $J_5/h$ \\
		\hline
			$R(\hat{\vec{z}},-\pi/2)$ & $0.096480$ & $-\pi/2$ & $1.1362$ & $30$ & $0.56070$ & $30$ & $0.54537$ \\
		\hline
			$R(\hat{\vec{z}},\pi/2)$ & $0.13734$ & $\pi/2$ & $2.6293$ & $0.59137$ & $30$ & $0.86896$ & $30$ \\
		\hline
			$R(\hat{\vec{z}},\pi)$ & $0.067969$ & $\pi$ & $1.0446$ & $30$ & $0.99351$ & $30$ & $0.37080$ \\
		\hline
		\end{tabular}
	\caption{Parameters for CO-II $z$ rotations by $-\pi/2$, $\pi/2$, and $\pi$.}
	\label{Tab:ZRots_COII}
\end{table*}

\subsubsection{$x$ rotations}
Next, we consider corrected $x$ rotations.  Unlike in the CUO case, we do not perform $x$ rotations with
single pulses.  Instead, we use our ``modified Ramon sequence'',
\begin{equation}
R(\hat{\vec{x}},\phi)=R(\theta,\chi)R(\theta',\alpha)R(\theta,\chi).
\end{equation}
As noted earlier, restrictions on experimental parameters required that $\theta\geq\frac{\theta'}{2}+\frac{\pi}{4}$.
As a special case, let us saturate the bound on $\theta$ for this sequence, i.e., we set $\theta=\frac{\theta'}{2}+\frac{\pi}{4}$.
In this case, for $\phi>0$, we find that $\alpha=2\pi-\phi$ (note that we have already added $2\pi$ to the
value obtained from Eq.\ \eqref{Eq:AlphaSolnX1}) and
\begin{equation}
\cos{\chi}=\frac{1+3\cos{\phi}-(1-\cos{\phi})\sin{\theta'}}{3+\cos{\phi}+(1-\cos{\phi})\sin{\theta'}},
\end{equation}
while, for $\phi<0$, we simply get $\alpha=2\pi+\phi$ and $\chi=\pi$.  We thus find that, for
negative $\phi$, the ``Ramon'' sequence reduces to the Hadamard-$z$-Hadamard sequence.  This
special case will in fact be the na\"ive sequence that we base our corrected sequence on.

We correct this identity in a manner similar to that of our $z$ rotation sequence (i.e., we split the rotation
by $\alpha$ into two rotations by $\pi+\frac{1}{2}\alpha$ and insert a level $5$ uncorrected identity), and
present the parameters for the corrected rotations in Table \ref{Tab:XRots_Ramon_COII}.
\begin{table*}
	\centering
		\begin{tabular}{| c | c | c | c | c | c | c | c | c |}
		\hline
			Operation & $J$ & $\phi$ & $J_1$ & $J_2$ & $J_3$ & $J_4$ & $J_5$ \\
		\hline
			$R(\hat{\vec{x}},-\pi/2)$ & $30$ & $-\pi/2$ & $0.70104$ & $14.803$ & $1.0934$ & $30$ & $0.45532$ \\
		\hline
			$R(\hat{\vec{x}},\pi/2)$ & $2.8086$ & $\pi/2$ & $0.89829$ & $30$ & $0.83484$ & $30$ & $0.48661$ \\
		\hline
			$R(\hat{\vec{x}},\pi)$ & $6.3521$ & $\pi$ & $0.84298$ & $30$ & $0.89938$ & $30$ & $0.60238$ \\
		\hline
		\end{tabular}
	\caption{Parameters for CO-II $x$ rotations by $-\pi/2$, $\pi/2$, and $\pi$.  Here, we take $\theta=\frac{\theta'}{2}+\frac{\pi}{4}$
	and the parameter $J=\cot{\theta'}$.}
	\label{Tab:XRots_Ramon_COII}
\end{table*}

\subsubsection{$y$ rotations}
We now consider the $y$ rotations.  In this case, our sequence for performing these rotations is asymmetric,
so that we will not be able to use an uncorrected identity of a level less than $6$.  In fact, we will employ
here a level $7$ identity of the form,
\begin{equation}
\tilde{I}^{(7)}=U(\theta_7,\pi-\gamma)\tilde{I}^{(6)}U(\theta_7,\pi+\gamma),
\end{equation}
where $\tilde{I}^{(6)}$ is a symmetric level $6$ identity.  We insert this identity between the third and
fourth rotation operators:
\begin{equation}
U\left (\theta_1,\frac{\pi}{2}\right )U(\theta_2,\pi)\tilde{I}^{(7)}U(\theta_3,\phi)U(\theta_2,\pi)U\left (\theta_1,\frac{3\pi}{2}\right ).
\end{equation}
Here, we will assume that $\theta_1=\theta_3=\theta$, so that $\theta_2=\theta+\frac{\pi}{4}$.  We provide
all of the parameters for the corrected rotations in Table \ref{Tab:YRots_COII}.
\begin{table*}
	\centering
		\begin{tabular}{| c | c | c | c | c | c | c | c | c | c | c |}
		\hline
			Operation & $\theta/\pi$ & $\phi$ & $J_1$ & $J_2$ & $J_3$ & $J_4$ & $J_5$ & $J_6$ & $J_7$ & $\gamma$ \\
		\hline
			$R(\hat{\vec{y}},-\pi/2)$ & $0.1525$ & $3\pi/2$ & $0.79604$ & $30$ & $0.99065$ & $1.2187$ & $0.11114$ & $30$ & $0.81728$ & $1.5905$ \\
		\hline
			$R(\hat{\vec{y}},\pi/2)$ & $\tfrac{1}{2}-(\arctan{30})/\pi$ & $\pi/2$ & $1.0391$ & $10.188$ & $1.4257$ & $30$ & $0.21446$ & $30$ & $11.565$ & $-1.1069$ \\
		\hline
			$R(\hat{\vec{y}},\pi)$ & $0.15085$ & $\pi$ & $30$ & $1.5408$ & $0.22841$ & $30$ & $0.61838$ & $30$ & $12.287$ & $2.7737$ \\
		\hline
		\end{tabular}
	\caption{Parameters for CO-II $y$ rotations by $-\pi/2$, $\pi/2$, and $\pi$.}
	\label{Tab:YRots_COII}
\end{table*}

\subsubsection{The $x-z$ rotation}
We now consider the rotation by $\pi$ about the axis given by $\hat{\vec{x}}-\hat{\vec{z}}$.  Our sequence
for these rotations is symmetric, and thus we can perform error correction using only a level $5$ symmetric
uncorrected identity.  We insert this identity in the same way as we did with our $z$ and $x$ rotation
sequences---we simply split the middle rotation into two rotations by $\pi+\frac{1}{2}\phi$ and insert the
identity between the two halves of this rotation:
\begin{eqnarray}
&&U_C(\hat{\vec{x}}-\hat{\vec{z}},-\phi)=U(\theta_1,\pi)U\left (2\theta_1+\frac{\pi}{4},\pi+\tfrac{1}{2}\phi\right )\tilde{I}^{(5)} \cr
&&\times U\left (2\theta_1+\frac{\pi}{4},\pi+\tfrac{1}{2}\phi\right )U(\theta_1,\pi).
\end{eqnarray}
We present the parameters for this rotation in Table \ref{Tab:XMinusZRot_COII}.
\begin{table*}
	\centering
		\begin{tabular}{| c | c | c | c | c | c | c | c |}
		\hline
			Operation & $\theta/\pi$ & $\phi$ & $J_1$ & $J_2$ & $J_3$ & $J_4$ & $J_5$ \\
		\hline
			$R(\hat{\vec{x}}-\hat{\vec{z}},\pi)$ & $0.021086$ & $\pi$ & $2.0552$ & $0.54844$ & $30$ & $0.80575$ & $30$ \\
		\hline
		\end{tabular}
	\caption{Parameters for the CO-II $\hat{\vec{x}}-\hat{\vec{z}}$ rotation by $\pi$.  Note that, up to a minus sign,
	a rotation by $-\pi$ is the same as a rotation by $\pi$.}
	\label{Tab:XMinusZRot_COII}
\end{table*}

\subsubsection{Other rotations}
Finally, we consider the remaining rotations, which are all based on the sequence,
\begin{equation}
R(\vec{n},\phi)=R(\theta_1,2\pi-\psi)R(\theta_2,\phi)R(\theta_1,\psi).
\end{equation}
We make use of an asymmetric level $7$ uncorrected identity of the same form as that
used for the corrected $y$ rotations.  We insert this identity between the first two
rotation operators:
\begin{equation}
U_C(\vec{n},\phi)=U(\theta_1,2\pi-\psi)U(\theta_2,\phi)\tilde{I}^{(7)}U(\theta_1,\psi).
\end{equation}
We present the parameters for the remaining Clifford gates in Table \ref{Tab:OtherRots_COII}.
\begin{table*}
	\centering
		\begin{tabular}{| c | c | c | c | c | c | c | c | c | c | c | c | c |}
		\hline
			Operation & $\theta_1/\pi$ & $\theta_2/\pi$ & $\psi/\pi$ & $\phi$ & $J_1$ & $J_2$ & $J_3$ & $J_4$ & $J_5$ & $J_6$ & $J_7$ & $\gamma$ \\
		\hline
			$R(\hat{\vec{x}}+\hat{\vec{y}},\pi)$ & $0.36572$ & $0.088911$ & $0.62370$ & $\pi$ & $1.9137$ & $0.78946$ & $30$ & $0.89791$ & $30$ & $1.0856$ & $30$ & $1.2867$ \\
		\hline
			$R(\hat{\vec{x}}-\hat{\vec{y}},\pi)$ & $0.31911$ & $0.022346$ & $1.3428$ & $\pi$ & $1.4287$ & $30$ & $1.0830$ & $30$ & $0.64235$ & $30$ & $3.0553$ & $1.6360$ \\
		\hline
			$R(\hat{\vec{y}}+\hat{\vec{z}},\pi)$ & $0.010606$ & $0.26078$ & $1.4894$ & $\pi$ & $1.7257$ & $1.1462$ & $30$ & $0.60456$ & $30$ & $0.88605$ & $1.6688$ & $1.1494$ \\
		\hline
			$R(-\hat{\vec{y}}+\hat{\vec{z}},\pi)$ & $0.010606$ & $0.26078$ & $0.51060$ & $\pi$ & $1.6614$ & $30$ & $0.50802$ & $2.7467$ & $1.0111$ & $30$ & $0.51060$ & $-1.1494$ \\
		\hline
			$R(\hat{\vec{x}}+\hat{\vec{y}}+\hat{\vec{z}},2\pi/3)$ & $0.010606$ & $0.30726$ & $1.7445$ & $2\pi/3$ & $1.7657$ & $0.62197$ & $30$ & $1.1774$ & $30$ & $0.64309$ & $1.9760$ & $-2.5331$ \\
		\hline
			$R(\hat{\vec{x}}+\hat{\vec{y}}+\hat{\vec{z}},4\pi/3)$ & $0.010606$ & $0.30726$ & $1.7445$ & $4\pi/3$ & $1.4124$ & $0.85916$ & $30$ & $0.87090$ & $30$ & $0.71657$ & $2.5246$ & $-2.6369$ \\
		\hline
			$R(-\hat{\vec{x}}-\hat{\vec{y}}+\hat{\vec{z}},4\pi/3)$ & $0.010606$ & $0.32225$ & $0.75513$ & $4\pi/3$ & $1.0888$ & $30$ & $0.63115$ & $2.4528$ & $0.83733$ & $30$ & $0.90701$ & $-1.9200$ \\
		\hline
			$R(-\hat{\vec{x}}-\hat{\vec{y}}+\hat{\vec{z}},2\pi/3)$ & $0.010606$ & $0.32225$ & $0.75513$ & $2\pi/3$ & $1.8354$ & $30$ & $0.46602$ & $2.8446$ & $0.88736$ & $30$ & $0.48343$ & $-1.8961$ \\
		\hline
			$R(\hat{\vec{x}}-\hat{\vec{y}}+\hat{\vec{z}},2\pi/3)$ & $0.010606$ & $0.30726$ & $0.25548$ & $2\pi/3$ & $2.1355$ & $30$ & $0.43106$ & $2.6247$ & $0.91504$ & $30$ & $0.34196$ & $-1.5958$ \\
		\hline
			$R(\hat{\vec{x}}-\hat{\vec{y}}+\hat{\vec{z}},4\pi/3)$ & $0.010606$ & $0.30726$ & $0.25548$ & $4\pi/3$ & $0.91817$ & $30$ & $1.2022$ & $3.6850$ & $0.27003$ & $30$ & $1.7613$ & $0.66640$ \\
		\hline
			$R(-\hat{\vec{x}}+\hat{\vec{y}}+\hat{\vec{z}},2\pi/3)$ & $0.010606$ & $0.32225$ & $1.2449$ & $2\pi/3$ & $1.9034$ & $30$ & $0.65056$ & $6.2506$ & $0.90090$ & $30$ & $0.83256$ & $-2.2863$ \\
		\hline
			$R(-\hat{\vec{x}}+\hat{\vec{y}}+\hat{\vec{z}},4\pi/3)$ & $0.016871$ & $0.33304$ & $1.242$ & $4\pi/3$ & $1.0864$ & $30$ & $0.40950$ & $1.2284$ & $0.58329$ & $30$ & $0.19263$ & $2.7855$ \\
		\hline
		\end{tabular}
	\caption{Parameters for the remaining CO-II Clifford gates.  We write the vectors specifying the axes of rotation in
	unnormalized form for the sake of brevity.}
	\label{Tab:OtherRots_COII}
\end{table*}

\section{Evaluation of sequence sets} \label{Sec:Evaluation}
Now that we have described the five sets of gates that we will be working with, let us compare their
relative merits.  We will begin by comparing the three sets of uncorrected sequences, the UCUO, UCO-I,
and UCO-II sets.  The main part of our evaluation will be in comparing the times required to perform
the pulse sequences in these three sets.  We consider all of the non-trivial gates here.  In all three
sets, the identity gate and the rotation by $\pi$ about the axis, $\hat{\vec{x}}+\hat{\vec{z}}$, can
be done with a single pulse, so there is no difference in timing for these rotations among the three
sets.

\subsection{$z$ rotations}
Let us begin by considering the $z$ rotations.  In both the UCO-I and UCO-II sets, we use the $\theta$-$2\theta$-$\theta$
sequence to perform these rotations:
\begin{equation}
R(\hat{\vec{z}},\phi)=-R(\theta,\pi)R(2\theta,\phi)R(\theta,\pi).
\end{equation}

Using Eq.\ \eqref{Eq:PulseTiming}, we find that the total time required to perform this sequence is
\begin{equation}
ht_{\text{total}}=\pi\sin{\theta}+\tfrac{1}{2}\phi\sin{2\theta}. \label{Eq:GateTime_vs_Angle}
\end{equation}

\begin{figure}[ht]
\includegraphics[width=\columnwidth]{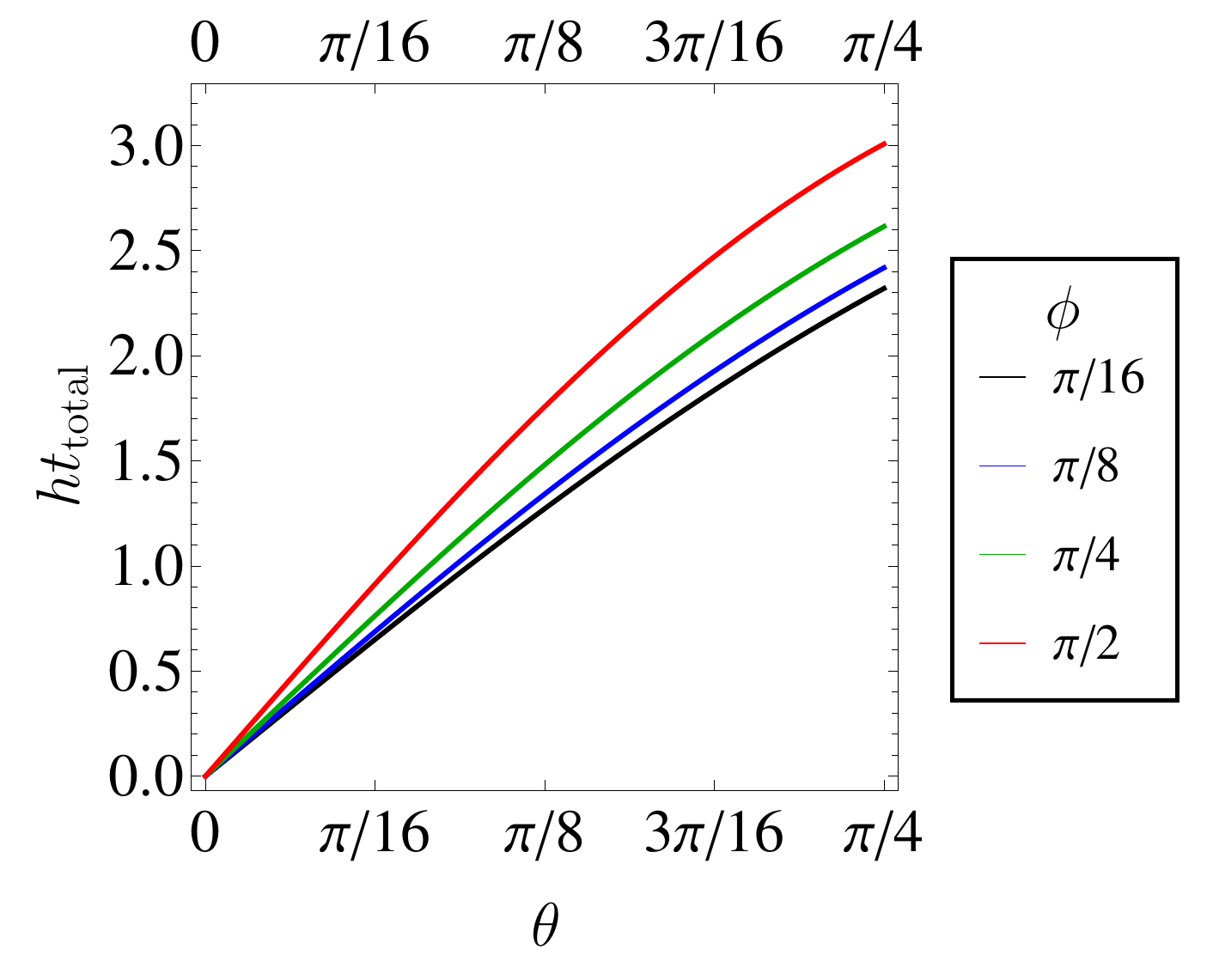}
\caption{Plots of the time required to execute our $\theta$-$2\theta$-$\theta$ sequence, Eq.\ \eqref{Eq:ZRotRT}
(solid lines), and the Ramon sequence, Eq.\ \eqref{Eq:ZRotGR} (dashed lines), as a function of $\theta$ for positive
values of $\phi$.}
\label{Fig:GateTime_vs_Angle_RT_vs_GR}
\end{figure}

We will now compare our sequence to the Ramon sequence, which is used in the UCUO set,
\begin{equation}
R(\hat{\vec{z}},\phi)=R(\theta,\chi)R(\hat{\vec{x}},\alpha)R(\theta,\chi),
\end{equation}
where $\alpha$ is given by Eq.\ \eqref{Eq:AlphaSoln1} if $\phi$ is positive or by Eq.\ \eqref{Eq:AlphaSoln2}
if $\phi$ is negative, and $\chi$ is given by Eq.\ \eqref{Eq:ChiSoln}.  The formula giving the total time
taken to perform this sequence is
\begin{equation}
ht_{\text{tot},R}=\chi\sin{\theta}+\tfrac{1}{2}\alpha.
\end{equation}
We plot this for $\phi=\frac{\pi}{16}$, $\frac{\pi}{8}$, $\frac{\pi}{4}$, and $\frac{\pi}{2}$ in Fig.\
\ref{Fig:GateTime_vs_Angle_RT_vs_GR}.  We see that, for any given value of $\theta$, the $\theta$-$2\theta$-$\theta$
sequence is faster for any value of $\phi$.

\begin{figure}[ht]
\includegraphics[width=\columnwidth]{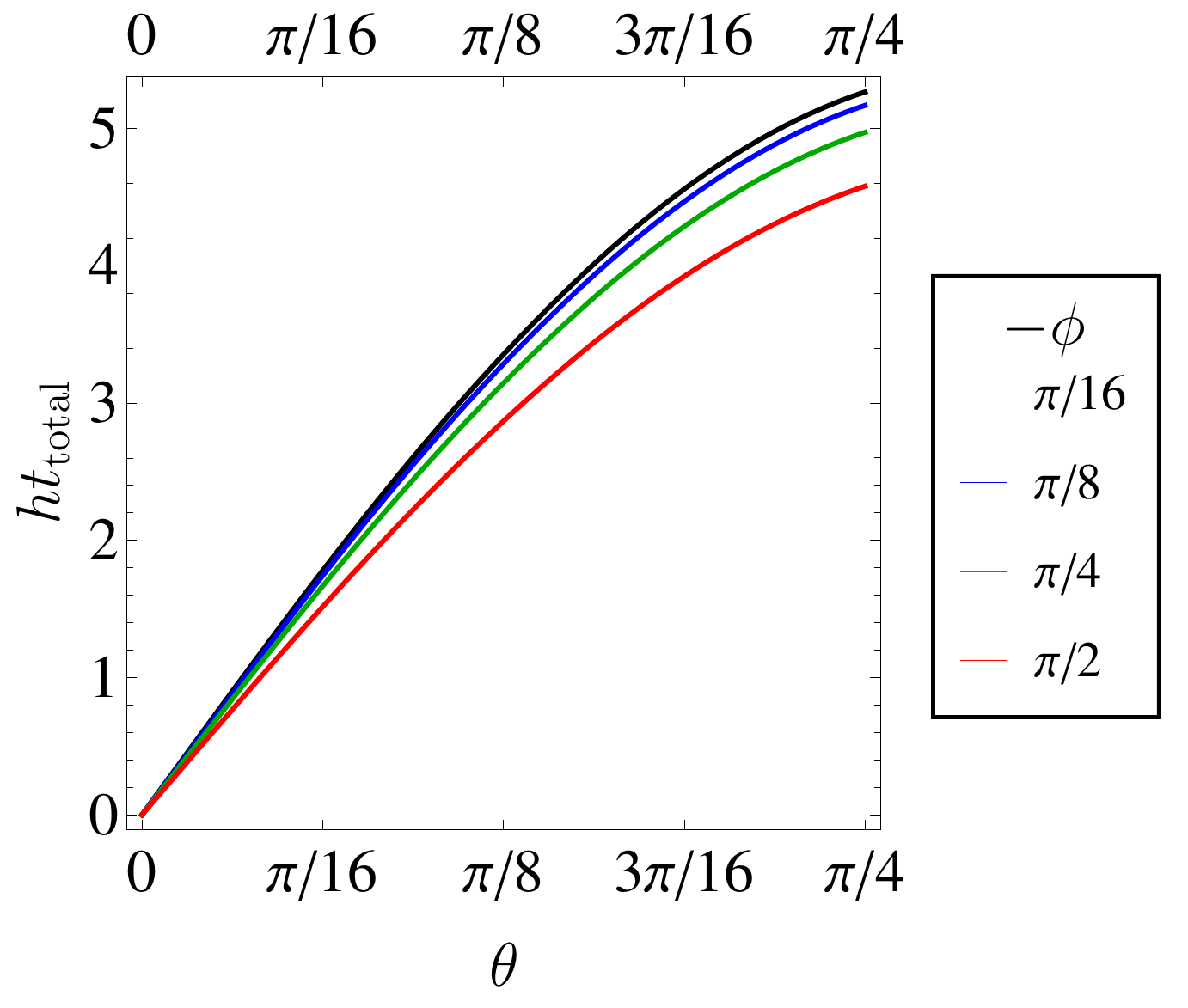}
\caption{Plots of the time required to execute our $\theta$-$2\theta$-$\theta$ sequence, Eq.\ \eqref{Eq:ZRotRT}
(solid lines), and the Ramon sequence, Eq.\ \eqref{Eq:ZRotGR} (dashed lines), as a function of $\theta$ for
negative values of $\phi$.}
\label{Fig:GateTime_vs_Angle_RT_vs_GR_NegPhi}
\end{figure}

We also find the time required to execute both the Ramon sequence and the $\theta$-$2\theta$-$\theta$ sequence
for negative values of $\phi$.  Because we cannot perform rotations by negative angles, we must instead perform
a rotation by $2\pi+\phi$ in the $\theta$-$2\theta$-$\theta$ sequence, so that the total time taken is now
\begin{equation}
ht_{\text{total}}=\pi\sin{\theta}+(\pi+\tfrac{1}{2}\phi)\sin{2\theta}.
\end{equation}
This additional $2\pi$ cancels out the overall minus sign that we would normally acquire and thus our sequence,
too, just performs a $z$ rotation.  We plot the time required to perform both sequences for each value of
$\phi$ in Fig.\ \ref{Fig:GateTime_vs_Angle_RT_vs_GR_NegPhi}.  We find that, once again, the $\theta$-$2\theta$-$\theta$
sequence is faster for given values of $\theta$ and $\phi$; in this case, however, we find that both sequences
become equally fast for $\theta=\frac{\pi}{4}$, since both sequences reduce to the Hadamard-$x$-Hadamard sequence
in this limit.

As we pointed out before, the Ramon sequence requires one to perform an $x$ rotation.  This, in turn, requires
that we be able to completely turn off the exchange coupling between the electrons, which is experimentally challenging.
As a result, we have also considerd a generalized version of this sequence, in which the $x$ rotation is replaced
by one about an axis at an angle $\theta'$ to the $z$ axis; we discuss this sequence in Appendix \ref{App:GenRamon}.

\subsection{$x$ rotations}
Now we turn our attention to the $x$ rotation sequences.  The UCUO sequences implement $x$ rotations in a single shot by switching off the exchange interaction for an interval of time, while  UCO-I sequences use a Hadamard-$z$-Hadamard
sequence to implement these rotations.  The UCO-I sequence implements the $z$ rotation using the $\theta$-$2\theta$-$\theta$
sequence, which we rewrite as
\begin{equation}
R(\hat{\vec{z}},\phi)=-R\left (\theta-\frac{\pi}{4},\pi\right )R\left (2\theta-\frac{\pi}{2},\phi\right )R\left (\theta-\frac{\pi}{4},\pi\right ).
\end{equation}
Here, we replace $\theta$ with $\theta-\frac{\pi}{4}$ to facilitate comparison with the corresponding UCO-II sequence;
the limits on $\theta$ will be the same for both. The UCO-II
set uses the ``modified Ramon sequence'',
\begin{equation}
R(\hat{\vec{x}},\phi)=R(\theta,\chi)R(\theta',\alpha)R(\theta,\chi),
\end{equation}
to implement $x$ rotations directly.

We now determine how fast these sequences are.  The time it takes to implement a UCUO $x$ rotation is $ht_{\text{total}}=\phi/2$, while the total time required to execute the full UCO-I sequence is
\begin{equation}
ht_{\text{total}}=\frac{\pi}{\sqrt{2}}+\pi\sin\left (\theta-\frac{\pi}{4}\right )+\tfrac{1}{2}\phi\sin\left (2\theta-\frac{\pi}{2}\right ). \label{Eq:GateTime_vs_Angle_XRotGH}
\end{equation}
The analogous formula for the UCO-II sequence, Eq.\ \eqref{Eq:XRotGGR}, is
\begin{equation}
ht_{\text{total}}=\chi\sin{\theta}+\tfrac{1}{2}\alpha\sin{\theta'}, \label{Eq:GateTime_vs_Angle_XRotGGR}
\end{equation}
where $\alpha$ and $\chi$ are determined by Eqs.\ \eqref{Eq:AlphaSolnX1}--\eqref{Eq:ChiSolnX}.

\begin{figure}[ht]
\includegraphics[width=0.49\columnwidth]{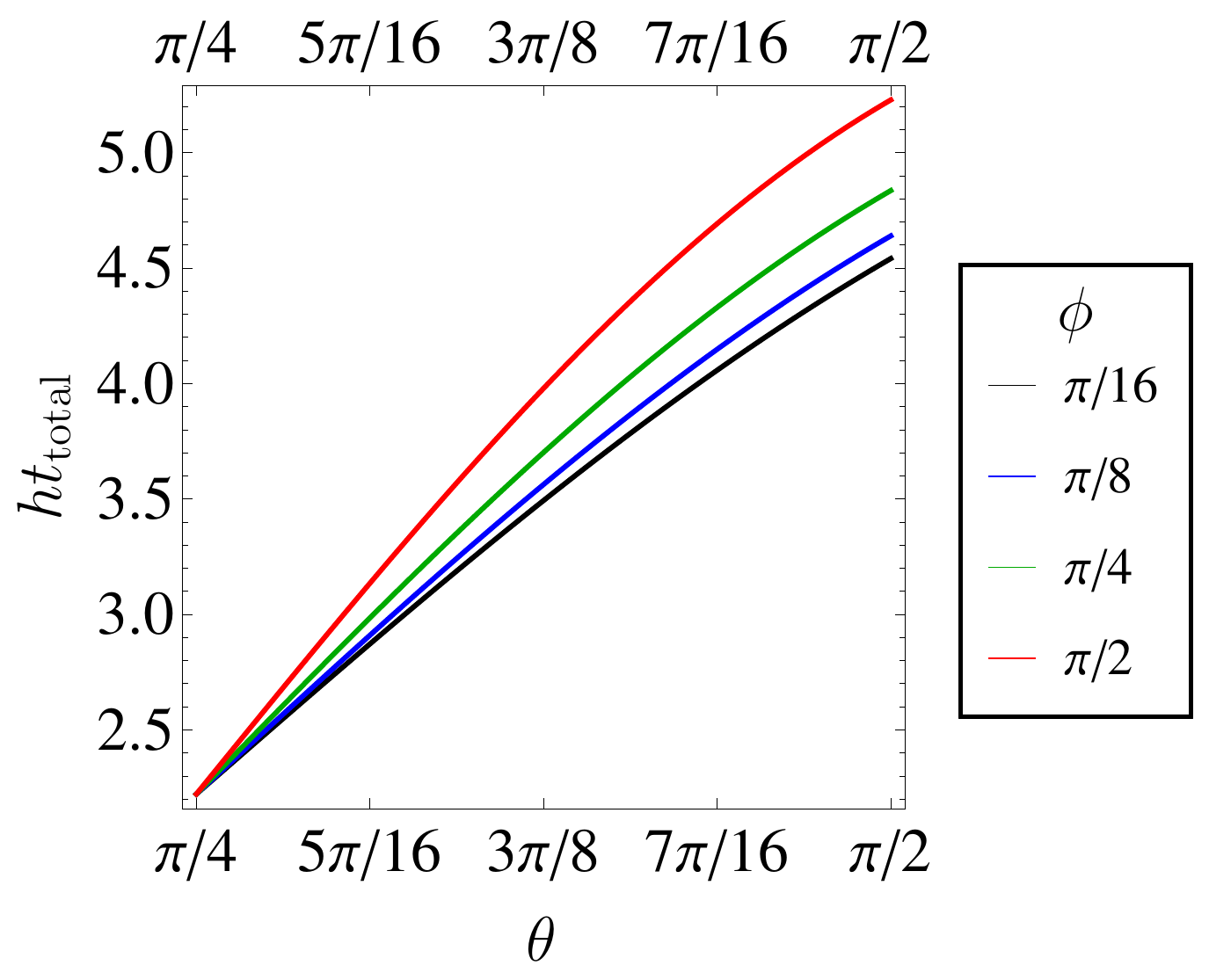}
\includegraphics[width=0.49\columnwidth]{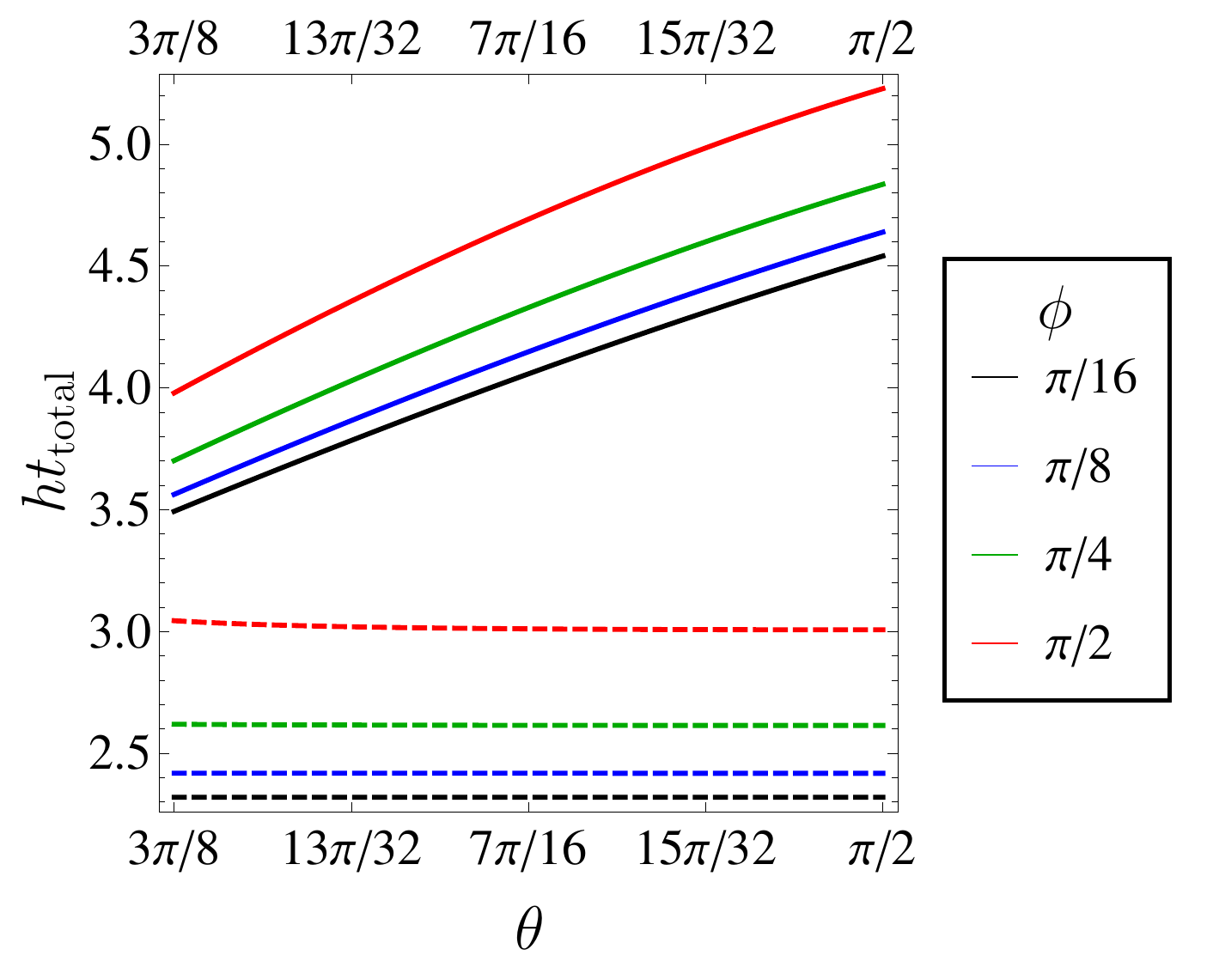}
\includegraphics[width=0.49\columnwidth]{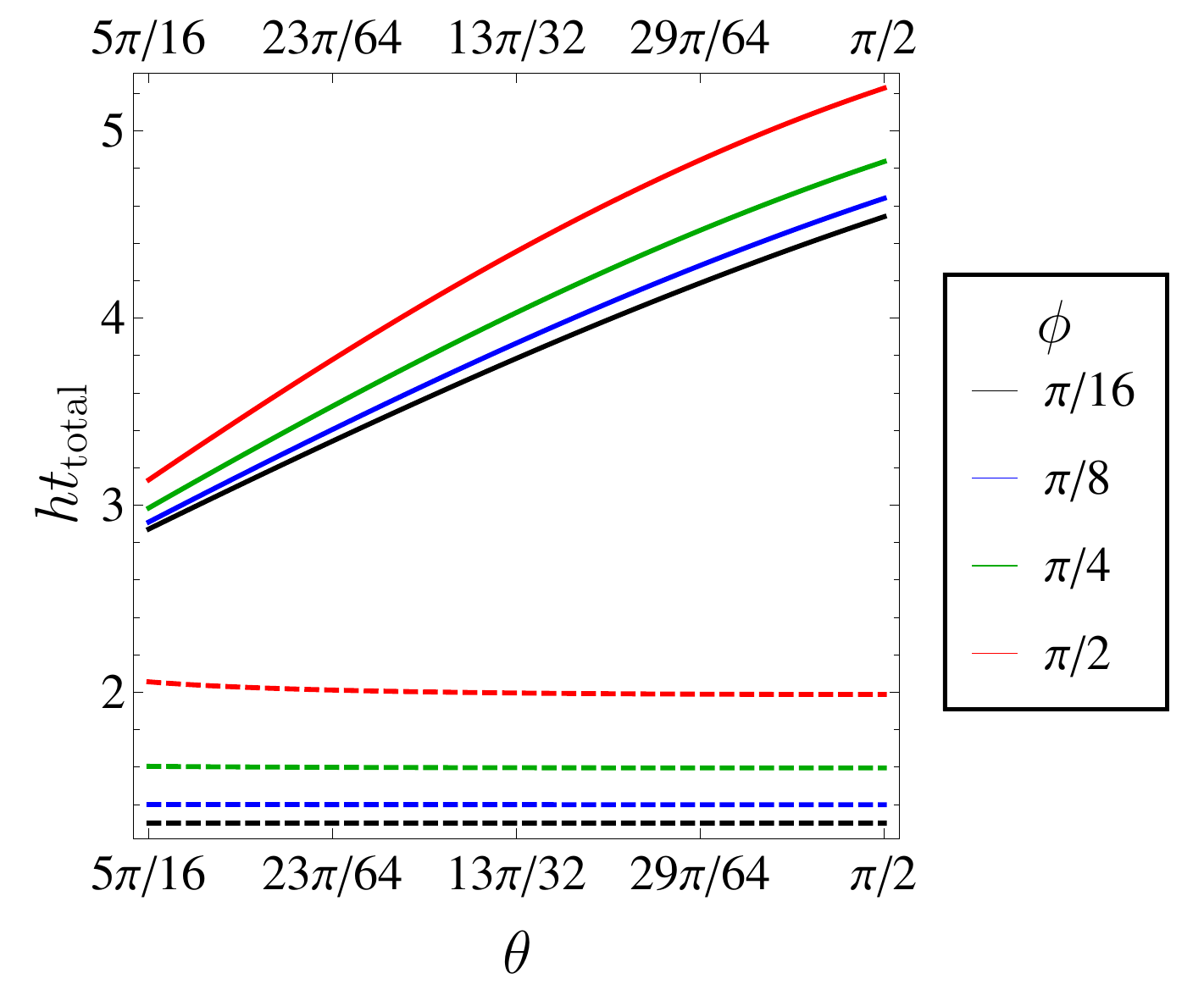}
\includegraphics[width=0.49\columnwidth]{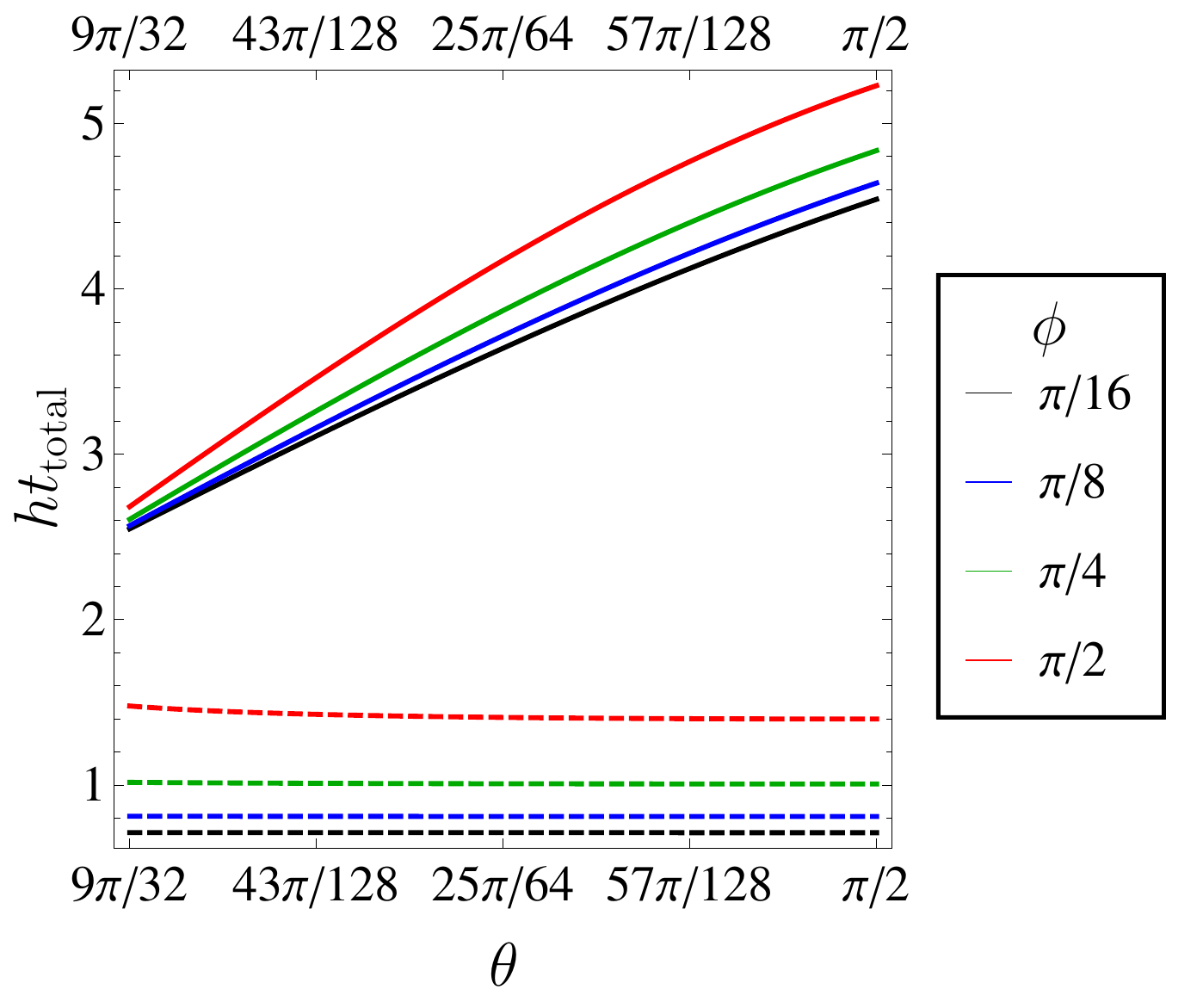}
\caption{(Top left) Plot of the time required to execute the UCO-I Hadamard-$z$-Hadamard sequence, Eq.\ \eqref{Eq:GateTime_vs_Angle_XRotGH}, for all possible values
of $\theta$, namely, $\pi/4 < \theta < \pi/2$.  The remaining plots compare the times required to execute this sequence
(solid lines) with the time required to execute the UCO-II modified Ramon sequence, Eq.\ \eqref{Eq:GateTime_vs_Angle_XRotGGR} (dashed lines),
for $\theta'=\frac{\pi}{4}$ (top right), $\frac{\pi}{8}$ (bottom left), and $\frac{\pi}{16}$ (bottom right).  All
plots shown here are for positive values of $\phi$.}
\label{Fig:tPlotGHvsMGGR_XRot_HZH}
\end{figure}

We will now consider four positive values of $\phi$, namely, $\frac{\pi}{2}$, $\frac{\pi}{4}$, $\frac{\pi}{8}$,
and $\frac{\pi}{16}$, as well as the corresponding negative values, and $\theta'=\frac{\pi}{4}$, $\frac{\pi}{8}$,
and $\frac{\pi}{16}$.  We plot the gate times for both sequences in Fig.\ \ref{Fig:tPlotGHvsMGGR_XRot_HZH}.  We
see that the modified Ramon sequence for performing a UCO-II $x$ rotation is always faster than the UCO-I Hadamard-$z$-Hadamard
sequence. UCUO implements the fastest $x$ rotations since it does not employ composite pulses for these rotations.

\begin{figure}[ht]
\includegraphics[width=0.49\columnwidth]{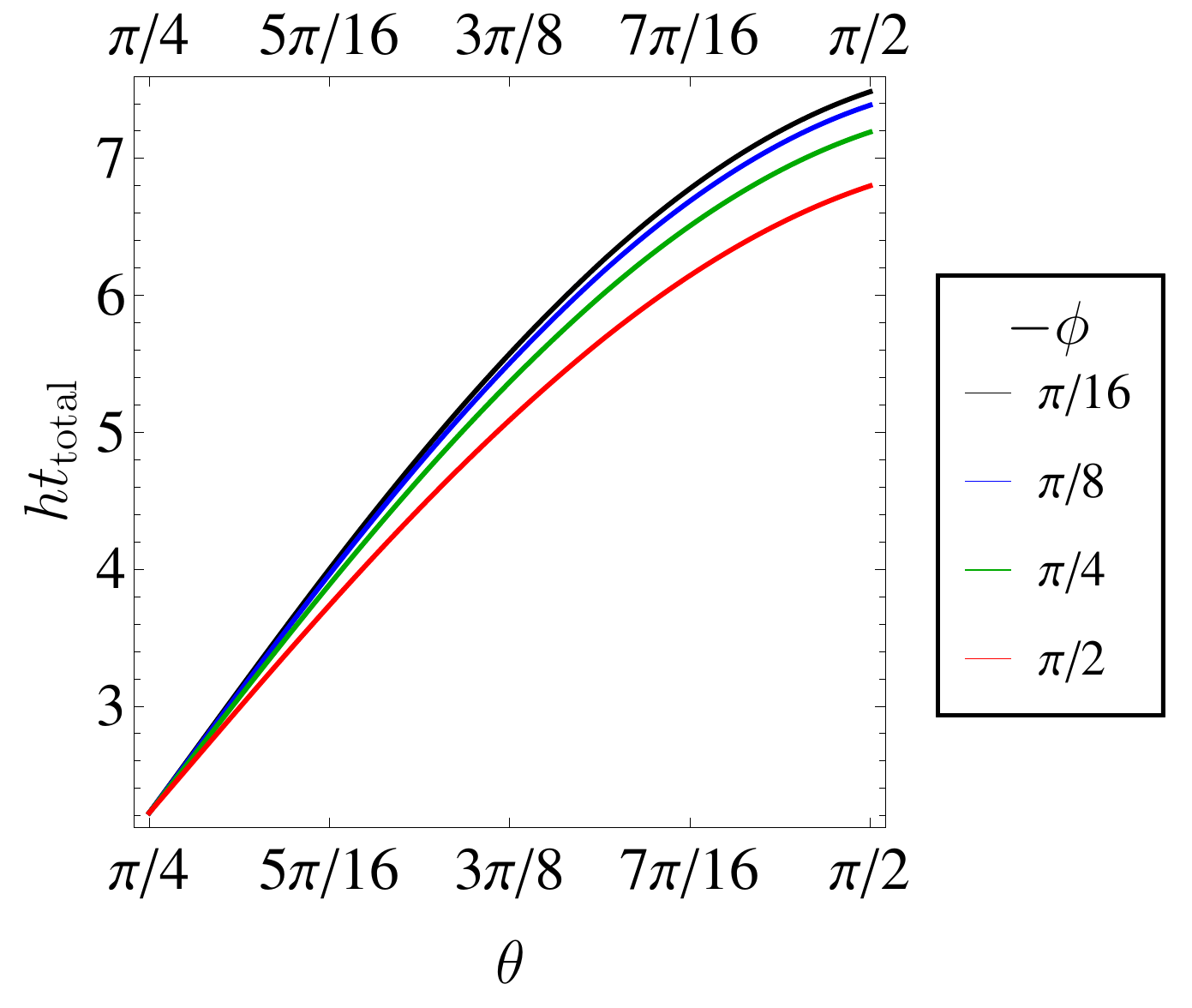}
\includegraphics[width=0.49\columnwidth]{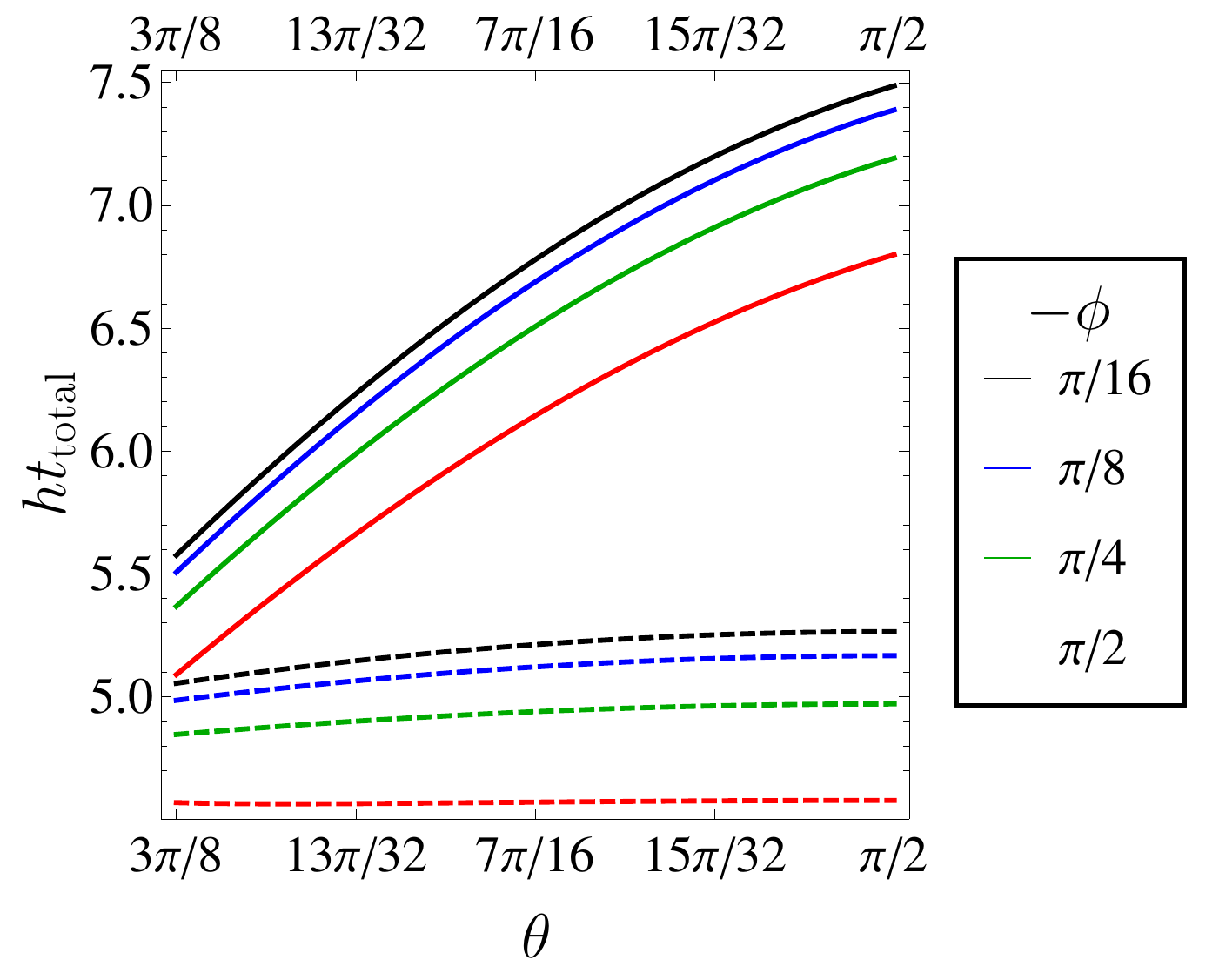}
\includegraphics[width=0.49\columnwidth]{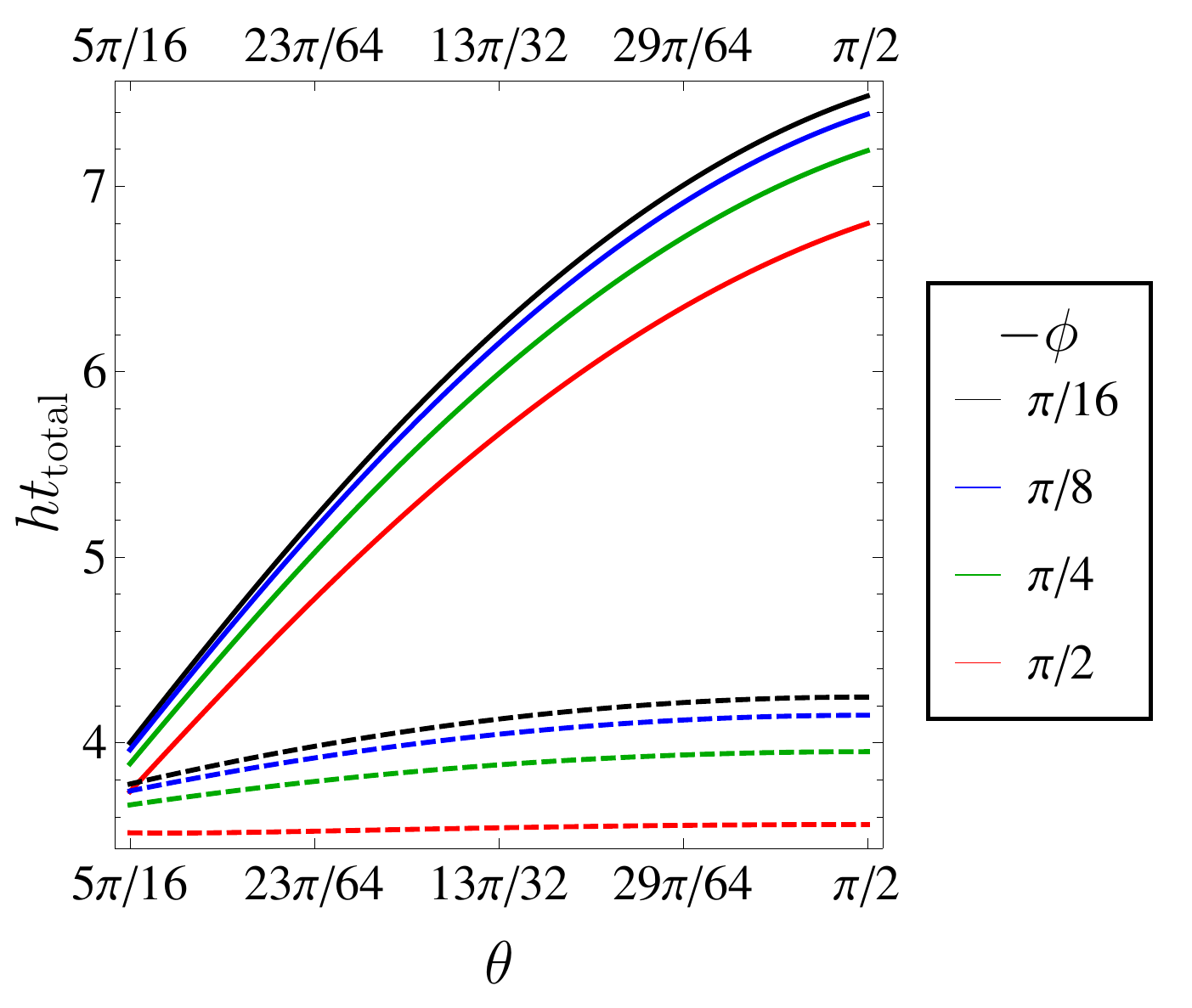}
\includegraphics[width=0.49\columnwidth]{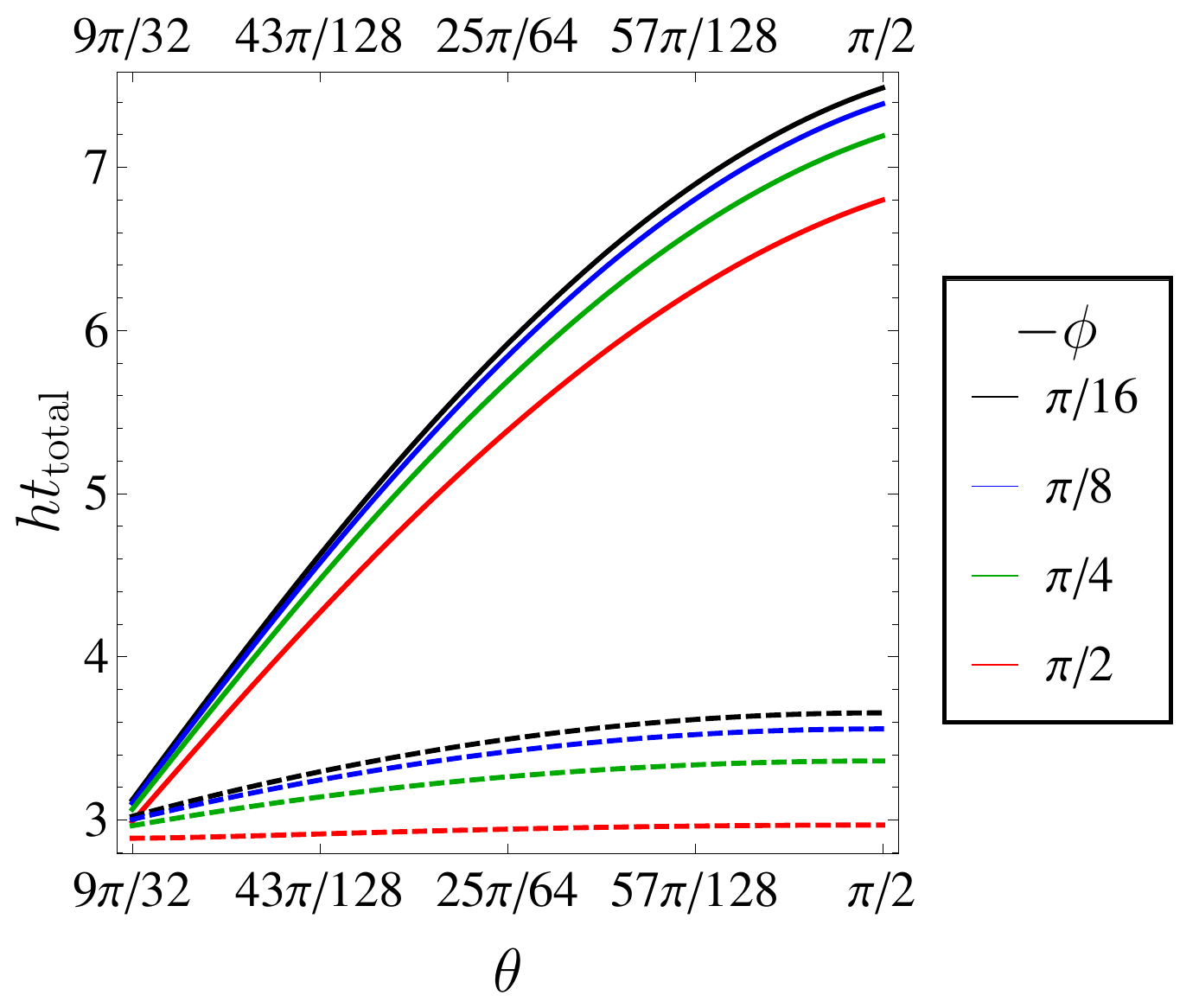}
\caption{Similar to Fig.\ \ref{Fig:tPlotGHvsMGGR_XRot_HZH}, but for negative values of $\phi$.}
\label{Fig:tPlotGHvsMGGR_XRot_NegPhi_HZH}
\end{figure}

Now let us consider rotations by negative angles.  In this case, as pointed out earlier, we must add $2\pi$
to the rotation by $\phi$ in our Hadamard-$z$-Hadamard sequence.  We plot the total times required to execute
the two sequences in Fig.\ \ref{Fig:tPlotGHvsMGGR_XRot_NegPhi_HZH}.  In this case, the UCO-II modified Ramon sequence
is once again faster than UCO-I.  We thus see that, for both positive and negative rotation angles, the modified
Ramon sequence is the faster of the two, while UCUO implements the fastest $x$ rotations overall.

We have also considered a modified version of the $\theta$-$2\theta$-$\theta$ sequence for performing $x$
rotations; we discuss it in Appendix \ref{App:GenHZH}.

\subsection{$y$ rotations}
Next, we consider the $y$ rotations.  The UCO-I set uses the $z$-$x$-$z$ sequence,
\begin{equation}
R(\hat{\vec{y}},\phi)=R\left (\hat{\vec{z}},\frac{\pi}{2}\right )R(\hat{\vec{x}},\phi)R\left (\hat{\vec{z}},\frac{3\pi}{2}\right ),
\end{equation}
to realize these rotations, using the $\theta$-$2\theta$-$\theta$ sequence for the $z$
rotations and the Hadamard-$z$-Hadamard sequence for $x$ rotations.  The UCUO sequence
is similar, except that the Ramon sequence is used to realize $z$ rotations, and $x$ rotations are implemented by turning off the exchange coupling. On the
other hand, $y$ rotations are realized in the UCO-II set with the five-pulse sequence,
\begin{equation}
R(\hat{\vec{y}},\phi)=R\left (\theta_1,\frac{\pi}{2}\right )R(\theta_2,\pi)R(\theta_3,\phi)R(\theta_2,\pi)R\left (\theta_1,\frac{3\pi}{2}\right ),
\end{equation}
where
\begin{equation}
\theta_2=\frac{\theta_1+\theta_3}{2}+\frac{\pi}{4}.
\end{equation}

\begin{figure}[ht]
\includegraphics[width=0.49\columnwidth]{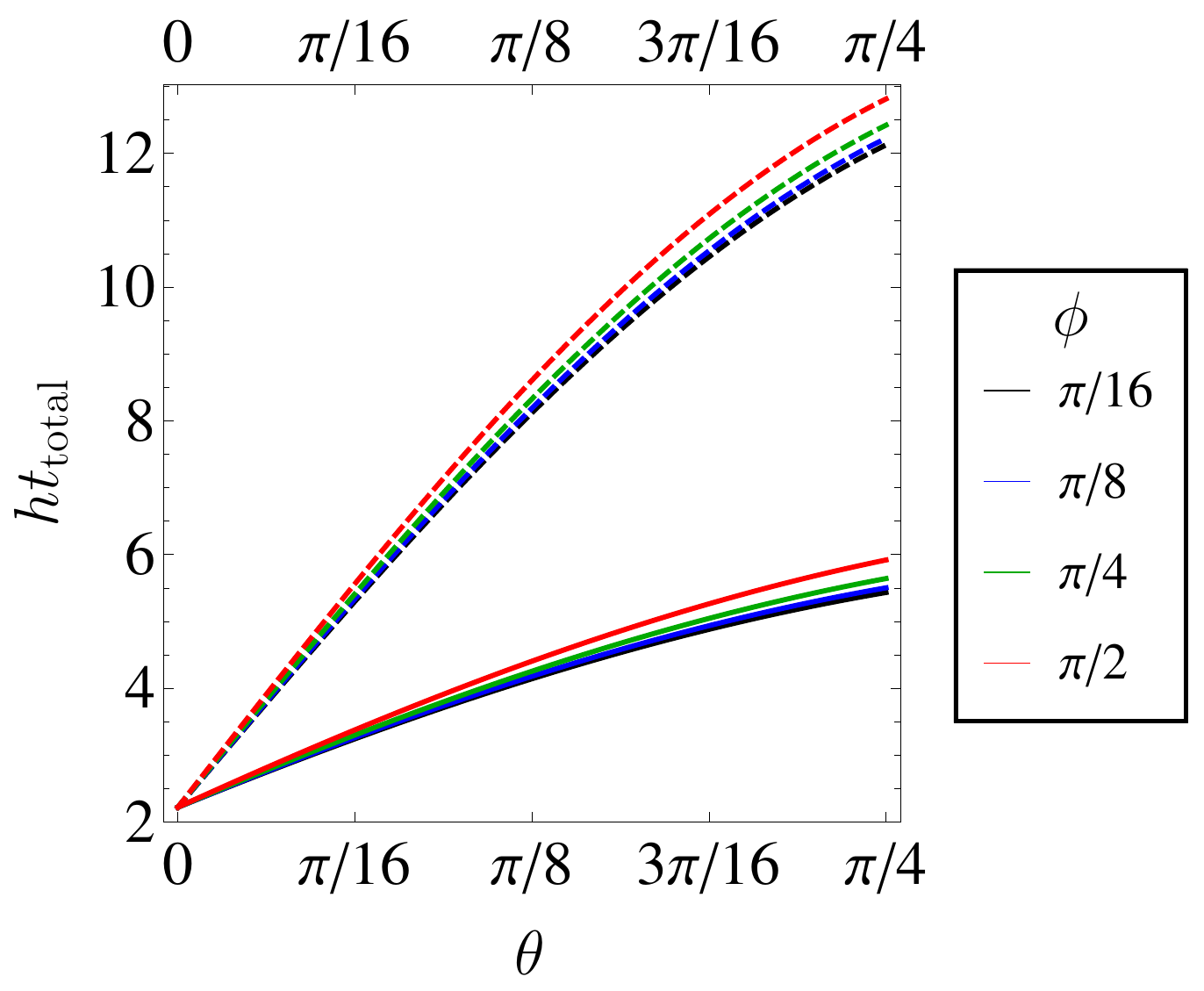}
\includegraphics[width=0.49\columnwidth]{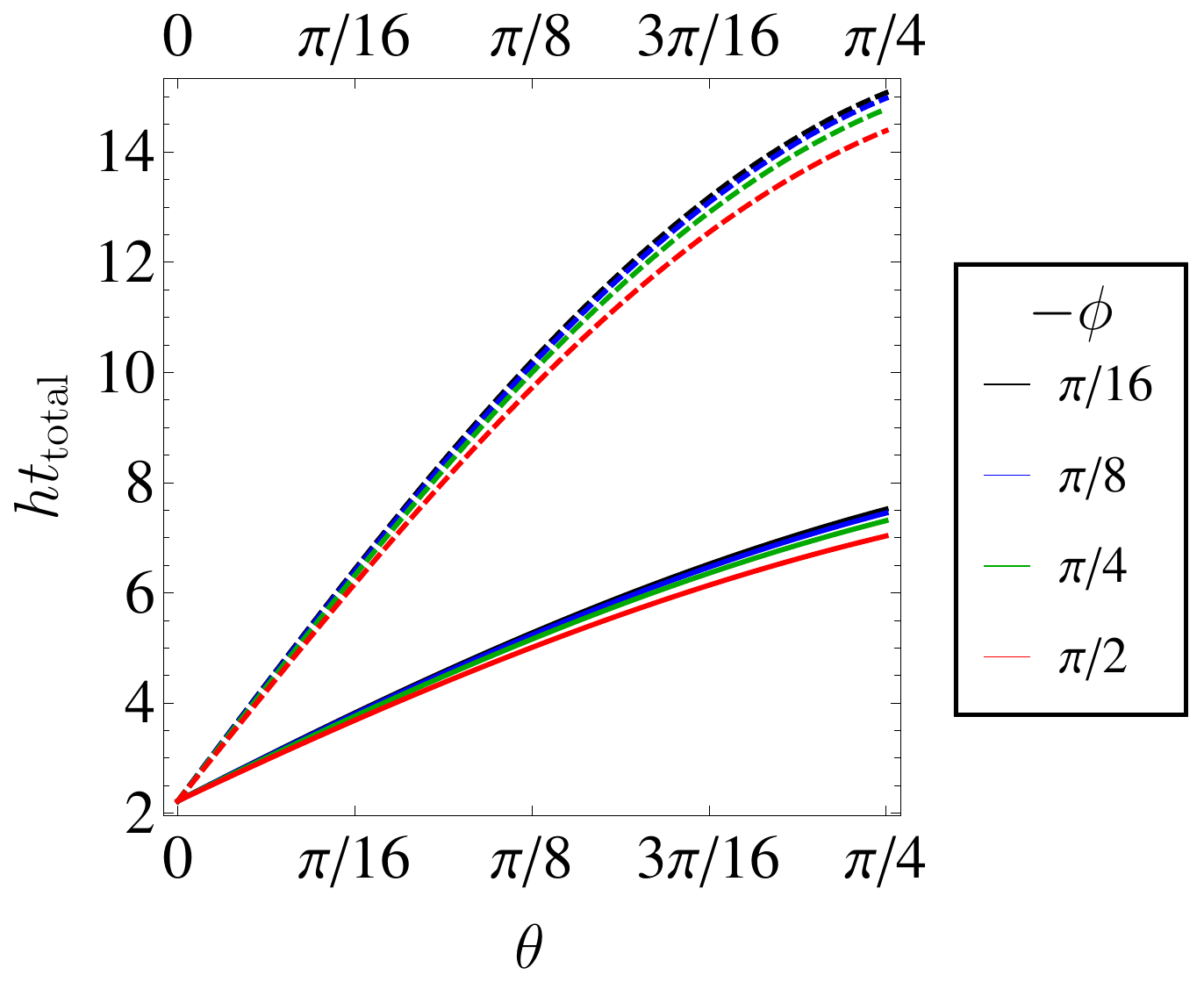}
\caption{Plots of $t_{\text{total}}$ for the five-pulse sequence (solid lines) and for the $z$-$x$-$z$ sequence (dashed
lines), given by Eqs.\ \eqref{Eq:GateTime_vs_Angle_EqTheta13} and \eqref{Eq:GateTime_vs_Angle_ZXZ}, respectively, for
positive $\phi$ (left) and negative $\phi$ (right).}
\label{Fig:tPlots_YRots}
\end{figure}

We now evaluate the performance of the UCUO, UCO-I, and UCO-II pulse sequences for $y$ rotations. The time required to perform the UCO-II five-pulse
sequence is
\begin{equation}
ht_{\text{total}}=\pi\sin{\theta_1}+\pi\sin\left (\frac{\theta_1+\theta_3}{2}+\frac{\pi}{4}\right )+\tfrac{1}{2}\phi\sin{\theta_3}.
\end{equation}
To compare this to the UCUO and UCO-I $z$-$x$-$z$ sequences, we will consider the special case, $\theta_1=\theta_3=\theta$.  In
this case, the time reduces to
\begin{equation}
ht_{\text{total}}=\tfrac{1}{2}[(2+\sqrt{2})\pi+\phi]\sin{\theta}+\tfrac{1}{2}\sqrt{2}\pi\cos{\theta}. \label{Eq:GateTime_vs_Angle_EqTheta13}
\end{equation}
We now give the expression for the UCO-I $z$-$x$-$z$ rotation.  We will assume that $\theta$ takes the
same value in all three of the $z$ rotations.  If we do this, then the total duration of this sequence
will be
\begin{equation}
ht_{\text{total}}=3\pi\sin{\theta}+\tfrac{1}{2}(\phi+2\pi)\sin{2\theta}+\frac{\pi}{\sqrt{2}}. \label{Eq:GateTime_vs_Angle_ZXZ}
\end{equation}
We provide plots of these two gate times, Eqs.\ \eqref{Eq:GateTime_vs_Angle_EqTheta13} and \eqref{Eq:GateTime_vs_Angle_ZXZ},
in Fig.\ \ref{Fig:tPlots_YRots}.  We see that the UCO-II five-pulse sequence is faster than the UCO-I $z$-$x$-$z$ sequence
except in the limit $\theta\to 0$, for which they are equal.  The total time for a UCUO $y$ rotation is
\begin{equation}
ht_{\text{total}}=(1+\sqrt{2})\pi+\tfrac{1}{2}\phi.
\end{equation}
One may easily show that the UCO-I and UCO-II sequences can be made faster than the UCUO sequences by as much as a factor of
more than $3$, depending on the choice of parameters.

\subsection{Other rotations}
Finally, we consider the other rotations.  The rotation by $\pi$ about $\hat{\vec{x}}-\hat{\vec{z}}$ is performed in
the UCO-II set with the sequence,
\begin{equation}
R(\hat{\vec{x}}-\hat{\vec{z}},-\phi)=-R(\theta,\pi)R\left (2\theta+\frac{\pi}{4},\phi\right )R(\theta,\pi).
\end{equation}
In the UCUO and UCO-I sets, we simply use the $z$-$x$-$z$ decomposition, as we do with the $y$ rotation:
\begin{equation}
R(\hat{\vec{x}}-\hat{\vec{z}},\pi)=R\left (\hat{\vec{z}},\frac{\pi}{2}\right )R\left (\hat{\vec{x}},\frac{3\pi}{2}\right )R\left (\hat{\vec{z}},\frac{\pi}{2}\right ),
\end{equation}
For the remaining rotations, we also use the $z$-$x$-$z$ decomposition,
\begin{equation}
R(\vec{n},\phi)=R(\hat{\vec{z}},\phi_3)R(\hat{\vec{x}},\phi_2)R(\hat{\vec{z}},\phi_1),
\end{equation}
for the UCUO and UCO-I sets.  In the UCO-II set, we use the formula,
\begin{equation}
R(\hat{\vec{x}}-\hat{\vec{z}},-\phi)=-R(\theta,\pi)R\left (2\theta+\frac{\pi}{4},\phi\right )R(\theta,\pi),
\end{equation}
for rotations about $\hat{\vec{x}}-\hat{\vec{z}}$, and
\begin{equation}
R(\hat{\vec{n}},\phi)=-R(\theta_1,2\pi-\psi)R(\theta_2,\phi)R(\theta_1,\psi),
\end{equation}
where
\begin{eqnarray}
\hat{\vec{n}}&=&[\sin{\theta_1}\cos(\theta_1-\theta_2)-\cos{\theta_1}\sin(\theta_1-\theta_2)\cos{\psi}]\hat{\vec{x}} \cr
&+&\sin(\theta_1-\theta_2)\sin{\psi}\hat{\vec{y}} \cr
&+&[\cos{\theta_1}\cos(\theta_1-\theta_2)+\sin{\theta_1}\sin(\theta_1-\theta_2)\cos{\psi}]\hat{\vec{z}}, \nonumber \\
\end{eqnarray}
for all other rotations.  We present a comparison of the time required to perform the Clifford gates not so far considered
using these three sets of pulse sequences in Table \ref{Tab:OtherRots_UCOIvsUCOII}.  We see that the UCO-II sequences are
the fastest of the three sets in all cases.
\begin{table*}
	\centering
		\begin{tabular}{| c | c | c | c | c | c | c | c | c | c | c |}
		\hline
			Operation & $\phi_1$ & $\phi_2$ & $\phi_3$ & $ht_{\text{UCUO}}$ & $ht_{\text{UCO-I}}$ & $\theta_1/\pi$ & $\phi$ & $\theta_2/\pi$ & $\psi/\pi$ & $ht_{\text{UCO-II}}$ \\
		\hline
			$R(\hat{\vec{x}}-\hat{\vec{z}},\pi)$ & $\pi/2$ & $3\pi/2$ & $\pi/2$ & $8.3699$ & $2.7969$ & $0.010606$ & $\pi$ & $0.27121$ & $1$ & $1.2869$ \\
		\hline
			$R(\hat{\vec{x}}+\hat{\vec{y}},\pi)$ & $\pi$ & $\pi/2$ & $0$ & $6.7991$ & $2.9015$ & $0.31119$ & $\pi$ & $0.010606$ & $0.66225$ & $1.6729$ \\
		\hline
			$R(\hat{\vec{x}}-\hat{\vec{y}},\pi)$ & $0$ & $\pi/2$ & $\pi$ & $6.7991$ & $2.6923$ & $0.31119$ & $\pi$ & $0.010606$ & $1.3377$ & $2.6572$ \\
		\hline
			$R(\hat{\vec{y}}+\hat{\vec{z}},\pi)$ & $3\pi/2$ & $\pi$ & $0$ & $8.3699$ & $2.7969$ & $0.010606$ & $\pi$ & $0.26078$ & $1.4894$ & $1.2524$ \\
		\hline
			$R(-\hat{\vec{y}}+\hat{\vec{z}},\pi)$ & $0$ & $\pi/2$ & $\pi$ & $6.7991$ & $2.6923$ & $0.010606$ & $\pi$ & $0.26078$ & $0.51060$ & $1.2524$ \\
		\hline
			$R(\hat{\vec{x}}+\hat{\vec{y}}+\hat{\vec{z}},2\pi/3)$ & $\pi/2$ & $\pi/2$ & $0$ & $6.0137$ & $2.6400$ & $0.010606$ & $2\pi/3$ & $0.30726$ & $1.7445$ & $0.96568$ \\
		\hline
			$R(\hat{\vec{x}}+\hat{\vec{y}}+\hat{\vec{z}},4\pi/3)$ & $0$ & $3\pi/2$ & $3\pi/2$ & $9.1553$ & $2.8492$ & $0.010606$ & $4\pi/3$ & $0.30726$ & $1.7445$ & $1.8267$ \\
		\hline
			$R(-\hat{\vec{x}}-\hat{\vec{y}}+\hat{\vec{z}},4\pi/3)$ & $0$ & $3\pi/2$ & $\pi/2$ & $7.5845$ & $2.7446$ & $0.010606$ & $4\pi/3$ & $0.32225$ & $0.75512$ & $1.8809$ \\
		\hline
			$R(-\hat{\vec{x}}-\hat{\vec{y}}+\hat{\vec{z}},2\pi/3)$ & $3\pi/2$ & $\pi/2$ & $0$ & $7.5845$ & $2.7446$ & $0.010606$ & $2\pi/3$ & $0.32225$ & $0.75512$ & $0.99279$ \\
		\hline
			$R(\hat{\vec{x}}-\hat{\vec{y}}+\hat{\vec{z}},2\pi/3)$ & $0$ & $3\pi/2$ & $\pi/2$ & $7.5845$ & $2.7446$ & $0.010606$ & $2\pi/3$ & $0.30726$ & $0.25548$ & $0.96568$ \\
		\hline
			$R(\hat{\vec{x}}-\hat{\vec{y}}+\hat{\vec{z}},4\pi/3)$ & $3\pi/2$ & $3\pi/2$ & $0$ & $9.1553$ & $2.8492$ & $0.010606$ & $4\pi/3$ & $0.30726$ & $0.25548$ & $1.8267$ \\
		\hline
			$R(-\hat{\vec{x}}+\hat{\vec{y}}+\hat{\vec{z}},2\pi/3)$ & $0$ & $\pi/2$ & $3\pi/2$ & $7.5845$ & $2.7446$ & $0.010606$ & $2\pi/3$ & $0.32225$ & $1.2449$ & $0.99279$ \\
		\hline
			$R(-\hat{\vec{x}}+\hat{\vec{y}}+\hat{\vec{z}},4\pi/3)$ & $\pi/2$ & $3\pi/2$ & $0$ & $7.5845$ & $2.7446$ & $0.010606$ & $4\pi/3$ & $0.32225$ & $1.2449$ & $1.8809$ \\
		\hline
		\end{tabular}
	\caption{Parameters for the remaining Clifford gates.  We write the vectors specifying the axes of rotation in unnormalized
	form for the sake of brevity.  In the $\theta$-$2\theta$-$\theta$ sequence used to implement the $z$ rotations in the UCO-I
	sequences, we always set $\theta=\frac{\pi}{2}-\arctan{30}\approx 0.010606$.}
	\label{Tab:OtherRots_UCOIvsUCOII}
\end{table*}

\subsection{Corrected pulse sequences}
Now we consider the two sets of dynamically-corrected pulse sequences, the CUO and CO-II sets.  As stated before, the CUO sequences
are those presented in Ref.\ \onlinecite{WangPRA2014} and reviewed in Sec. \ref{Sec:CUOSeq}, while the CO-II sets are dynamically-corrected
versions of the UCO-II sequences described earlier.  To determine the timing of these sequences, we simply use the fact that, for a
given pulse, $J=h\cot{\theta}$, where $\theta$ is the angle that the rotation axis makes with the $z$ axis, and that the duration of
a single pulse is $ht=\tfrac{1}{2}\phi\sin{\theta}$, where $\phi$ is the rotation angle.  We give our results for all $24$ Clifford
gates in Table \ref{Tab:Timing_CUOvsCOII}.  We see that the CO-II sequences are always faster than the CUO sequences.
\begin{table*}
	\centering
		\begin{tabular}{| c | c | c || c | c | c |}
		\hline
			Operation & $ht_{\text{CUO}}$ & $ht_{\text{CO-II}}$ & Operation & $ht_{\text{CUO}}$ & $ht_{\text{CO-II}}$ \\
		\hline
			$I$ & $15.462$ & $12.384$ & $R(\hat{\vec{x}}+\hat{\vec{y}},\pi)$ & $28.672$ & $13.455$ \\
		\hline
			$R(\hat{\vec{x}}+\hat{\vec{z}},\pi)$ & $14.381$ & $12.326$ & $R(\hat{\vec{x}}-\hat{\vec{y}},\pi)$& $31.858$ & $12.427$ \\
		\hline
			$R(\hat{\vec{z}},-\pi/2)$ & $16.243$ & $12.139$ & $R(\hat{\vec{y}}+\hat{\vec{z}},\pi)$ & $33.166$ & $13.331$ \\
		\hline
			$R(\hat{\vec{z}},\pi/2)$ & $20.688$ & $11.816$ & $R(-\hat{\vec{y}}+\hat{\vec{z}},\pi)$ & $26.577$ & $13.585$ \\
		\hline
			$R(\hat{\vec{z}},\pi)$ & $22.338$ & $12.346$ & $R(\hat{\vec{x}}+\hat{\vec{y}}+\hat{\vec{z}},2\pi/3)$ & $25.342$ & $13.034$ \\
		\hline
			$R(\hat{\vec{x}},-\pi/2)$ & $14.920$ & $12.882$ & $R(\hat{\vec{x}}+\hat{\vec{y}}+\hat{\vec{z}},4\pi/3)$ & $35.566$ & $14.129$ \\
		\hline
			$R(\hat{\vec{x}},\pi/2)$ & $21.036$ & $13.084$ & $R(-\hat{\vec{x}}-\hat{\vec{y}}+\hat{\vec{z}},4\pi/3)$ & $36.473$ & $14.919$ \\
		\hline
			$R(\hat{\vec{x}},\pi)$ & $14.049$ & $13.161$ & $R(-\hat{\vec{x}}-\hat{\vec{y}}+\hat{\vec{z}},2\pi/3)$ & $26.702$ & $13.276$ \\
		\hline
			$R(\hat{\vec{y}},-\pi/2)$ & $28.168$ & $20.434$ & $R(\hat{\vec{x}}-\hat{\vec{y}}+\hat{\vec{z}},2\pi/3)$ & $31.120$ & $13.133$ \\
		\hline
			$R(\hat{\vec{y}},\pi/2)$ & $30.445$ & $12.445$ & $R(\hat{\vec{x}}-\hat{\vec{y}}+\hat{\vec{z}},4\pi/3)$ & $32.738$ & $14.080$ \\
		\hline
			$R(\hat{\vec{y}},\pi)$ & $28.312$ & $13.260$ & $R(-\hat{\vec{x}}+\hat{\vec{y}}+\hat{\vec{z}},2\pi/3)$ & $29.581$ & $12.003$ \\
		\hline
			$R(\hat{\vec{x}}-\hat{\vec{z}},\pi)$ & $26.748$ & $12.110$ & $R(-\hat{\vec{x}}+\hat{\vec{y}}+\hat{\vec{z}},4\pi/3)$ & $35.956$ & $17.133$ \\
		\hline
		\end{tabular}
	\caption{Times required to perform the Clifford gates using the CUO and CO-II pulse sequences.  The parameters used to determine
	these sequence durations are those presented in the tables in Sec.\ \ref{Sec:PulseSequences} in the relevant subsections.}
	\label{Tab:Timing_CUOvsCOII}
\end{table*}

\section{Randomized benchmarking} \label{Sec:RBResults}
We now evaluate the above sets of pulse sequences using randomized benchmarking.  Before doing so, let us quickly review the randomized
benchmarking procedure\cite{MagesanPRL2011,MagesanPRA2012}.  The basic idea behind randomized benchmarking is to generate random sequences
of Clifford gates of varying lengths (i.e., number of individual gates) and determine the state- and sequence-averaged fidelity for that
sequence length.  We may define the state- and sequence-dependent fidelity as follows:
\begin{equation}
F(O,\psi)=\left |\bra{\psi}O^{\dagger}_{\text{ideal}}O\ket{\psi}\right |^2,
\end{equation}
where $\ket{\psi}$ is the initial state of the system, $O_{\text{ideal}}$ is the operation that the gate sequence would ideally perform on
the system, and $O$ is the actual operation that is performed.  In other words, the fidelity is the probability that, upon measuring the
system, we will find it in the state that we expect the operation $O$ to leave it in if there is no noise present.  The average fidelity is
then the average over gate sequences of a given length and over all (distinct) starting states.  We can then plot this average fidelity as
a function of sequence length and determine a characteristic number of gates $n_0$, which provides us with a measure of how many gates can
reliably be performed.  We can also plot the average fidelity as a function of time and extract the effective decoherence time, $T_2$,
characterizing the decay of the average fidelity.  We will now proceed to do this for two different noise models---quasistatic noise and
$1/f^{\alpha}$ noise.

\subsection{Quasistatic noise}
\begin{figure}[ht]
\includegraphics[width=\columnwidth]{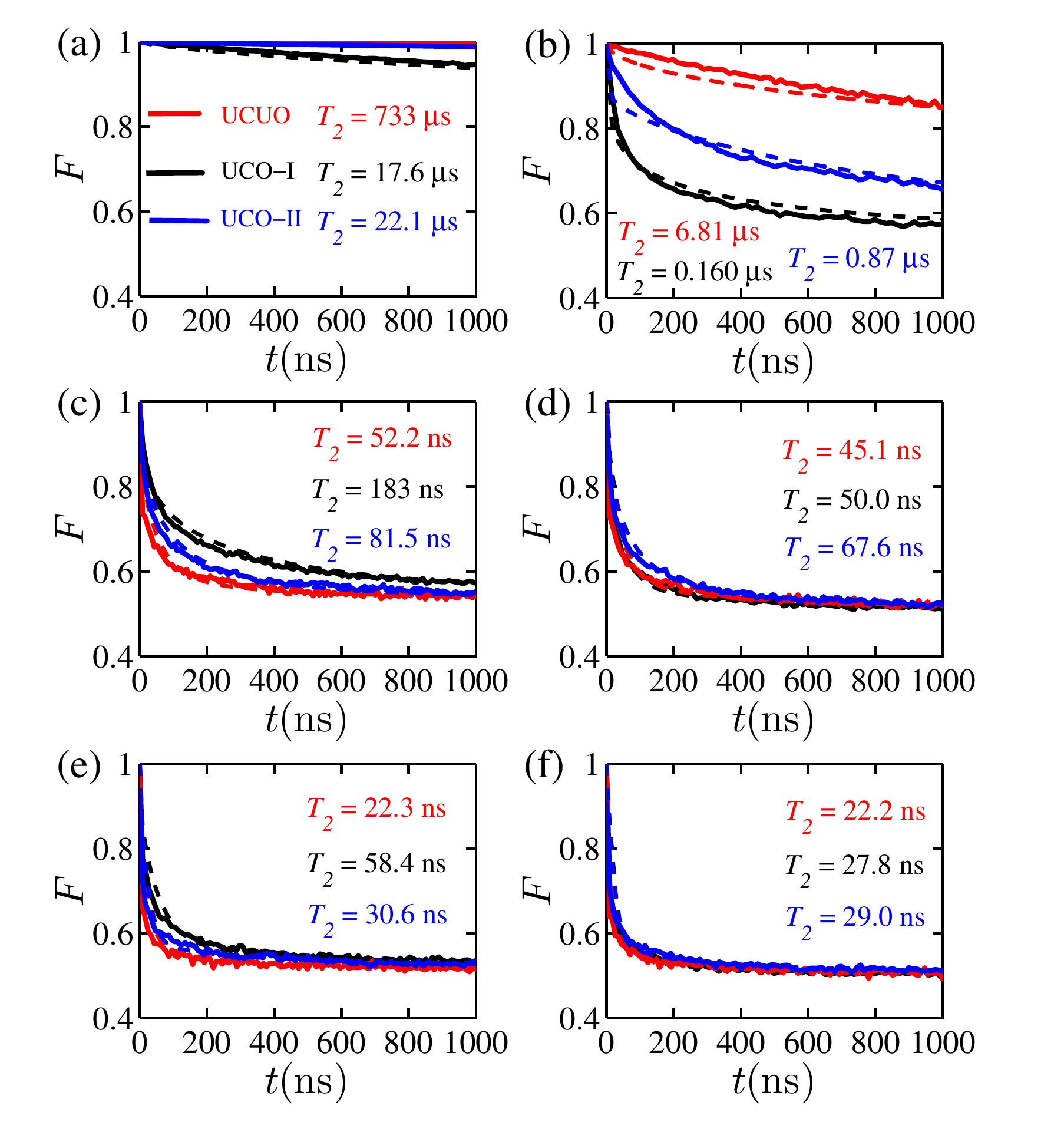}
\caption{Plots of average fidelity extracted from randomized benchmarking simulations as a function of time for our uncorrected sequence
sets (UCUO, UCO-I, UCO-II) in the presence of quasistatic noise.  The left column (a, c, e) represents a quantum dot operated with barrier
control\cite{ReedPRL2016,MartinsPRL2016}, while the right column (b, d, f) uses tilt control.  These two methods of control correspond to charge noise
strengths of $\sigma_J=0.00426J$ and $\sigma_J=0.0563J$, respectively\cite{MartinsPRL2016}.  The solid curves are data, and the dashed curves
are fits to Eq.\ \eqref{Eq:FvsTimeFit}.  The top row (a, d) represents a system with no field noise (e.g., isotopically purified Si) and
magnetic field gradient $h_0=23\text{ MHz}$, the middle row (b, e) a system with $\sigma_h=11.5\text{ MHz}$ and $h_0=40\text{ MHz}$, and
the bottom row (c, f) a system with $\sigma_h=23\text{ MHz}$ and $h_0=40\text{ MHz}$.  For the UCO-II pulse sequences, we use the ``na\"ive''
versions of the corresponding CO-II pulse sequences, with the same parameters.}
\label{Fig:RB_QS_Uncorrected}
\end{figure}

We first consider the simpler of the two models, quasistatic noise.  In the quasistatic noise model, we assume that the noise may be
approximated as a constant-in-time stochastic shift of the exchange coupling and the magnetic field gradient, each of which is drawn from a Gaussian
distribution.  This model is often used\cite{ZhangPRL2017,WangPRA2014} because fluctuations due to noise in our qubit tend to be slow
compared to gate times; gate times are on the order of ns, while noise fluctuation frequencies are on the order of kHz \cite{MedfordPRL2012,DialPRL2013}. For this reason,
we expect such a quasistatic model to agree well with experiment, which has been demonstrated in previous work, at least qualitatively\cite{MartinsPRL2016,Barnes_PRB2016}.
In our randomized benchmarking simulations, we draw random values for the magnetic field gradient $h$ and the exchange coupling $J$ from
Gaussian distributions for each realization (i.e., for each random gate sequence considered) with means $h_0$ and $J_0$ and standard deviations
$\sigma_h$ and $\sigma_J$, respectively.  From this point forward, we will refer to the standard deviations as the strengths of the respective
noises.  We plot our results comparing the performance of the uncorrected gate sets in Fig.\ \ref{Fig:RB_QS_Uncorrected}.  The $T_2$ values
given in the plot are extracted from fits to the equation,
\begin{equation}
F=\tfrac{1}{2}+\tfrac{1}{4}[e^{-(t/T_2)^{\gamma_1}}+e^{-(t/T_2)^{\gamma_2}}]. \label{Eq:FvsTimeFit}
\end{equation}

From these results, we see that, in the absence of field noise, the UCUO pulse sequences have the longest $T_2$ of all of the uncorrected sets
by an order of magnitude.  This is because the UCUO sequences include segments during which the exchange coupling is completely turned off, and
thus there is zero charge noise for the duration of these segments (recall that we assume that the charge noise strength is proportional to the
exchange coupling).  On the other hand, we never turn off the exchange coupling in the UCO-I and UCO-II sequences, and thus they are affected by
charge noise for the entire duration of the pulse sequence.  Once we introduce field noise, however, the UCUO sequences have the shortest $T_2$
of the three sets.  This is due to a combination of the fact that they require more time to execute than the corresponding UCO-I and UCO-II
sequences and the fact that they are, in this case, subject to the effects of noise for their entire durations.  We also note that the UCO-II
set is intermediate between the UCUO and UCO-I sets.  This is due to the fact that the UCO-II pulse sequences that we use in our simulation are
the ``na\"ive'' versions of the corresponding CO-II sequences, and thus some of the sequences never set $J$ to the maximum possible value.

\begin{figure}[ht]
\includegraphics[width=\columnwidth]{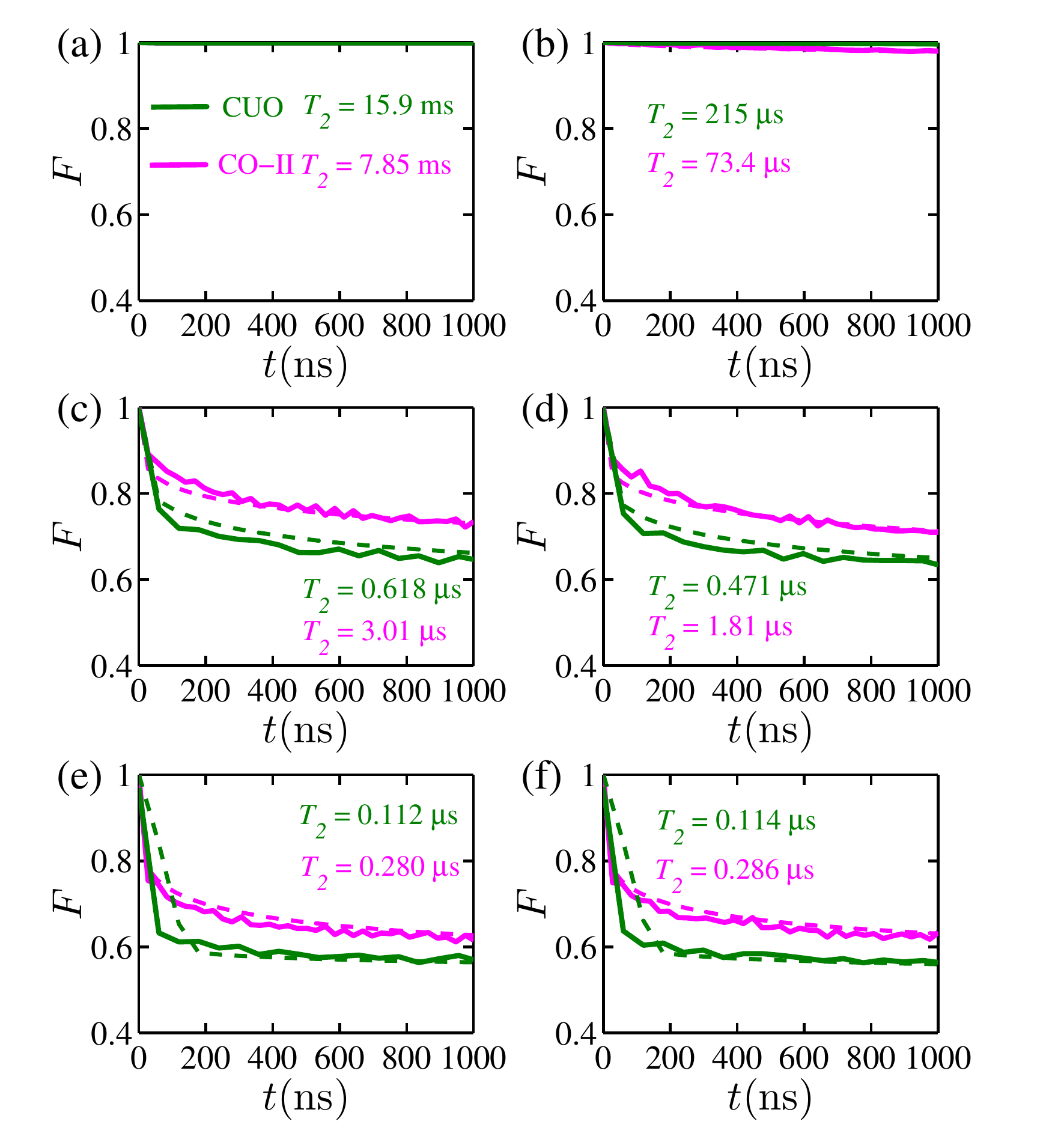}
\caption{Plots of average fidelity extracted from randomized benchmarking simulations as a function of time for our corrected sequence
sets (CUO, CO-II) in the presence of quasistatic noise.  The left column (a, c, e) represents a quantum dot operated with barrier
control\cite{ReedPRL2016,MartinsPRL2016}, while the right column (b, d, f) uses tilt control.  These two methods of control correspond to charge noise
strengths of $\sigma_J=0.00426J$ and $\sigma_J=0.0563J$, respectively\cite{MartinsPRL2016}.  The solid curves are data, and the dashed
curves are fits to Eq.\ \eqref{Eq:FvsTimeFit}.  The top row (a, d) represents a system with no field noise (e.g., isotopically purified
Si) and magnetic field gradient $h_0=23\text{ MHz}$, the middle row (b, e) a system with $\sigma_h=11.5\text{ MHz}$ and $h_0=40\text{ MHz}$,
and the bottom row (c, f) a system with $\sigma_h=23\text{ MHz}$ and $h_0=40\text{ MHz}$.}
\label{Fig:RB_QS_Corrected}
\end{figure}

We now turn our attention to the corrected sequence sets, the CUO and CO-II sets.  We compare the performance of these two sets in Fig.\ \ref{Fig:RB_QS_Corrected}.
We see a similar pattern in the values of $T_2$ for these two sets as for the uncorrected sets.  For zero field noise, we find that the CUO
sequences give better $T_2$ values.  This is for similar reasons for why $T_2$ is longest for the UCUO sequences out of all of the uncorrected
sequences.  However, once we turn on field noise, we find that the CO-II sequences have longer $T_2$.  We also note that the $T_2$ values obtained
from these corrected sequences are longer by at least an order of magnitude than any of the times obtained from the uncorrected sequences in the
corresponding situation, thus illustrating the effectiveness of \textsc{supcode} in reducing the effects of noise on our gates.

\begin{figure}[ht]
\includegraphics[width=\columnwidth]{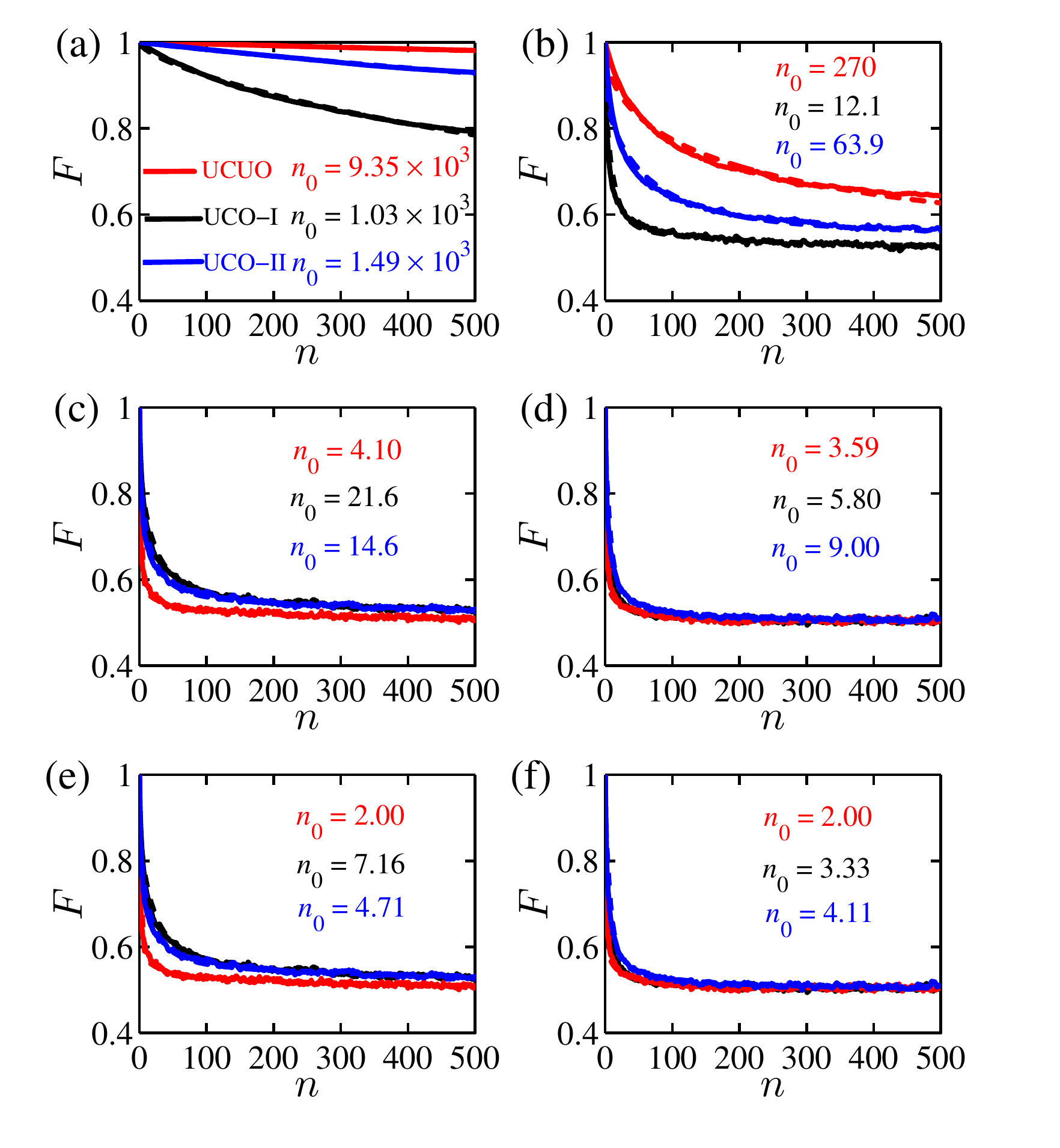}
\caption{Plots of average fidelity extracted from randomized benchmarking simulations as a function of number of gates for our uncorrected
sequence sets (UCUO, UCO-I, UCO-II) in the presence of quasistatic noise.  The left column (a, c, e) represents a quantum dot operated with
barrier control\cite{ReedPRL2016,MartinsPRL2016}, while the right column (b, d, f) uses tilt control.  These two methods of control correspond to charge
noise strengths of $\sigma_J=0.00426J$ and $\sigma_J=0.0563J$, respectively\cite{MartinsPRL2016}.  The solid curves are data, and the dashed curves
are fits to Eq.\ \eqref{Eq:FvsGateFit}.  The top row (a, d) represents a system with no field noise (e.g., isotopically purified Si) and
magnetic field gradient $h_0=23\text{ MHz}$, the middle row (b, e) a system with $\sigma_h=11.5\text{ MHz}$ and $h_0=40\text{ MHz}$, and
the bottom row (c, f) a system with $\sigma_h=23\text{ MHz}$ and $h_0=40\text{ MHz}$.  For the UCO-II pulse sequences, we use the ``na\"ive''
versions of the corresponding CO-II pulse sequences, with the same parameters.}
\label{Fig:RB_QS_Uncorrected_Gates}
\end{figure}

\begin{figure}[ht]
\includegraphics[width=\columnwidth]{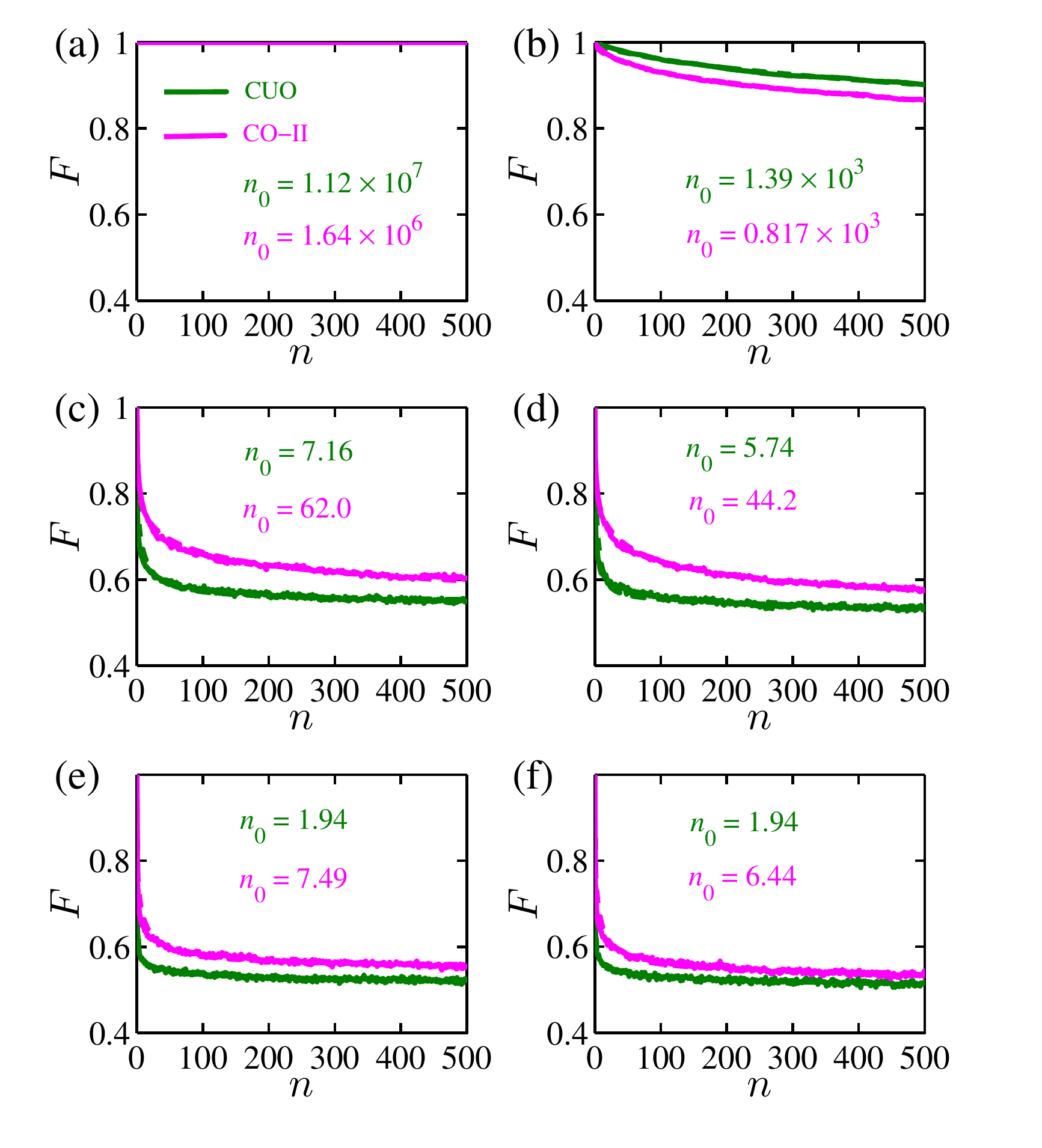}
\caption{Plots of average fidelity extracted from randomized benchmarking simulations as a function of number of gates for our corrected sequence
sets (CUO, CO-II) in the presence of quasistatic noise.  The left column (a, c, e) represents a quantum dot operated with barrier control\cite{ReedPRL2016,MartinsPRL2016},
while the right column (b, d, f) uses tilt control.  These two methods of control correspond to charge noise strengths of $\sigma_J=0.00426J$
and $\sigma_J=0.0563J$, respectively\cite{MartinsPRL2016}.  The solid curves are data, and the dashed curves are fits to Eq.\ \eqref{Eq:FvsGateFit}.
The top row (a, d) represents a system with no field noise (e.g., isotopically purified Si) and magnetic field gradient $h_0=23\text{ MHz}$, the
middle row (b, e) a system with $\sigma_h=11.5\text{ MHz}$ and $h_0=40\text{ MHz}$, and the bottom row (c, f) a system with $\sigma_h=23\text{ MHz}$
and $h_0=40\text{ MHz}$.}
\label{Fig:RB_QS_Corrected_Gates}
\end{figure}

We also plot the fidelities of the uncorrected and corrected sequence sets as functions of the number of gates in Figs.\ \ref{Fig:RB_QS_Uncorrected_Gates}
and \ref{Fig:RB_QS_Corrected_Gates}, respectively, and fit the data to a form similar to that used for the plots as a function of time:
\begin{equation}
F=\tfrac{1}{2}+\tfrac{1}{4}[e^{-(n/n_0)^{\gamma_1}}+e^{-(n/n_0)^{\gamma_2}}]. \label{Eq:FvsGateFit}
\end{equation}
This alternate way of plotting the results allows us to determine how many gates one can perform with each set before the fidelity decays
to an unacceptable level, characterized by the parameter $n_0$.  We see that these plots show a similar pattern as the plots with respect
to time, but some of the curves may be closer or further apart vertically than in the time plots, due simply to the fact that the sequences
from each set for a given gate take different amounts of time to complete.  Despite what our results so far may imply, we will see later
that achieving a longer $T_2$ with a given set of pulse sequences does not guarantee that we can perform more gates with them within that
longer $T_2$, especially if the sequences take longer to perform.

We present the parameters for all of the curves that we fit to our data in Table \ref{Tab:FitP_Quasistatic} in Appendix \ref{App:FittingP}.

\subsection{$1/f^\alpha$ noise}
We now consider $1/f^\alpha$ noise.  It is important to investigate more realistic models of noise that capture the time dependence of the noise
in semiconductor spin qubits because, as we will see shortly, we may find qualitatively different results from the quasistatic limit, especially if one or both
types of noise have significant high-frequency components.  The field and charge noise present in quantum dot spin qubits may, and in fact do, have different
exponents characterizing them, $\alpha_h$ and $\alpha_J$, respectively.  We will begin by fixing these exponents to their experimental values\cite{DialPRL2013,MedfordPRL2012},
$\alpha_h=2.6$ and $\alpha_J=0.7$.  Throughout this section, we determine the strength of the noise as follows.  We assume that the power spectrum
has the form,
\begin{equation}
S_{h,J}(\omega)=\frac{A_{h,J}}{(\omega/h)^{\alpha_{h,J}}},
\end{equation}
where the constants $A_{h,J}$ are determined by requiring that these spectra integrate to\cite{BarnesPRB2016}
\begin{equation}
\int_{\omega_{\text{ir}}}^{\omega_{\text{uv}}}d\omega\,S_h(\omega)=\pi\sigma_h^2
\end{equation}
and
\begin{equation}
\int_{\omega_{\text{ir}}}^{\omega_{\text{uv}}}d\omega\,S_J(\omega)=\pi\left (\frac{\sigma_J}{J/h}\right )^2.
\end{equation}
We will take $\sigma_h=23\text{ MHz}$ throughout and use the usual values, $\sigma_J=0.00426J$ and $\sigma_J=0.0563J$ for barrier and tilt control,
respectively.  We set the infrared and ultraviolet cutoffs\cite{BarnesPRB2016} for the field noise to $\omega_{\text{ir}}=10\text{ kHz}$ and
$\omega_{\text{uv}}=100\text{ kHz}$, and those for the charge noise to $\omega_{\text{ir}}=50\text{ kHz}$ and $\omega_{\text{uv}}=1\text{ MHz}$.

\begin{figure}[ht]
\includegraphics[width=\columnwidth]{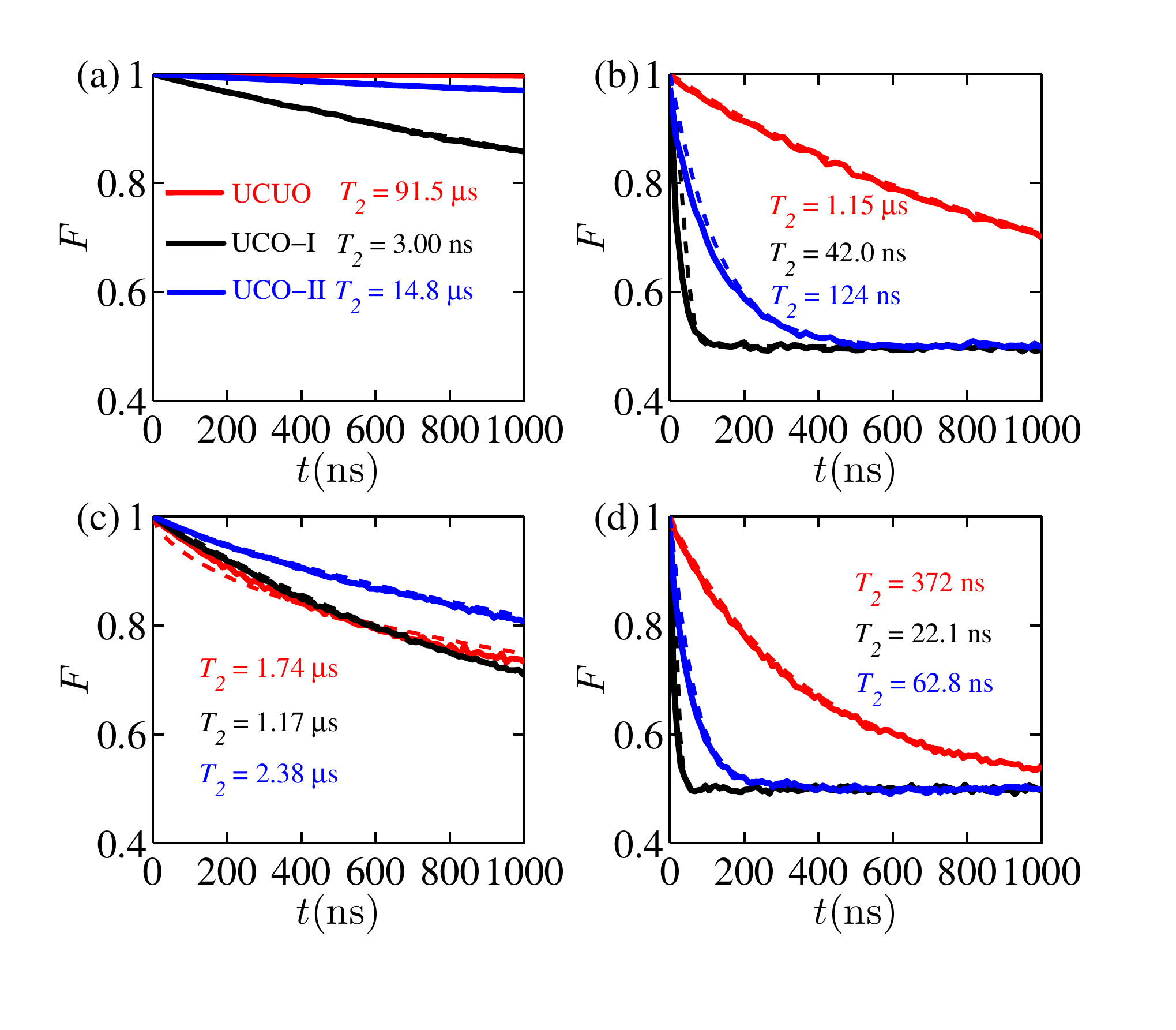}
\caption{Plots of average fidelity extracted from randomized benchmarking simulations as a function of time for our uncorrected sequence
sets (UCUO, UCO-I, UCO-II) in the presence of $1/f^\alpha$ noise.  The solid curves are data, and the dashed curves are fits to Eq.\
\eqref{Eq:FvsTimeFit}.  The left column (a, c) represents a quantum dot operated with barrier control\cite{ReedPRL2016,MartinsPRL2016}, while the right
column (b, d) uses tilt control.  The top row (a, b) represents a system with no field noise (e.g., isotopically purified Si) and magnetic
field gradient $h_0=23\text{ MHz}$, and the bottom row (c, d) a system with field noise and $h_0=40\text{ MHz}$.  For the UCO-II pulse
sequences, we use the ``na\"ive'' versions of the corresponding CO-II pulse sequences, with the same parameters.}
\label{Fig:RB_OneOverF_Uncorrected}
\end{figure}

\begin{figure}[ht]
\includegraphics[width=\columnwidth]{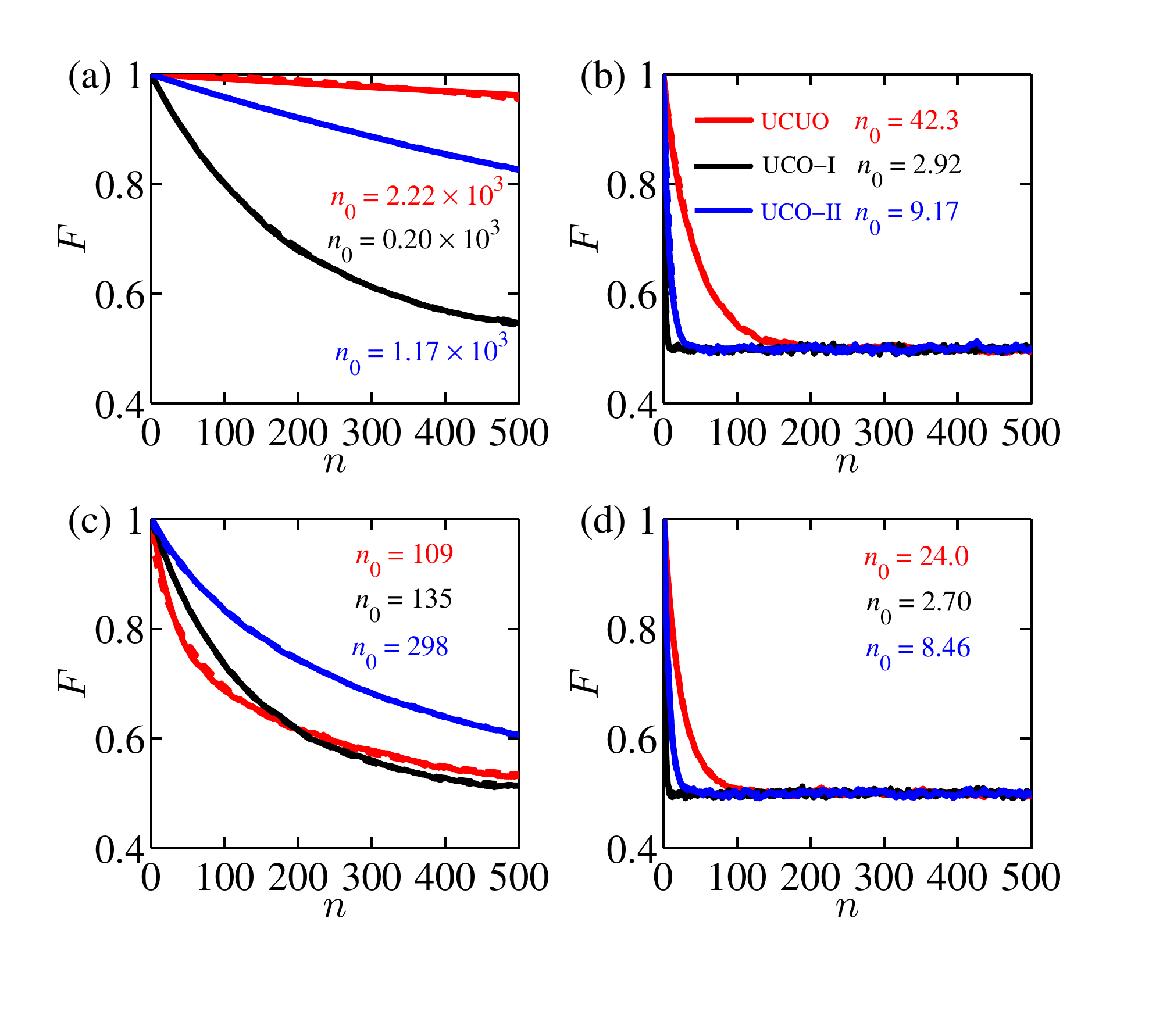}
\caption{Plots of average fidelity extracted from randomized benchmarking simulations as a function of number of gates for our uncorrected
sequence sets (UCUO, UCO-I, UCO-II) in the presence of $1/f^\alpha$ noise.  The solid curves are data, and the dashed curves are fits to Eq.\
\eqref{Eq:FvsTimeFit}.  The left column (a, c) represents a quantum dot operated with barrier control\cite{ReedPRL2016,MartinsPRL2016}, while the right
column (b, d) uses tilt control.  The top row (a, b) represents a system with no field noise (e.g., isotopically purified Si) and magnetic
field gradient $h_0=23\text{ MHz}$, and the bottom row (c, d) a system with field noise and $h_0=40\text{ MHz}$.  For the UCO-II pulse sequences,
we use the ``na\"ive'' versions of the corresponding CO-II pulse sequences, with the same parameters.}
\label{Fig:RB_OneOverF_Uncorrected_Gates}
\end{figure}

We now present our plots of the fidelity for the uncorrected sequence sets as a function of time in Fig.\ \ref{Fig:RB_OneOverF_Uncorrected}
and as a function of the number of gates in Fig.\ \ref{Fig:RB_OneOverF_Uncorrected_Gates}.  We find from both sets of plots that, in the absence
of field noise, the UCUO sequences both have the longest $T_2$ and allow one to perform the longest gate sequences of any of the sets.  This
also remains true even in the presence of field noise if one uses tilt control.  However, if one uses barrier control in the presence of field noise,
then the UCO-II sequences are best.

\begin{figure}[ht]
\includegraphics[width=\columnwidth]{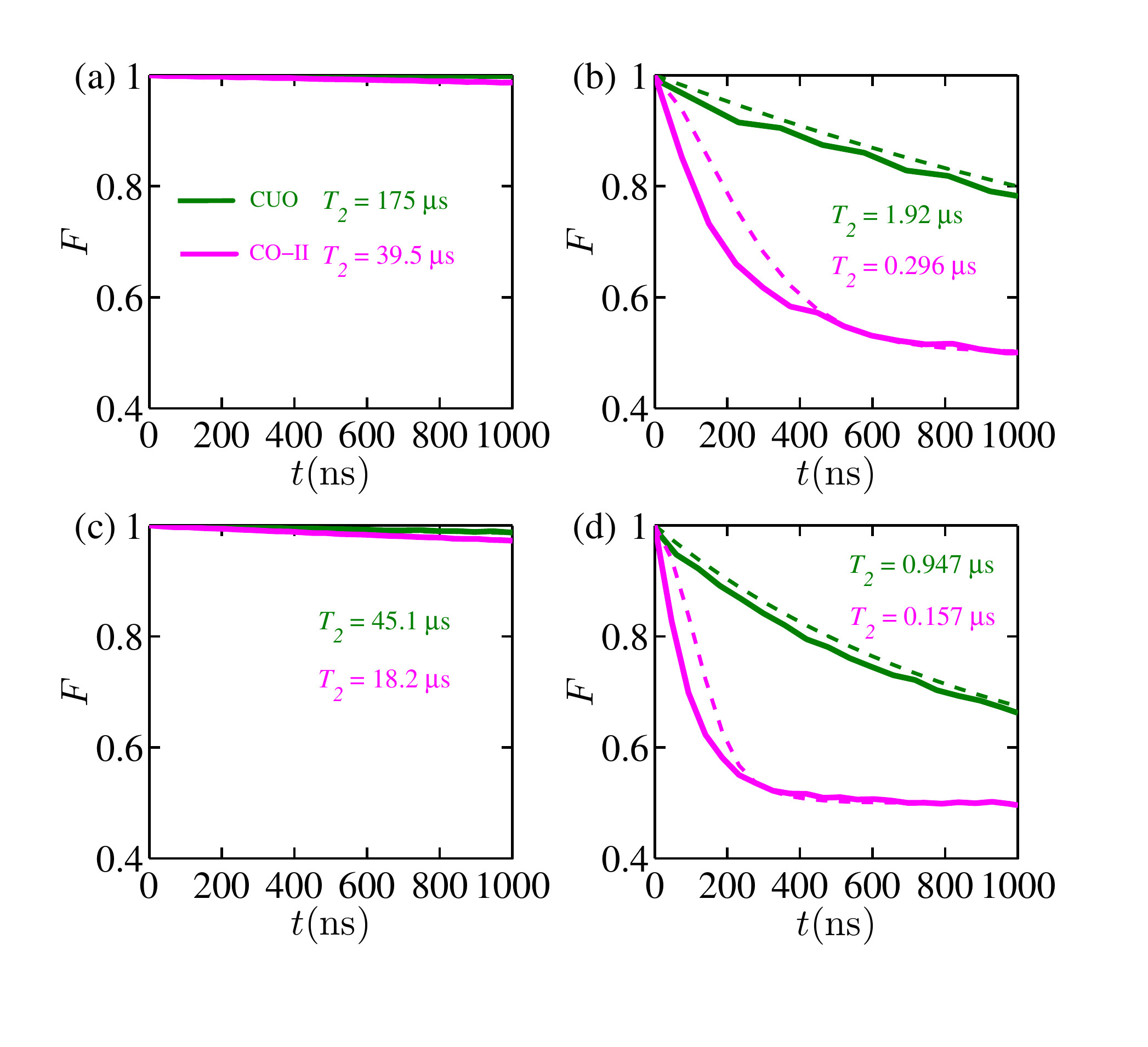}
\caption{Plots of average fidelity extracted from randomized benchmarking simulations as a function of time for our corrected sequence
sets (CUO, CO-II) in the presence of $1/f^\alpha$ noise.  The left column (a, c) represents a quantum dot operated with barrier
control\cite{ReedPRL2016,MartinsPRL2016}, while the right column (b, d) uses tilt control.  The top row (a, b) represents a system with no field noise
(e.g., isotopically purified Si) and magnetic field gradient $h_0=23\text{ MHz}$, and the bottom row (c, d) a system with field noise
and $h_0=40\text{ MHz}$.}
\label{Fig:RB_OneOverF_Corrected}
\end{figure}

\begin{figure}[ht]
\includegraphics[width=\columnwidth]{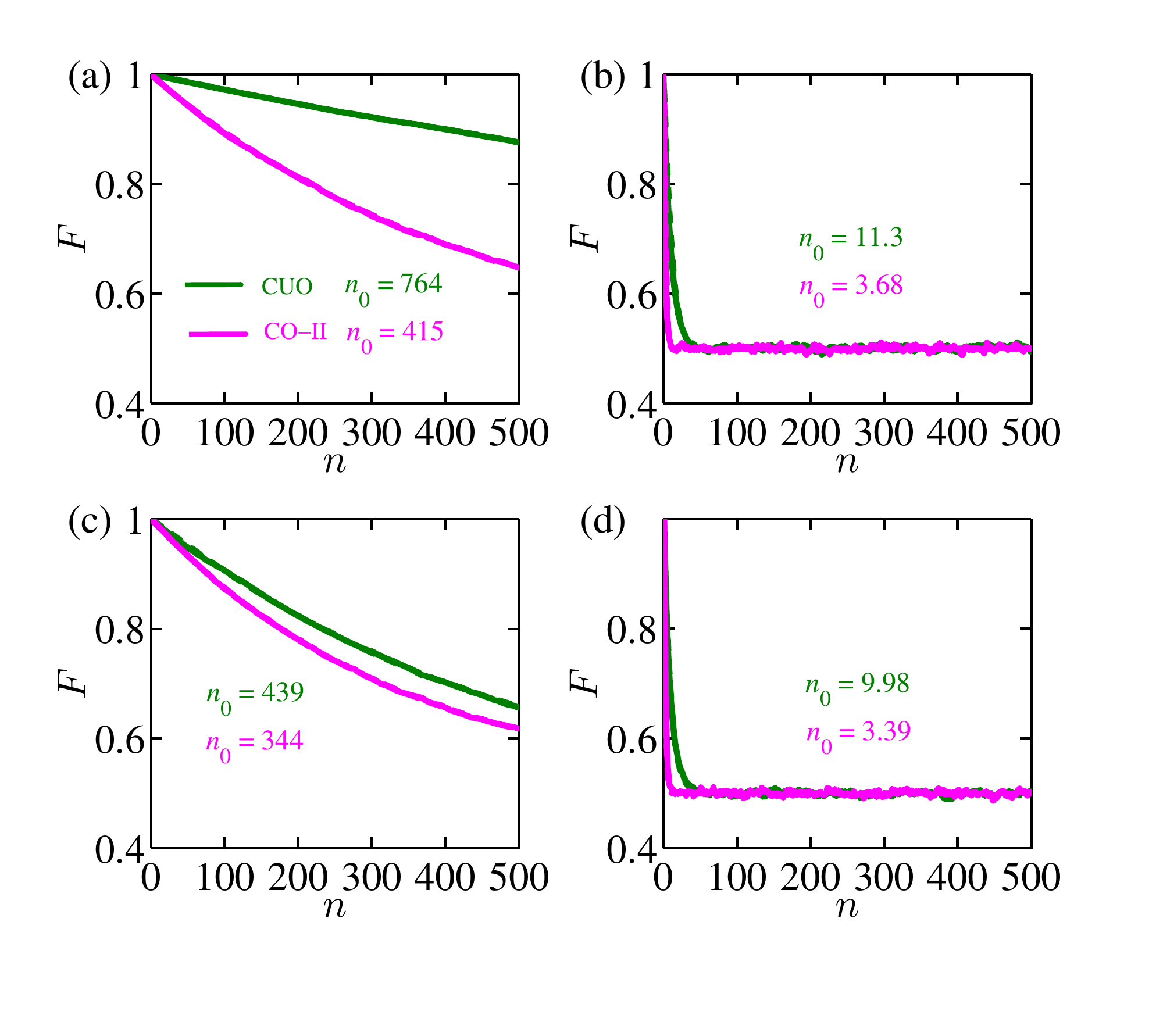}
\caption{Plots of average fidelity extracted from randomized benchmarking simulations as a function of number of gates for our corrected
sequence sets (CUO, CO-II) in the presence of $1/f^\alpha$ noise.  The left column (a, c) represents a quantum dot operated with barrier
control\cite{ReedPRL2016,MartinsPRL2016}, while the right column (b, d) uses tilt control.  The top row (a, b) represents a system with no field noise
(e.g., isotopically purified Si) and magnetic field gradient $h_0=23\text{ MHz}$, and the bottom row (c, d) a system with field noise
and $h_0=40\text{ MHz}$.}
\label{Fig:RB_OneOverF_Corrected_Gates}
\end{figure}

We plot the corresponding fidelities for the corrected sequence sets in Figs.\ \ref{Fig:RB_OneOverF_Corrected} and \ref{Fig:RB_OneOverF_Corrected_Gates},
respectively.  We see that, in all four combinations of types of control (barrier vs. tilt) and presence/absence of field noise, we obtain the
best $T_2$ using the CUO pulse sequences. This result is likely due to the fact that, in this case, significant high-frequency components are
present in the charge noise due to the low exponent $\alpha_J=0.7$; the fact that we set $J$ to larger values in the CO-II sequences than in the
CUO sequences results in a larger amount of noise.  However, we see that, in the case of barrier control in the presence of field noise, we can
perform roughly the same number of gates with either set of pulse sequences.  This is because the CO-II sequences are shorter in duration than
the corresponding CUO sequences, so that, even though $T_2$ may be shorter for the CO-II sequences, we can perform as many gates in that time span
as we can with the CUO sequences within their longer $T_2$.

The fitting parameters for results shown in Figs.\ \ref{Fig:RB_OneOverF_Uncorrected} through \ref{Fig:RB_OneOverF_Corrected_Gates} are summarized in
Table \ref{Tab:FitP_OneOverF} in Appendix \ref{App:FittingP}.


\begin{figure}[ht]
\includegraphics[width=\columnwidth]{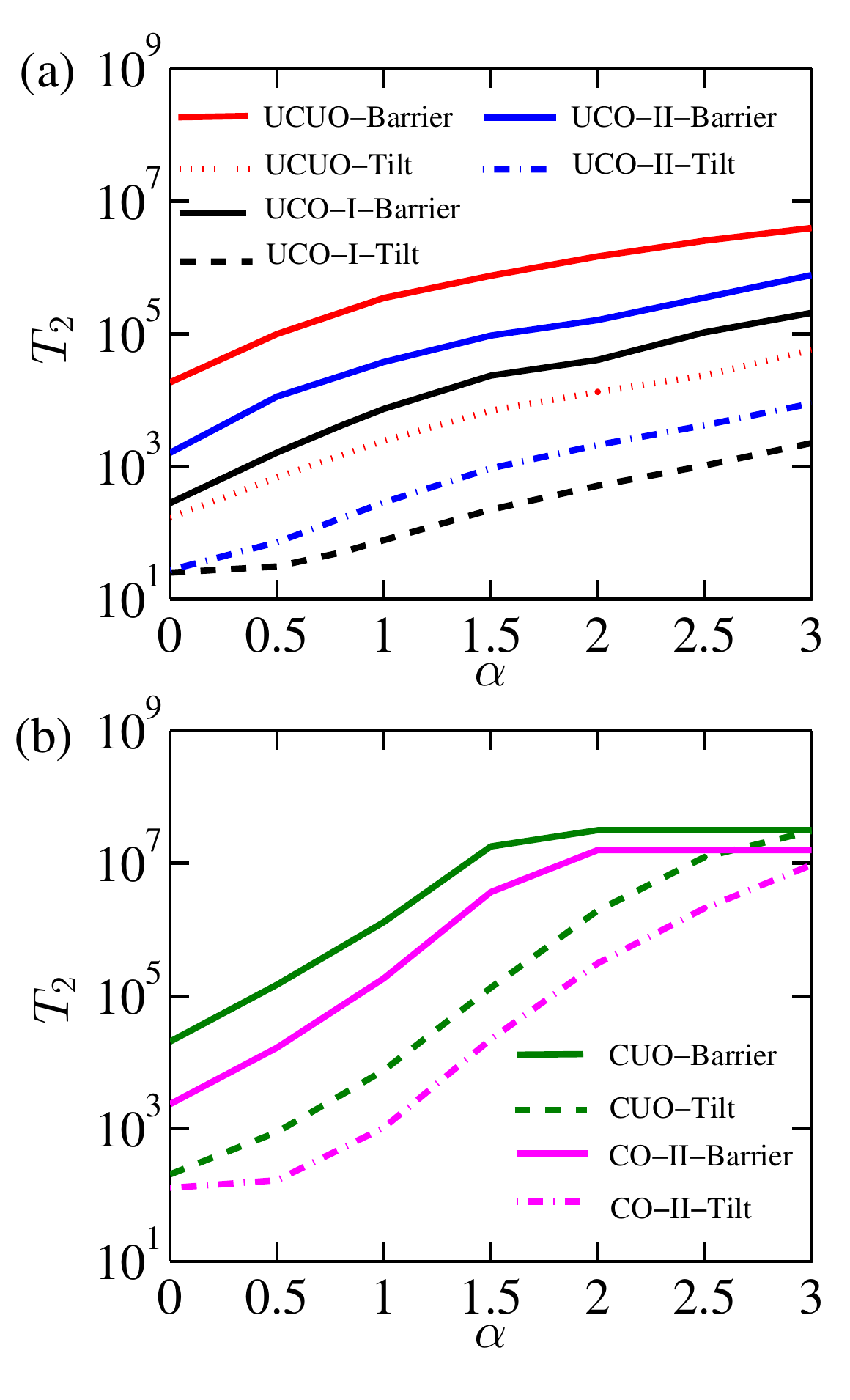}
\caption{Plots of $T_2$ extracted from randomized benchmarking simulations in the presence of $1/f^\alpha$ noise for (a) the uncorrected
sequence sets and (b) the corrected sets with no field noise as functions of $\alpha_J=\alpha$.}
\label{Fig:RB_OneOverF-noFieldNoise}
\end{figure}

\begin{figure}[ht]
\includegraphics[width=\columnwidth]{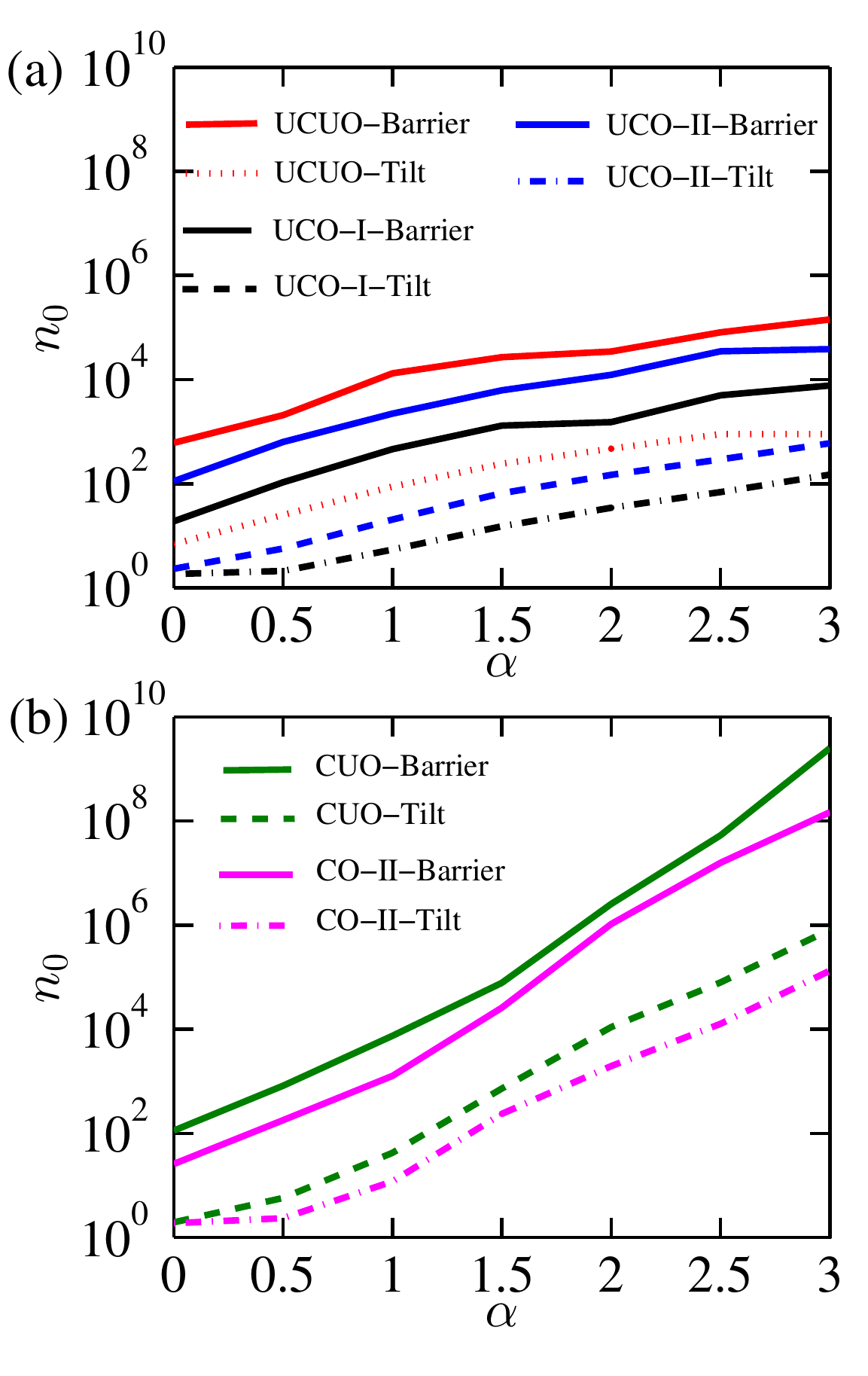}
\caption{Plots of $n_0$ extracted from randomized benchmarking simulations in the presence of $1/f^\alpha$ noise for (a) the uncorrected
sequence sets and (b) the corrected sets with no field noise as functions of $\alpha_J=\alpha$.}
\label{Fig:RB_OneOverF-noFieldNoise_Gates}
\end{figure}

We now consider $T_2$ and $n_0$ as functions of $\alpha_h$ and $\alpha_J$.  We will treat three cases here, one in which there is no field
noise and we vary $\alpha_J$, one in which we fix $\alpha_h=2.6$ and vary $\alpha_J$, and one in which we take $\alpha_h=\alpha_J=\alpha$
and vary $\alpha$.  We give our results for $T_2$ and $n_0$ in the first case in Figs.\ \ref{Fig:RB_OneOverF-noFieldNoise} and
\ref{Fig:RB_OneOverF-noFieldNoise_Gates}, respectively.  The results that we obtain here are relatively simple; by both metrics, the UCUO sequences outperform the other uncorrected sequences and the CUO sequences outperform the CO-II sequences.

\begin{figure}[ht]
\includegraphics[width=\columnwidth]{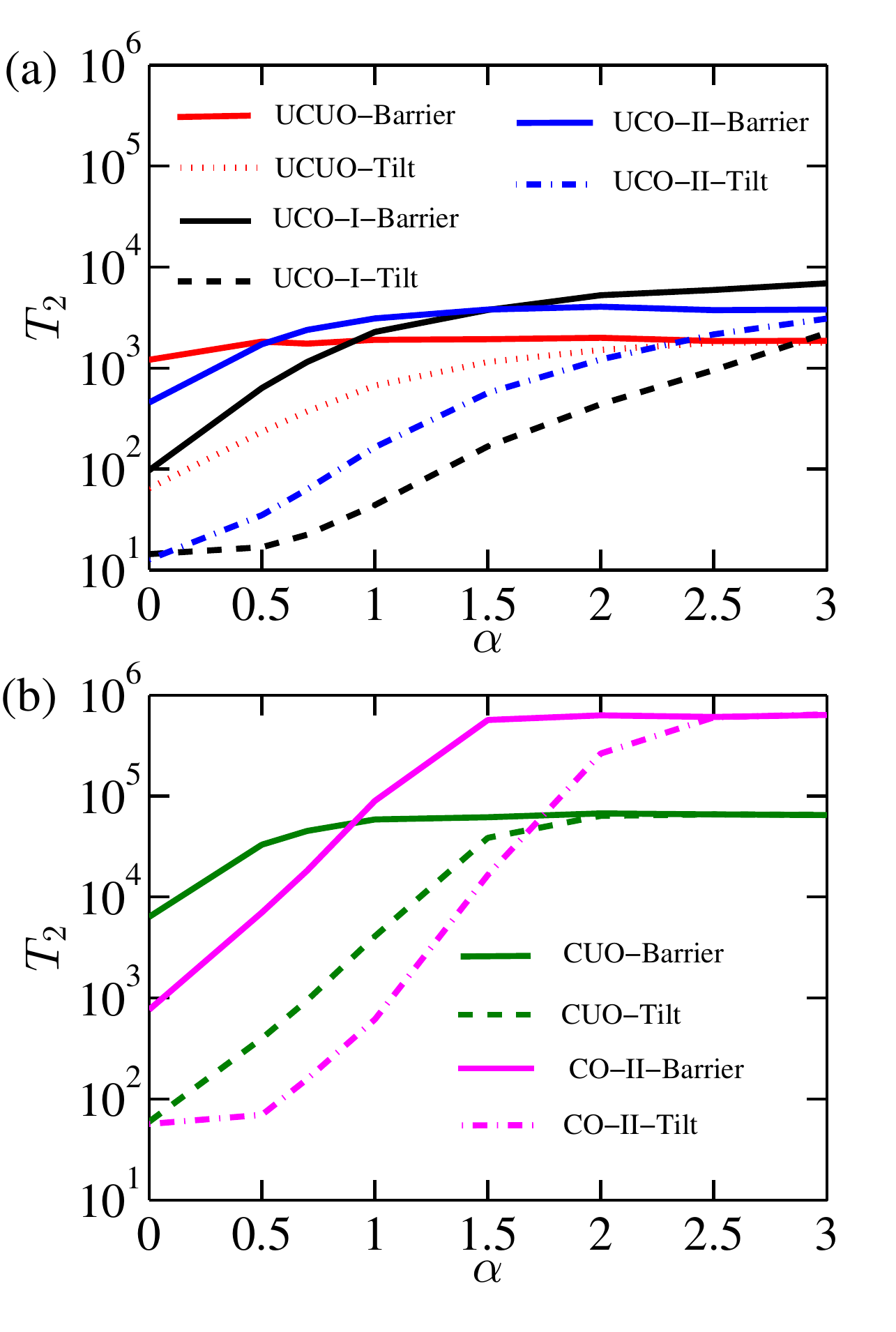}
\caption{Plots of $T_2$ extracted from randomized benchmarking simulations in the presence of $1/f^\alpha$ noise for (a) the uncorrected
sequence sets and (b) the corrected sets with $\alpha_h=2.6$ and as functions of $\alpha_J=\alpha$.}
\label{Fig:RB_OneOverF-alpha_h26}
\end{figure}

\begin{figure}[ht]
\includegraphics[width=\columnwidth]{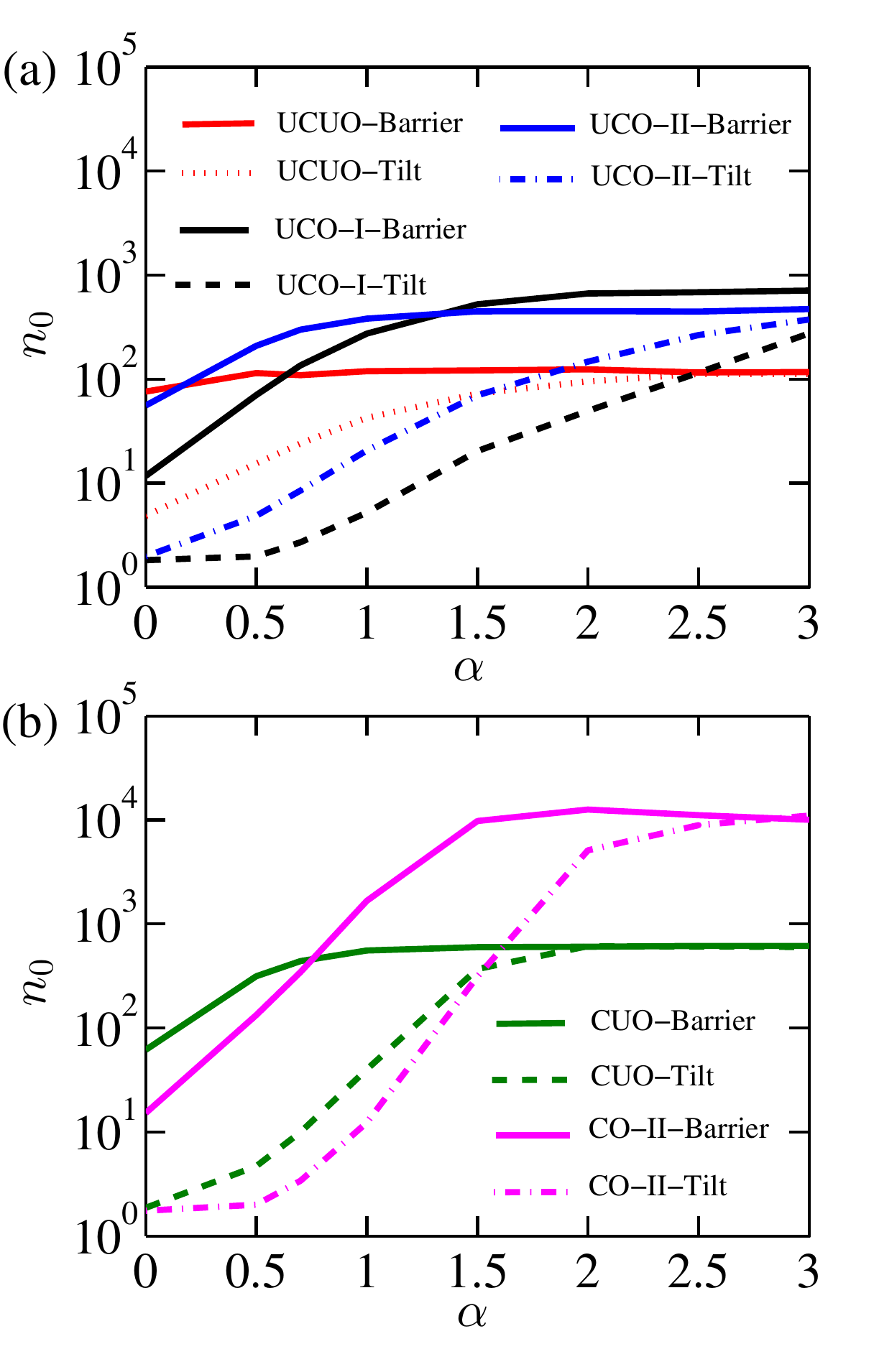}
\caption{Plots of $n_0$ extracted from randomized benchmarking simulations in the presence of $1/f^\alpha$ noise for (a) the uncorrected
sequence sets and (b) the corrected sets with $\alpha_h=2.6$ and as functions of $\alpha_J=\alpha$.}
\label{Fig:RB_OneOverF-alpha_h26_Gates}
\end{figure}

We present our plots of $T_2$ as a function of $\alpha_J$ with $\alpha_h=2.6$ in Fig.\ \ref{Fig:RB_OneOverF-alpha_h26} and the corresponding
plots of $n_0$ in Fig.\ \ref{Fig:RB_OneOverF-alpha_h26_Gates}.  We notice that, for barrier control, we see two points at which we transition
from one sequence set being better to a different set being superior.  For $\alpha_J<0.5$, we find that the UCUO sequences result in the longest
$T_2$.  Above this value, we instead find that the UCO-II sequences give the longest $T_2$.  Another transition occurs around $\alpha_J=1.5$,
above which the UCO-I sequences give the best $T_2$.  If we use tilt control, on the other hand, then we find that $T_2$ is shorter for all
values of $\alpha_J$ for any given sequence set than for barrier control, consistent with our previous results.  We also find only one transition
around $\alpha_J=2.2$, below which the UCUO sequences give the longest $T_2$, and above which the UCO-II sequences are best.  For the corrected
sequences, we see that there is a transition at $\alpha=0.9$ for barrier control and around $\alpha=1.7$ for tilt control, below (above) which the
CUO (CO-II) sequences give a longer $T_2$.

We find similar results in the plots of $n_0$, but the transitions happen at different values of $\alpha_J$ than in the plot of $T_2$---the
UCUO to UCO-II transition for barrier control now happens around $\alpha_J=0.2$ and the UCO-II to UCO-I transition around $\alpha_J=1.4$.
We also see that the UCUO-to-UCO-II transition has shifted to around $\alpha=1.5$.  This helps to emphasize our earlier point that a longer
$T_2$ will not necessarily mean that we can perform more gates; the time required to execute the individual pulse sequences matters as well.
This indicates that, for example, for $0.2<\alpha_J<0.5$, while $T_2$ is shorter for the UCO-II sequences than for UCUO, the fact that the
UCO-II sequences take less time to perform means that we can actually perform {\it more} gates reliably if we were to use the UCO-II sequences.
We see similar behavior for the corrected sequences; now the CO-II sequences outperform the CUO sequences for $\alpha\gtrsim 0.7$ when using
barrier control and for $\alpha\gtrsim 1.5$ for tilt control.

\begin{figure}[ht]
\includegraphics[width=\columnwidth]{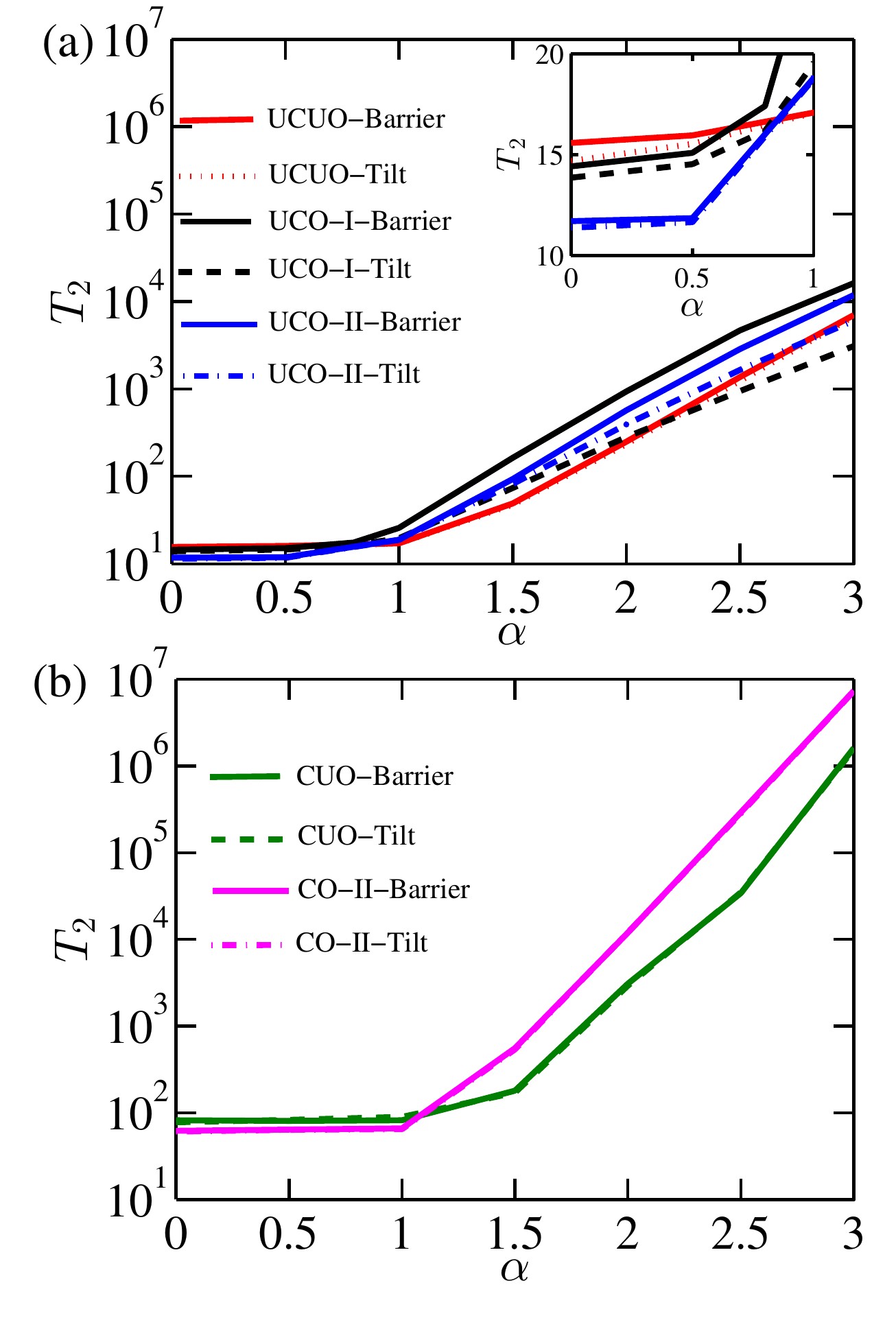}
\caption{Plots of $T_2$ extracted from randomized benchmarking simulations in the presence of $1/f^\alpha$ noise for (a) the uncorrected
sequence sets and (b) the corrected sets as functions of $\alpha_h=\alpha_J=\alpha$.}
\label{Fig:RB_OneOverF-alpha_hEqalpha_J}
\end{figure}

\begin{figure}[ht]
\includegraphics[width=\columnwidth]{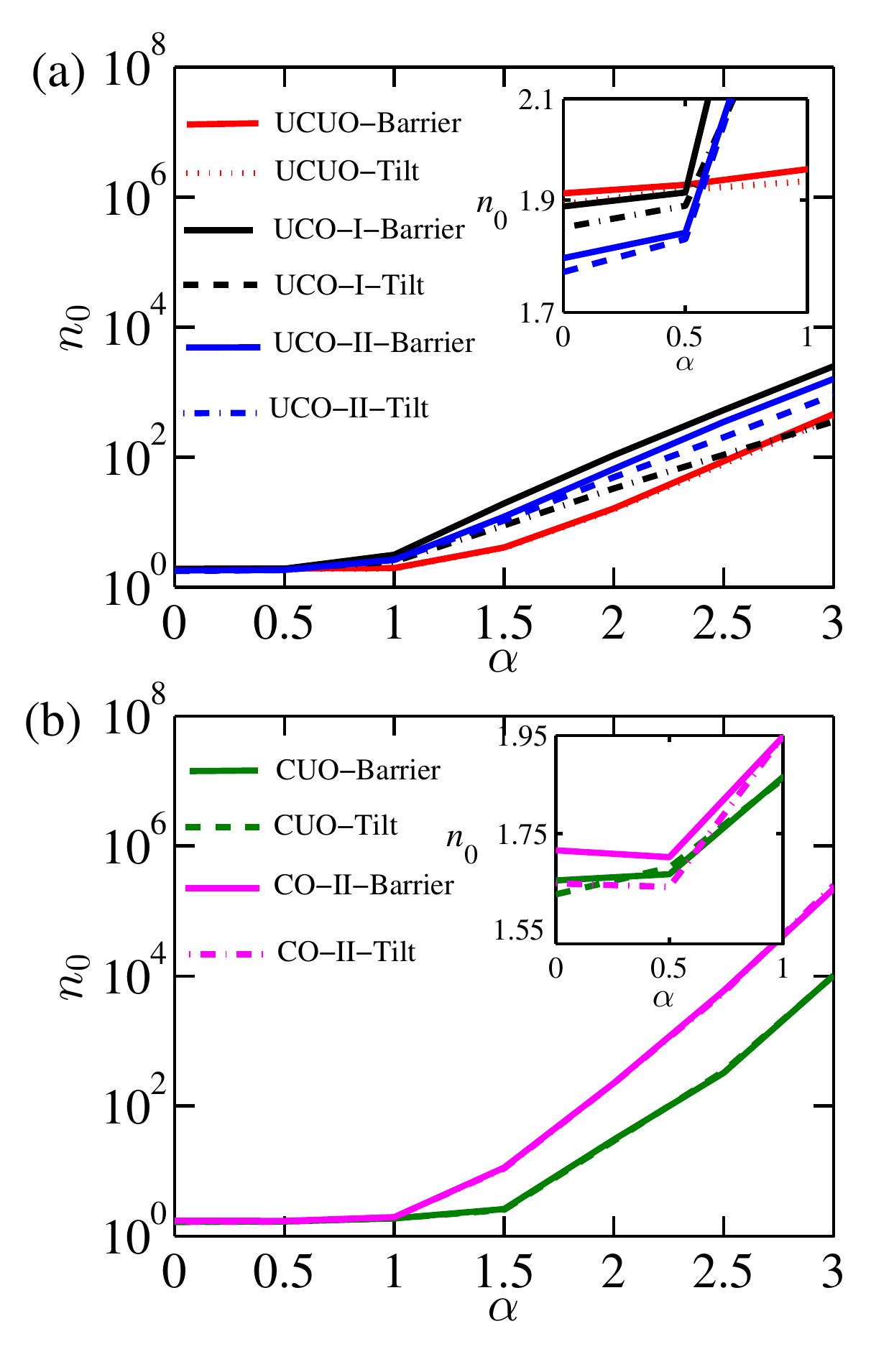}
\caption{Plots of $n_0$ extracted from randomized benchmarking simulations in the presence of $1/f^\alpha$ noise for (a) the uncorrected
sequence sets and (b) the corrected sets as functions of $\alpha_h=\alpha_J=\alpha$.}
\label{Fig:RB_OneOverF-alpha_hEqalpha_J_Gates}
\end{figure}

Now let us turn to the case, $\alpha_h=\alpha_J=\alpha$.  We plot our results for $T_2$ as a function of $\alpha$ in Fig.\ \ref{Fig:RB_OneOverF-alpha_hEqalpha_J}
and for $n_0$ in Fig.\ \ref{Fig:RB_OneOverF-alpha_hEqalpha_J_Gates}.  For the uncorrected sequence sets, we find that, in the case of barrier
control, there is a transition around $\alpha=0.6$, below which the UCUO sequences have the longest $T_2$ and above which the UCO-I sequences
do.  In the case of tilt control, we again find that the UCUO sequences give the longest $T_2$ below this transition, but it is the UCO-II
sequences that have the longest $T_2$ above it.  We see a similar transition in our plots of $n_0$, but it happens around $\alpha=0.5$.  For
the corrected gates, we find a transition around $\alpha=1.1$, below which we find that the CUO gates have a slightly longer $T_2$ than the
CO-II gates, and above which the CO-II gates have a longer $T_2$.  In our data for $n_0$, we find that the CO-II sequences do better for all
values of $\alpha$ when using barrier control; when using tilt control, the CUO sequences overtake the CO-II sequences for a small range of
$\alpha$ around $0.5$.  Overall, we see very little difference, either qualitatively or quantitatively, between barrier and tilt control in
this situation.

\section{Conclusion} \label{Sec:Conclusion}
We have presented a two-pronged approach to improving gate performance in singlet-triplet semiconductor double quantum dot spin qubits.
First, we reviewed existing sets of pulse sequences found in the literature and introduced new sets.  Two of the existing sets do not incorporate
dynamical error correction, while the third does via the \textsc{supcode} technique\cite{WangPRA2014}.  We introduced one new uncorrected set as
well as a corrected set based on these new sequences.  The uncorrected sets are the ``uncorrected unoptimized'' (UCUO) sequence, which implements
$z$ rotations via a sequence introduced by Guy Ramon\cite{RamonPRB2011}, and the ``uncorrected optimized, type I'' (UCO-I) sequence, which instead
uses a generalization of the Hadamard-$x$-Hadamard sequence, a ``$\theta$-$2\theta$-$\theta$'' sequence\cite{ZhangPRL2017}.  Both utilize the $z$-$x$-$z$
sequence to implement rotations, although UCO-I additionally makes use of the $\theta$-$2\theta$-$\theta$ sequence to perform the $x$ rotation to
avoid requiring that the exchange coupling be completely switched off.  Our new uncorrected sequences are dubbed the ``uncorrected optimized, type
II'' (UCO-II) sequences.  The $z$ rotations are implemented using the same sequence as in the UCO-I set, but we also introduced new sequences for
implementing the other rotations, all of which are shorter in terms of the number of single-pulse rotations that they use.  We show that, furthermore,
these new sequences can be made up to twice as fast in terms of the time required to perform them than the other two sets.  Similarly, the corrected
sets are as follows.  The existing set is the ``corrected unoptimized'' (CUO) set, which assumes that $x$ rotations may be done with a single pulse,
i.e., it is possible to set the exchange coupling exactly to zero.  The $z$ rotation is then implemented via the Hadamard-$x$-Hadamard sequence, and
all others using an $x$-$z$-$x$ decomposition.  These sequences are then corrected using the \textsc{supcode} method, in which one inserts a sequence
of pulses that, ideally, would perform an identity operation and is arranged in such a way that the error in this ``identity'' exactly cancels the
error in the base pulse sequence to first order.  The new set of corrected sequences is the ``corrected optimized, type II'' (CO-II) set, which we
obtained by applying the same idea to the UCO-II sequences.  We demonstrated that these sequences, similarly to the corresponding uncorrected set,
are up to three times faster than the CUO sequences.

We then evaluated the performance of these sets in the presence of (magnetic) field and charge noise using randomized benchmarking\cite{MagesanPRL2011,MagesanPRA2012}.
This procedure consists of determining the state-averaged fidelity of random sequences of gates of a given length, allowing one to determine the
fidelity as a function of the number of gates or of time.  We did exactly this for two different models of noise, quasistatic (Gaussian) and $1/f^\alpha$,
and for two types of control of the qubit, barrier and tilt, which correspond to having two different amounts of charge noise\cite{MartinsPRL2016}.  We then
performed fits to this data to determine the effective decoherence time $T_2$ and an analogous parameter $n_0$ characterizing how quickly the fidelity
decays with the number of gates performed.  In the case of quasistatic noise, we found that, in the complete absence of field noise, the UCUO sequences
give the longest $T_2$ and the highest values of $n_0$.  However, in the presence of field noise, we found that the ``optimized'' sequences do better---in
the case of barrier control, the UCO-I sequences do best, while all of the sequences have roughly equal performance for strong field noise and for tilt
control.  We note that the UCO-II sequences that are used in these simulations are not the fastest versions of these sequences, but rather the sequences
that the CO-II sequences are based on, some of which have no segments during which the exchange coupling is set to its (experimental) largest possible
value.  It turns out that attempting to correct the fastest versions of the UCO-II sequences via \textsc{supcode} requires us to insert uncorrected
identities with segments during which the exchange coupling exceeds its maximum value or becomes negative.  We believe that, if we were to use the
fastest versions of the UCO-II sequences, then they would likely yield the best results among the uncorrected sequences. We found similar results from the corrected sets; if there is
no field noise, then the CUO sequences outperform the CO-II sequences, having longer $T_2$ and a higher $n_0$.  However, once we introduce field noise,
the CO-II sequences are better.

We then considered the case of $1/f^\alpha$ noise.  We first considered the experimentally relevant case\cite{DialPRL2013,MedfordPRL2012} for which the
exponent for the field noise is $\alpha_h=2.6$ and that for the charge noise is $\alpha_J=0.7$.  Among the uncorrected pulse sequences, we find that the UCUO
sequences once again do best in the absence of field noise, and even in its presence when tilt control is used.  When we use barrier control instead of
tilt control in the presence of field noise, however, the UCO-II sequences do best.  In the case of the corrected sequences, we similarly find that the
CUO sequences are better in most cases, except when using barrier control in the presence of field noise, in which case the number of gates that can be
performed is roughly the same for both the CUO and CO-II sequences.  We then considered $T_2$ and $n_0$ as we vary the exponents for two cases, one in which
we fix $\alpha_h=2.6$ and vary $\alpha_J$, and another in which we set $\alpha_h=\alpha_J=\alpha$ and vary $\alpha$.  The results for the latter case are
relatively simple.  In the absence of field noise, we find that the ``unoptimized'' sequences always do better, regardless of the value of $\alpha$.  When we
turn on field noise, on the other hand, then we find that there is a transition from the ``unoptimized'' sequences being best to an ``optimized''
set being best (UCO-I for the uncorrected sequences, CO-II for the corrected sequences).  The results from the first case, in which we fix $\alpha_h$,
are more complex, featuring two transitions among the three uncorrected sequences; again, we see that ``unoptimized'' sequences do best for small $\alpha$
and ``optimized'' sequences tend to be best when $\alpha$ is small.

Overall, our results imply that which set of pulse sequences will work best depends on how much, and what type of, noise is present in the system
and, in the case of power law noise, the exponent characterizing the frequency dependence.  We see that, for example, the ``unoptimized'' sequences,
both corrected and uncorrected, seem to do best when there is little field noise, such as in isotopically-purified Si\cite{WitzelPRL2010}, and in
situations in which there are significant high-frequency components to the noise, even if it is only in the charge noise.  The ``optimized'' sequences,
however, do best in cases that better approximate the quasistatic limit and in the presence of field noise, and thus they would perform better in,
for example, natural Si or GaAs.  Our results suggest two avenues that one may pursue to improve control of singlet-triplet qubits.  One would be to
achieve exactly zero exchange coupling.  Existing results suggest that this may be possible in high magnetic fields or when we introduce more than
two electrons into the quantum dots\cite{HuPRA2001,NielsenPRB2013,MartinsArXiv2017}; it is found that the exchange coupling could become zero or
negative in these situations.  Another would be to reduce the high-frequency components present in the charge noise.  We find that, if this were to
be done, then one can achieve a longer $T_2$ and perform a larger number of gates with sufficiently high fidelity using the CO-II pulse sequences in
this situation.  In general, one must have some information about the type of noise prevalent in the system in order to best optimize the required
pulse sequences for singlet-triplet spin qubit gate operations in semiconductor quantum dots.

\acknowledgements
C.\ Z., X.-C.\ Y., and X.\ W.\ are supported by the Research Grants Council of the Hong Kong Special Administrative Region, China (No.\
CityU 21300116), the National Natural Science Foundation of China (No.\ 11604277), and the Guangdong Innovative and Entrepreneurial Research
Team Program (No.\ 2016ZT06D348). E.B. is supported by the US Army Research Office (W911NF-17-0287).  R.\ E.\ T.\ and S.\ D.\ S.\ are supported
by LPS-MPO-CMTC.

\appendix
\section{Generalized Ramon sequence for $z$ rotations} \label{App:GenRamon}
We note one major problem with the Ramon sequence---it requires one to perform an $x$ rotation.  This, in turn,
requires that we be able to completely turn off the exchange coupling between the electrons, which is experimentally challenging.  We therefore consider a generalized version of this sequence, in which the $x$ rotation is replaced
by one about an axis at an angle $\theta'$ to the $z$ axis:
\begin{equation}
R(\hat{\vec{z}},\phi)=R(\theta,\chi)R(\theta',\alpha)R(\theta,\chi). \label{Eq:ZRotGGR}
\end{equation}
If we now multiply out the right-hand side and set both sides equal as before, we once again find two solutions for
$\alpha$,
\begin{equation}
\alpha=-2\arcsin\left [\frac{\sin{\theta}}{\sin(\theta'-\theta)}\sin\left (\frac{\phi}{2}\right )\right ] \label{Eq:AlphaSolnGen1}
\end{equation}
and
\begin{equation}
\alpha=2\pi+2\arcsin\left [\frac{\sin{\theta}}{\sin(\theta'-\theta)}\sin\left (\frac{\phi}{2}\right )\right ]. \label{Eq:AlphaSolnGen2}
\end{equation}
The solutions for $\chi$ are
\begin{widetext}
\begin{equation}
\chi=\arccos\left\{\frac{\pm\cos\left (\frac{\phi}{2}\right )\sqrt{\sin^2(\theta'-\theta)-\sin^2\left (\frac{\phi}{2}\right )\sin^2{\theta}}-\sin^2\left (\frac{\phi}{2}\right )\cos(\theta'-\theta)\sin{\theta}\cos{\theta}}{\left [\cos^2\left (\frac{\phi}{2}\right )+\sin^2\left (\frac{\phi}{2}\right )\cos^2{\theta}\right ]\sin(\theta'-\theta)}\right\}, \label{Eq:ChiSolnGen}
\end{equation}
\end{widetext}
where, once again, our choice of sign for $\pm$ depends on which solution for $\alpha$ we use; we choose the plus sign
if we use Eq.\ \eqref{Eq:AlphaSolnGen1} and the minus sign if we use Eq.\ \eqref{Eq:AlphaSolnGen2}.  We use the first set
of solutions, i.e., we choose the plus sign in Eq.\ \eqref{Eq:ChiSolnGen} and use Eq.\ \eqref{Eq:AlphaSolnGen1} to obtain
$\alpha$, if $\phi$ is positive, while we use the second set if $\phi$ is negative.  One may verify that we recover the
equations for the Ramon sequence if we set $\theta'=\frac{\pi}{2}$.  Looking at our solution for $\alpha$, we see that we
can only obtain a real-valued solution for any value of $\phi$ if $\theta\leq \frac{1}{2}\theta'$.  We note that, with this
restriction on $\theta$ in place, Eq.\ \eqref{Eq:AlphaSolnGen1} will always give us a negative value of $\alpha$.  In this
case, we may, as with the Ramon sequence, add $2\pi$ to the solution, at the cost of introducing an overall minus sign to
the rotation.  The time required to perform this sequence may be determined similarly to that of the Ramon sequence, and
is given by
\begin{equation}
ht_{\text{tot},R}=\chi\sin{\theta}+\tfrac{1}{2}\alpha\sin{\theta'}.
\end{equation}

\begin{figure}[ht]
\includegraphics[width=0.49\columnwidth]{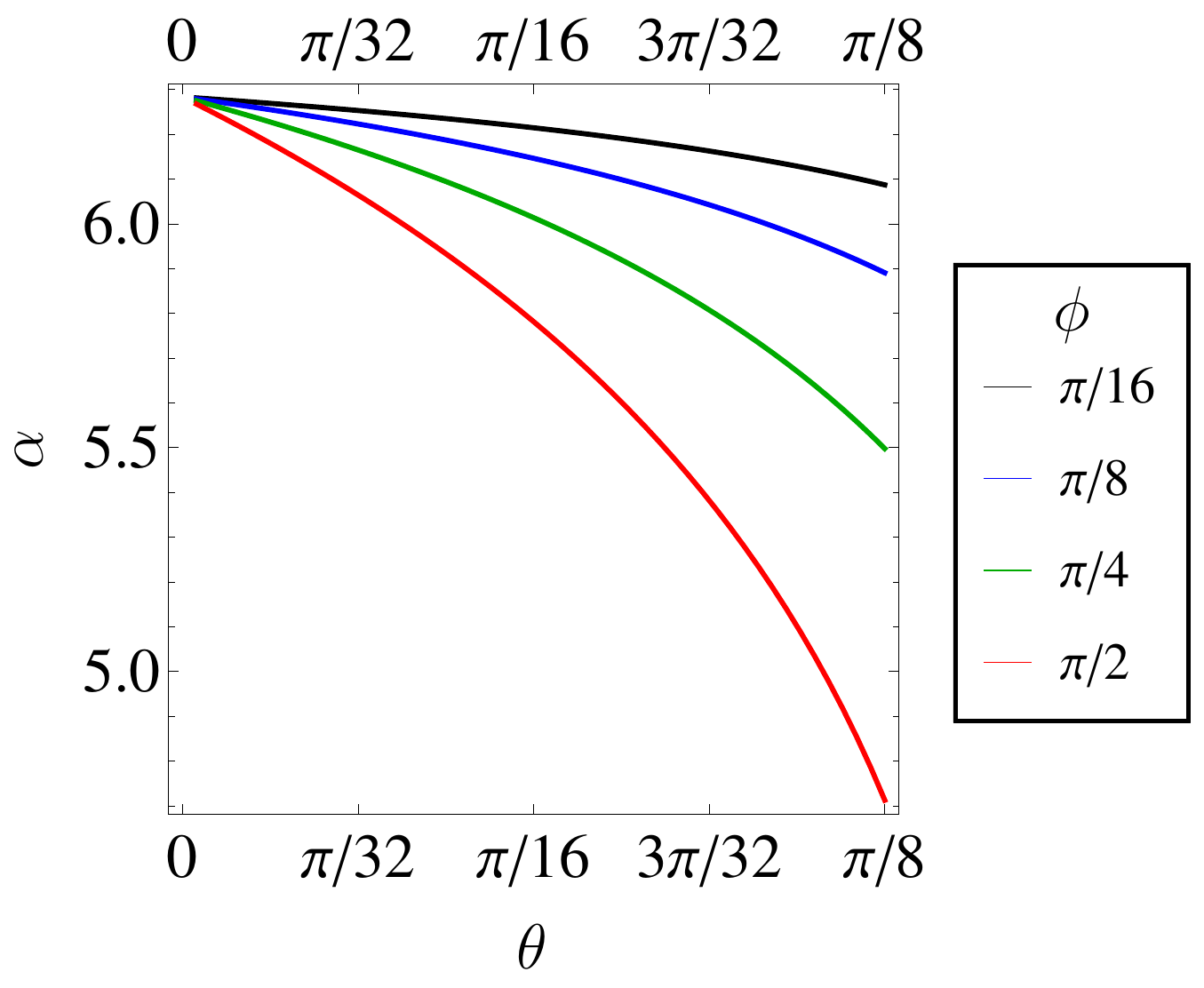}
\includegraphics[width=0.49\columnwidth]{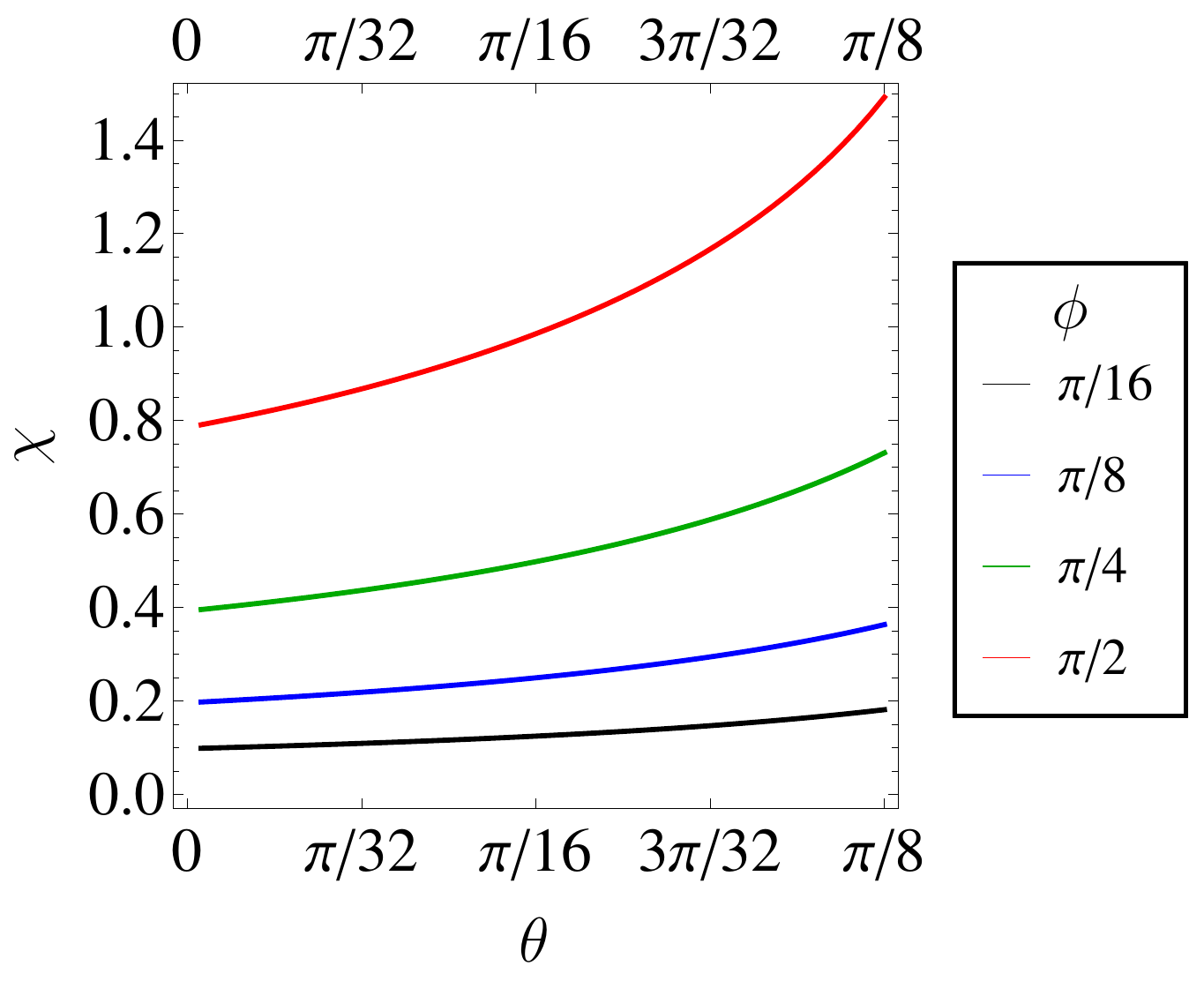}
\includegraphics[width=0.49\columnwidth]{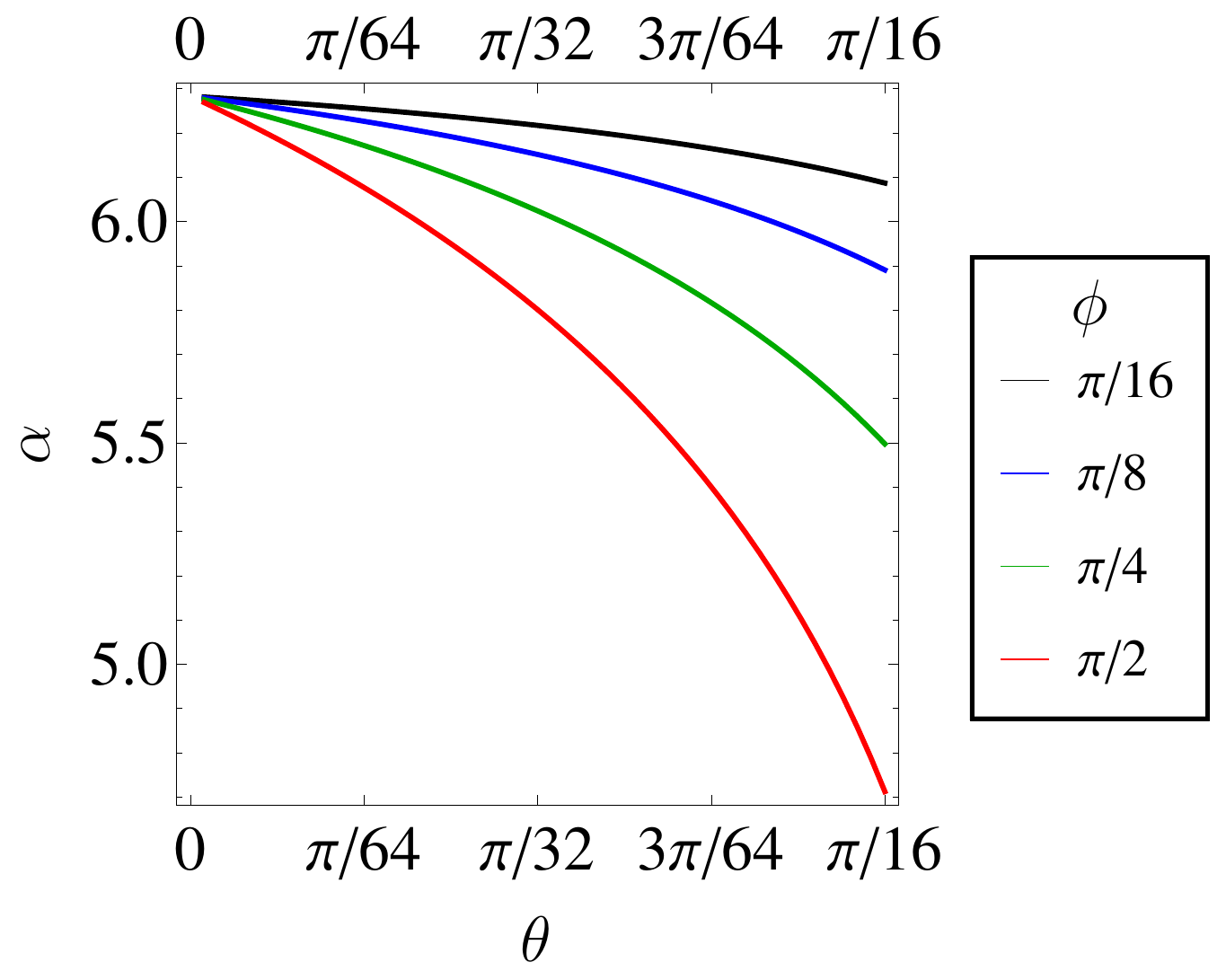}
\includegraphics[width=0.49\columnwidth]{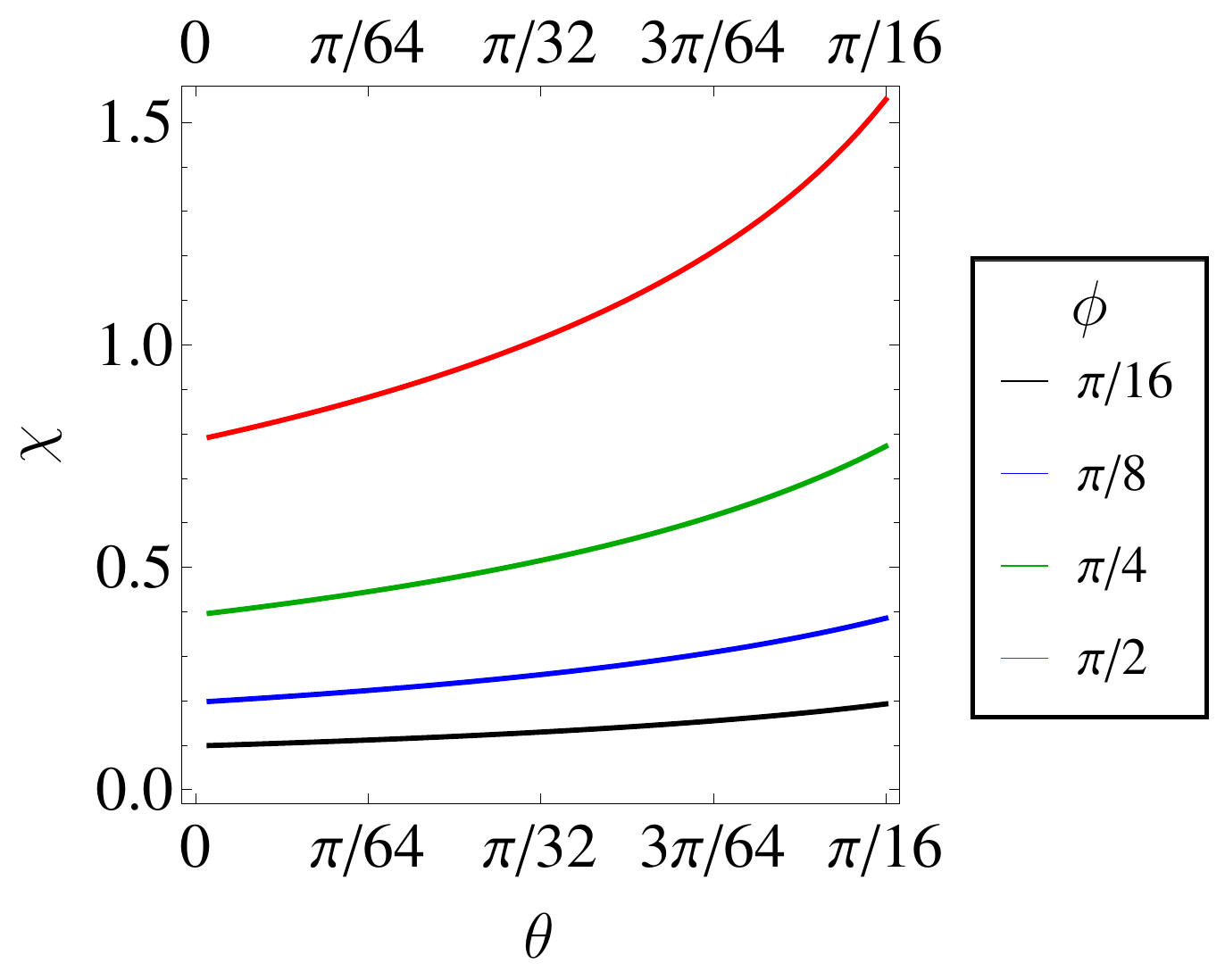}
\includegraphics[width=0.49\columnwidth]{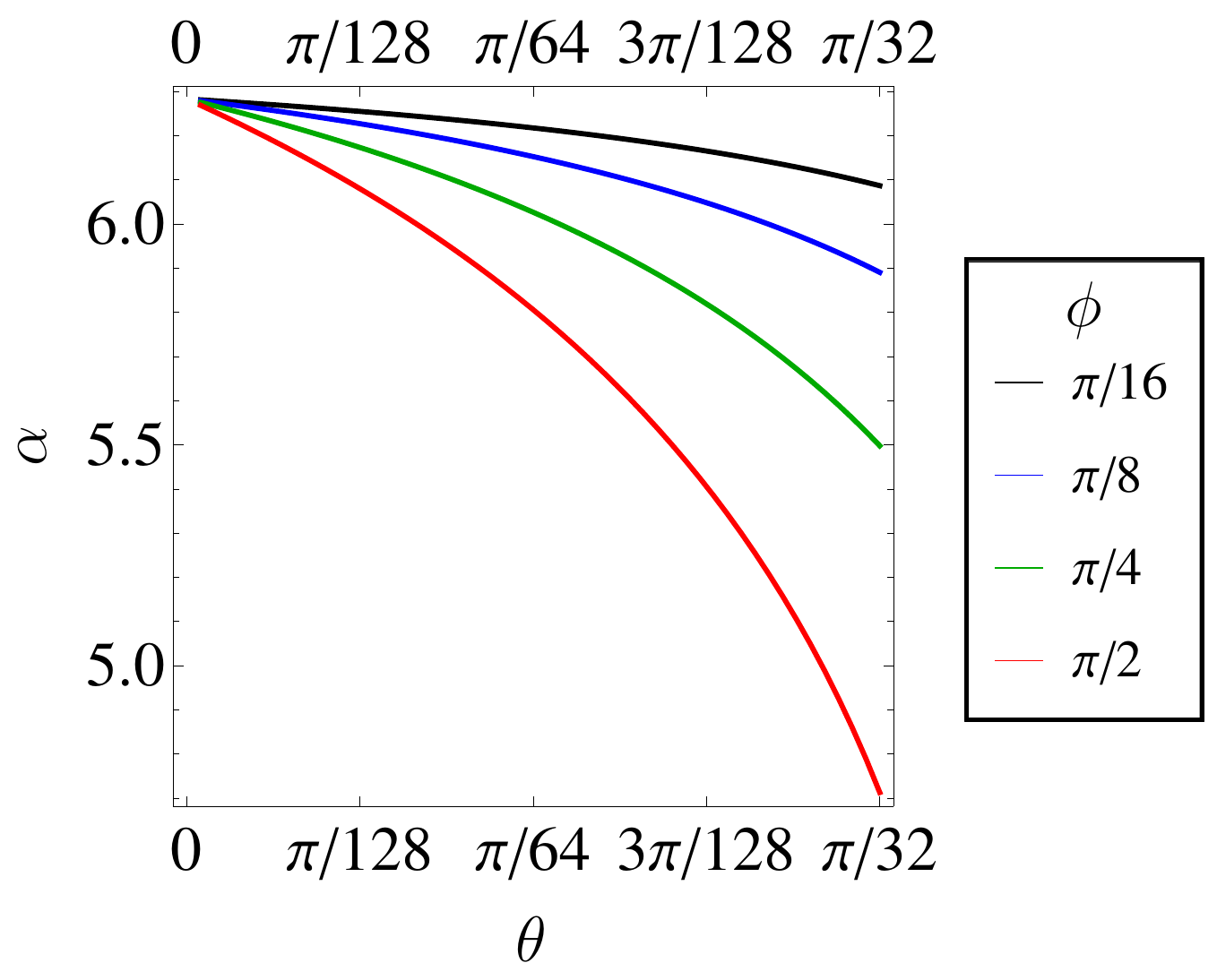}
\includegraphics[width=0.49\columnwidth]{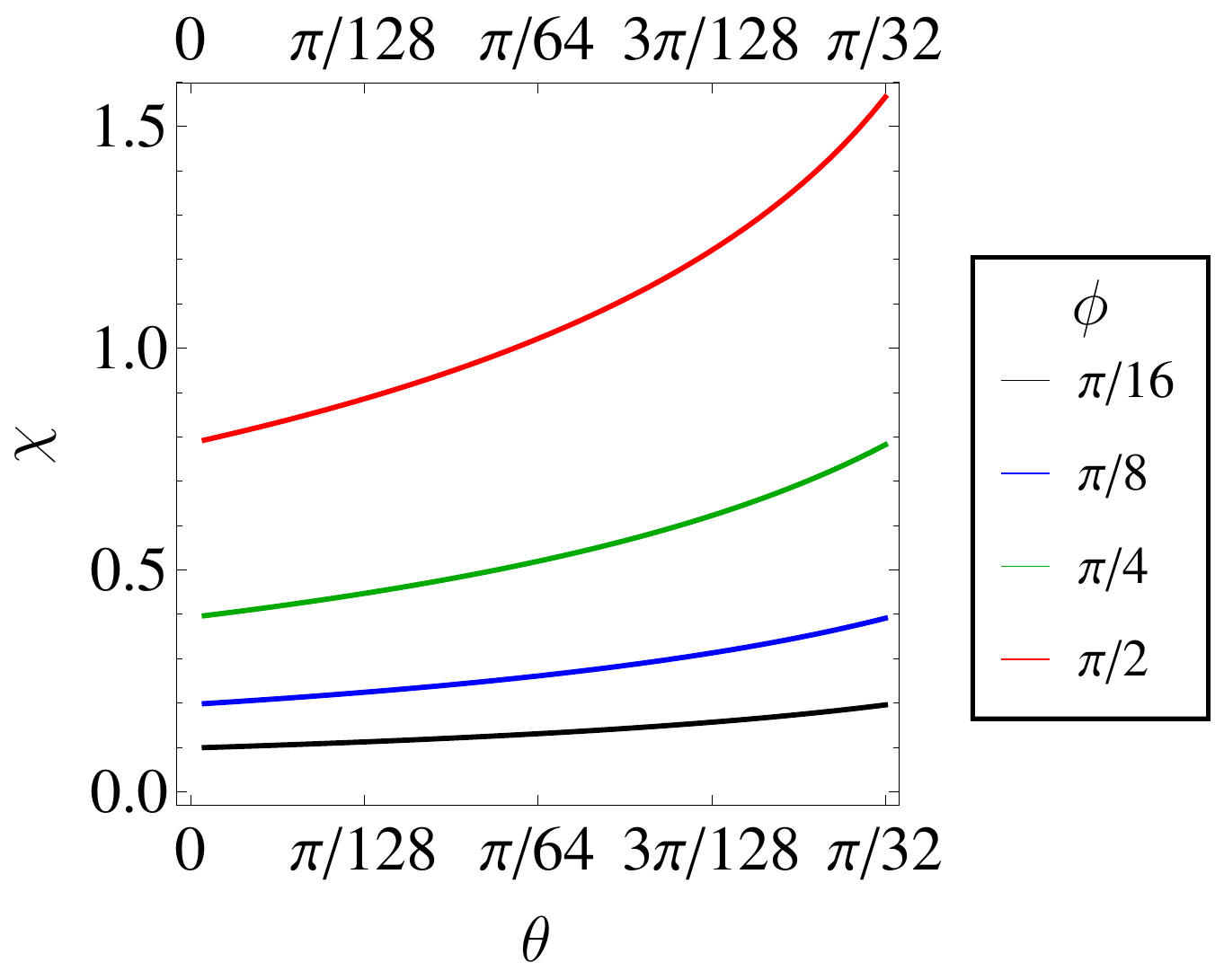}
\caption{Plots of $\alpha$ (left) and $\chi$ (right) for the ``generalized Ramon sequence'' with $\theta'=\frac{\pi}{4}$
(top row), $\frac{\pi}{8}$ (middle row), and $\frac{\pi}{16}$ (bottom row) and for positive values of $\phi$.}
\label{Fig:Plots_AlphaChiGRS}
\end{figure}

\begin{figure}[ht]
\includegraphics[width=0.49\columnwidth]{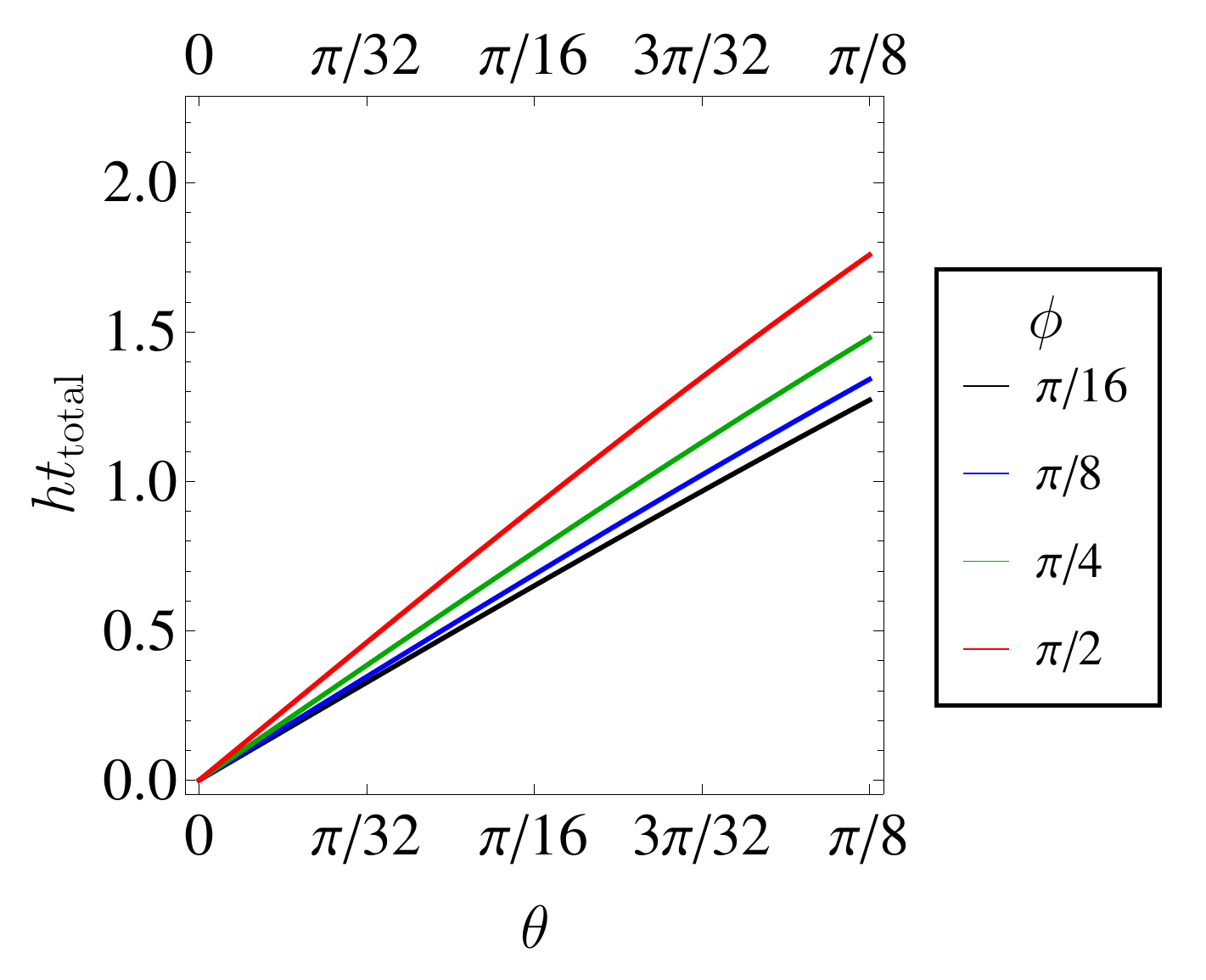}
\includegraphics[width=0.49\columnwidth]{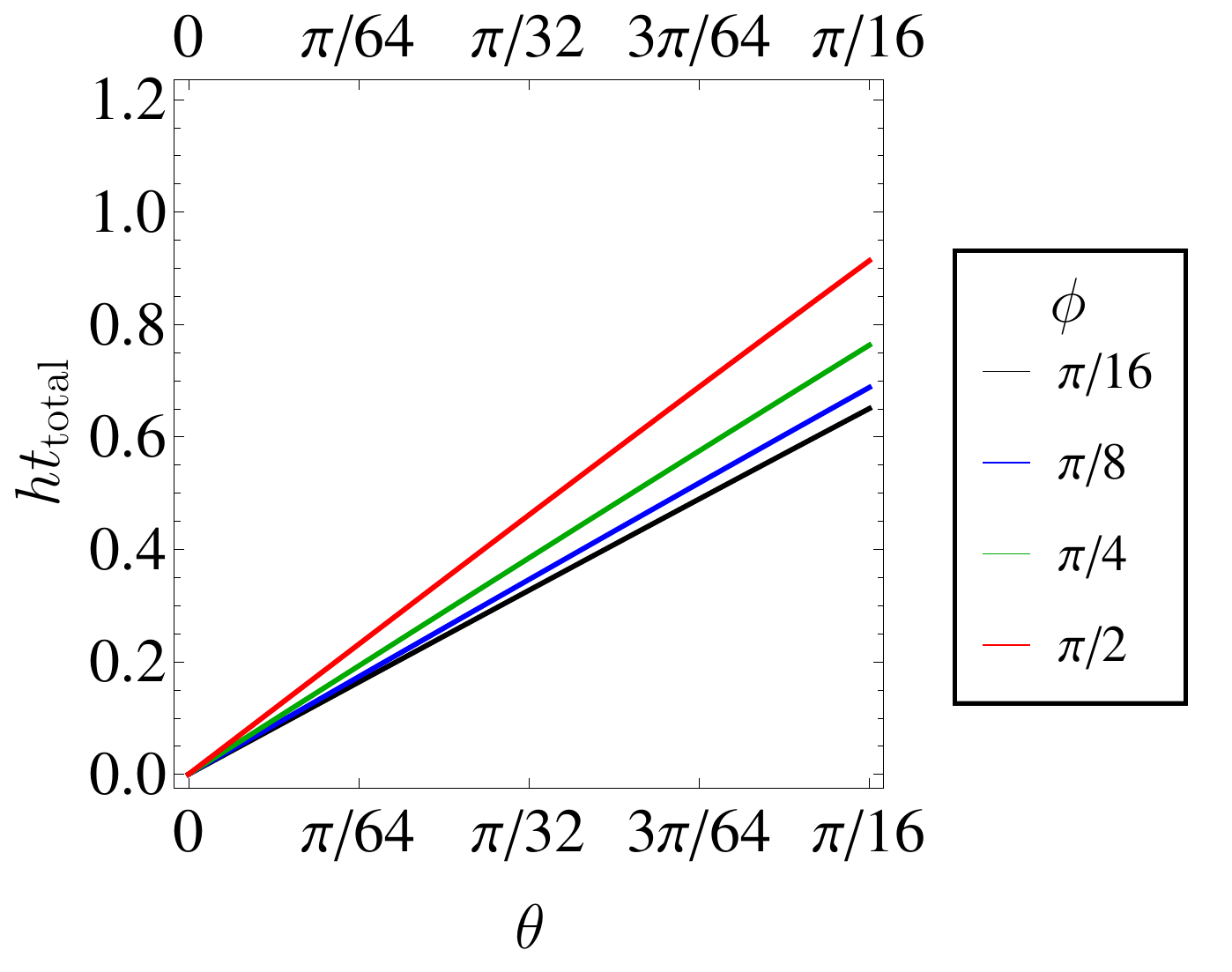}
\includegraphics[width=0.49\columnwidth]{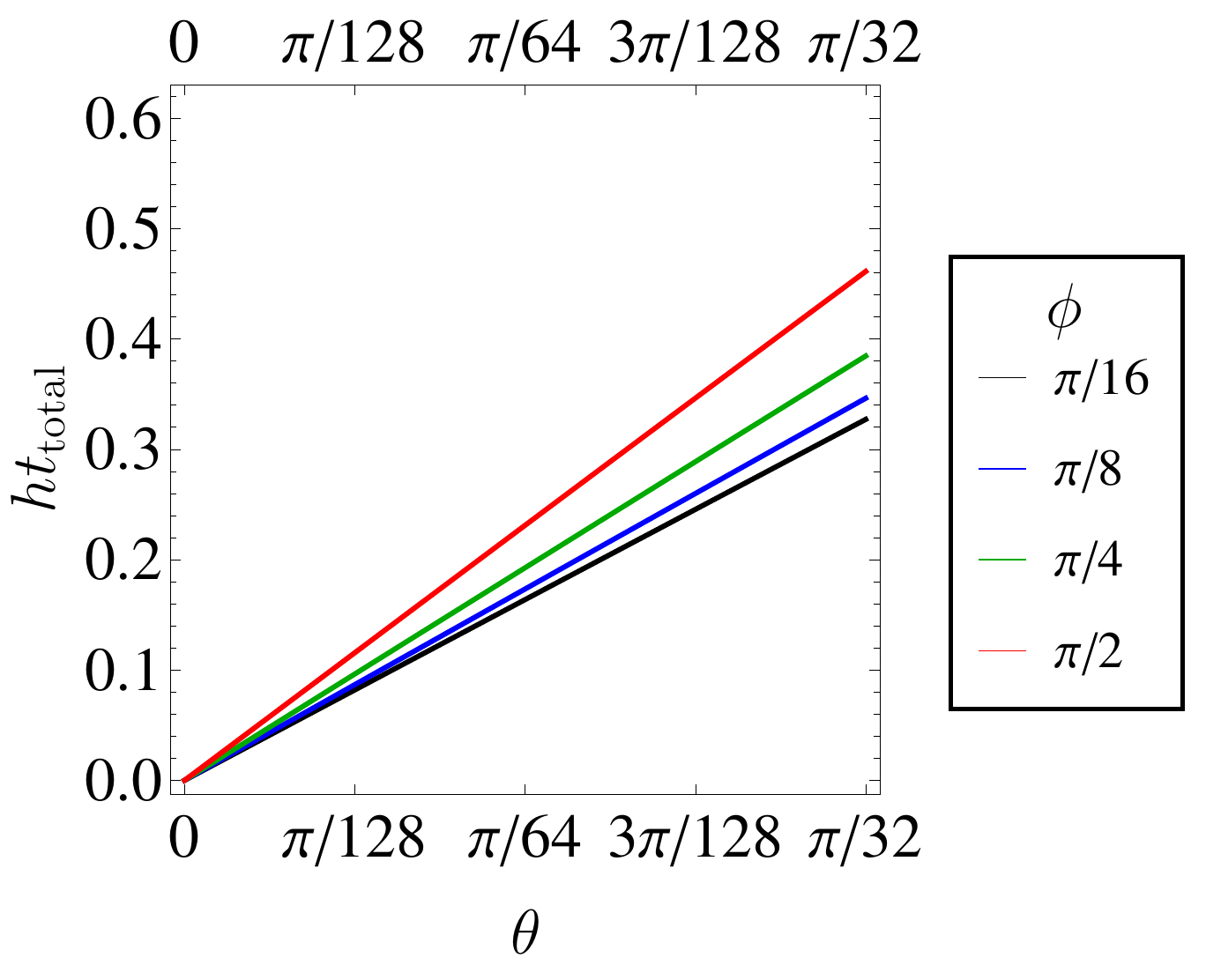}
\caption{Plots of times required to execute the $\theta$-$2\theta$-$\theta$ sequence, Eq.\ \eqref{Eq:ZRotRT} (solid lines),
and the ``generalized Ramon sequence'', Eq.\ \eqref{Eq:ZRotGGR} (dashed lines), with $\theta'=\frac{\pi}{4}$ (top left),
$\frac{\pi}{8}$ (top right), and $\frac{\pi}{16}$ (bottom) and for positive values of $\phi$.}
\label{Fig:Plots_tRTvsGRS}
\end{figure}

\begin{figure}[ht]
\includegraphics[width=0.49\columnwidth]{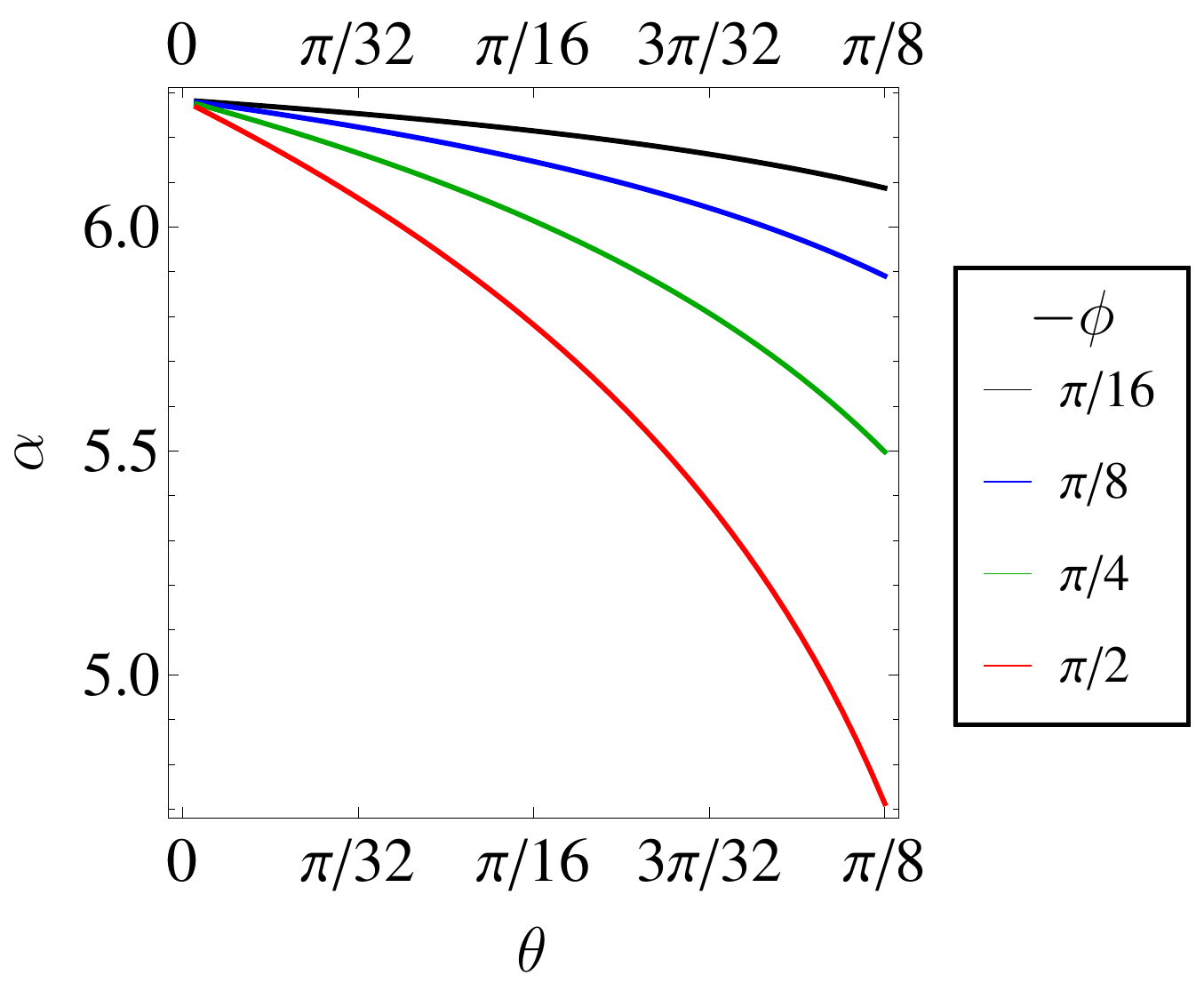}
\includegraphics[width=0.49\columnwidth]{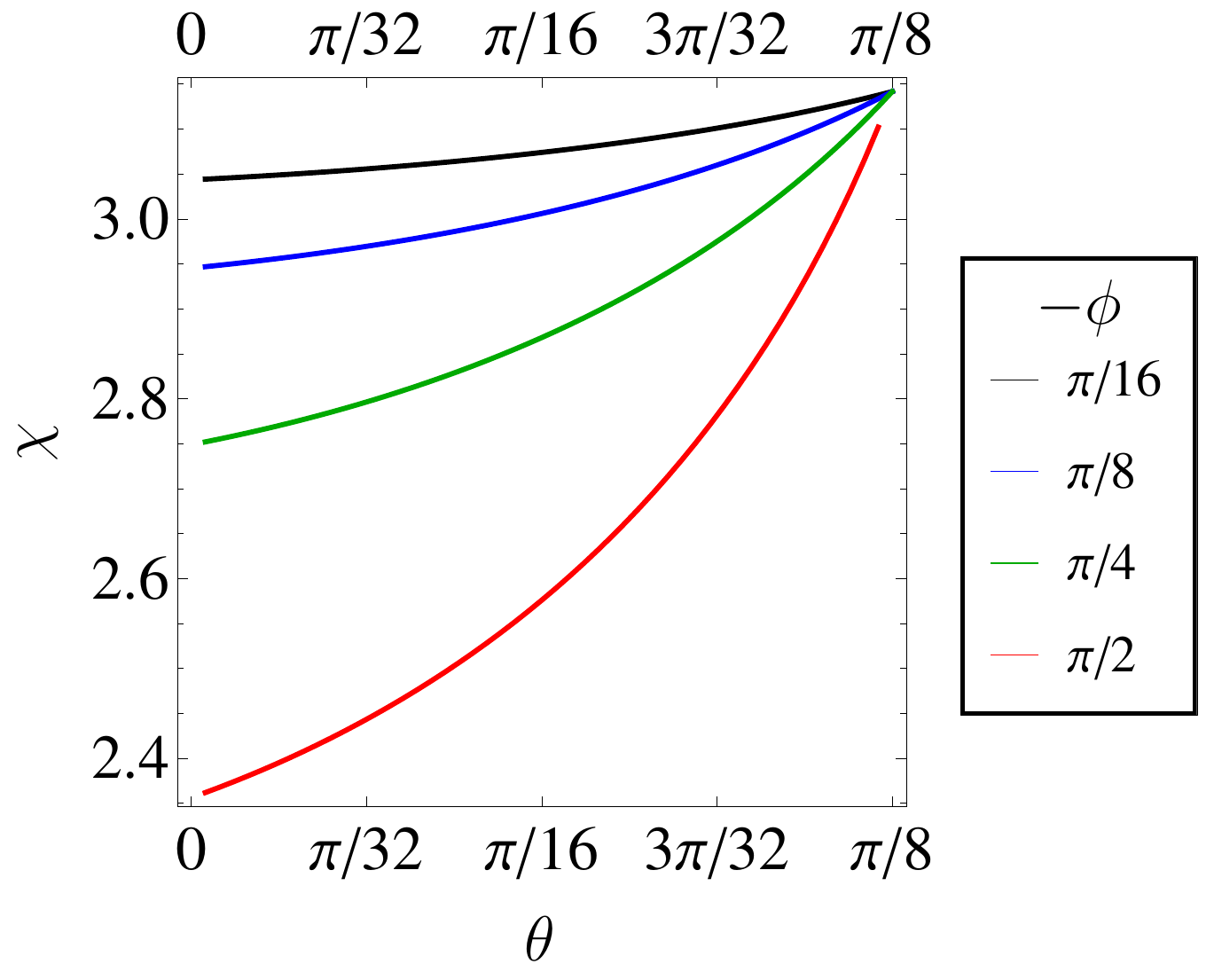}
\includegraphics[width=0.49\columnwidth]{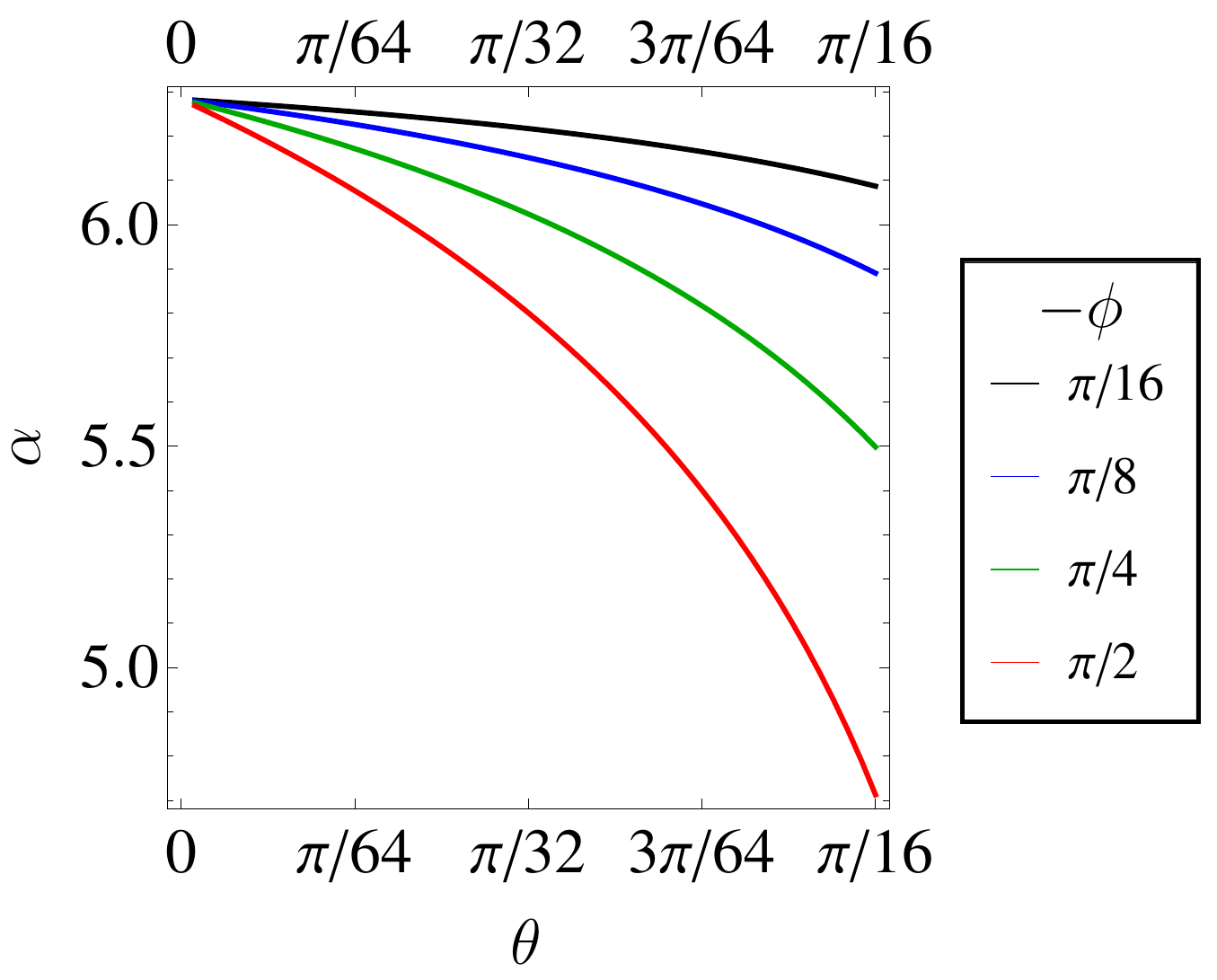}
\includegraphics[width=0.49\columnwidth]{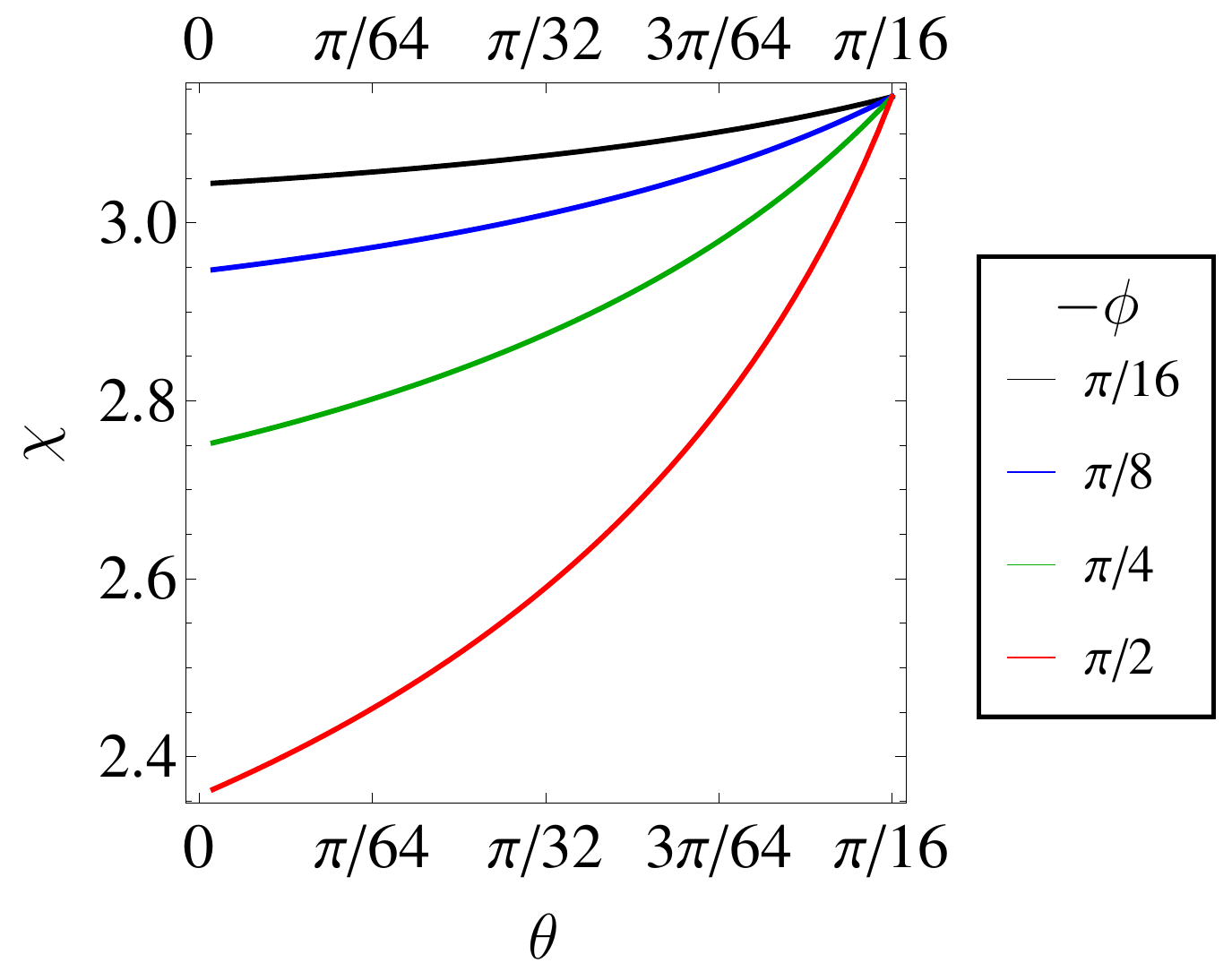}
\includegraphics[width=0.49\columnwidth]{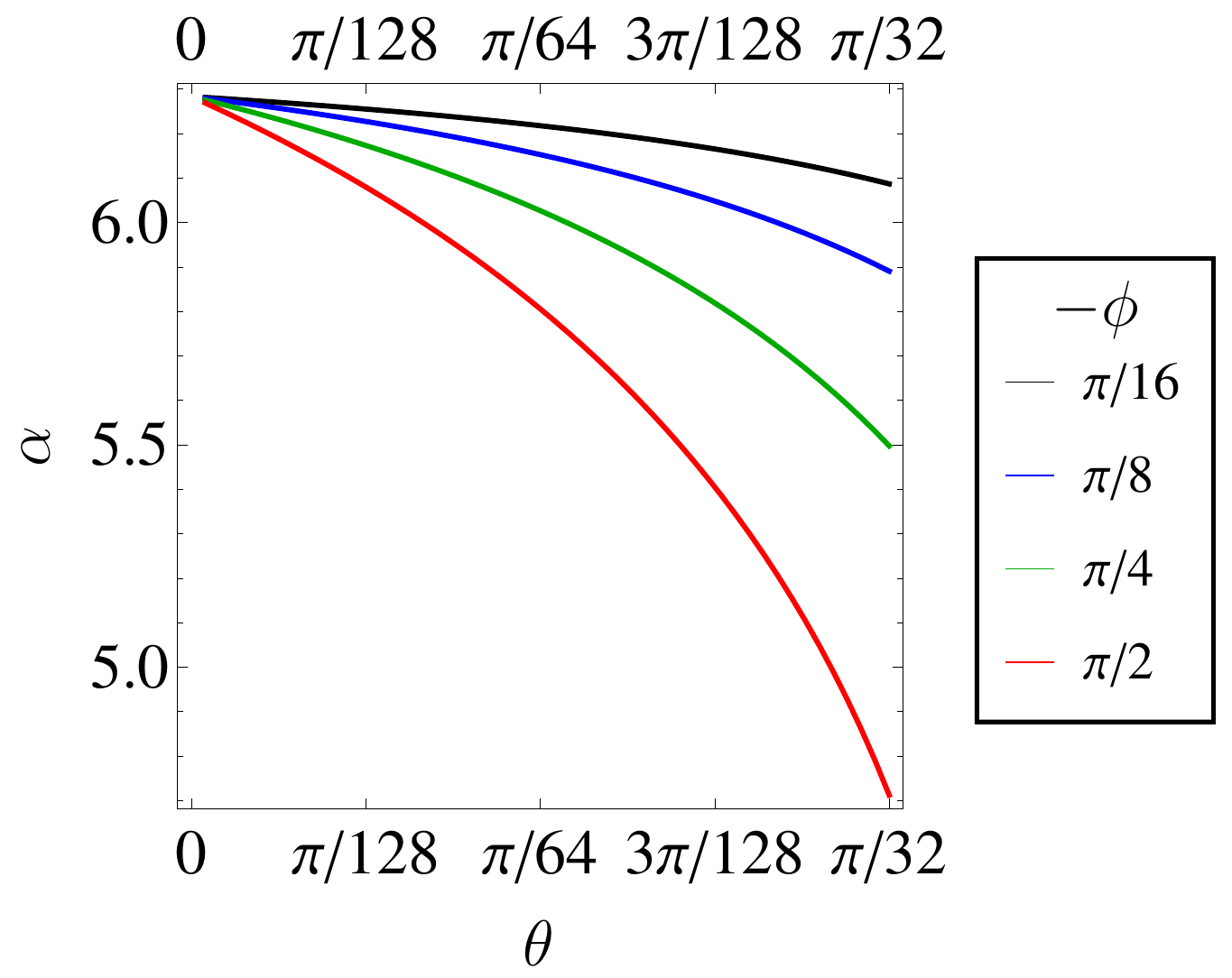}
\includegraphics[width=0.49\columnwidth]{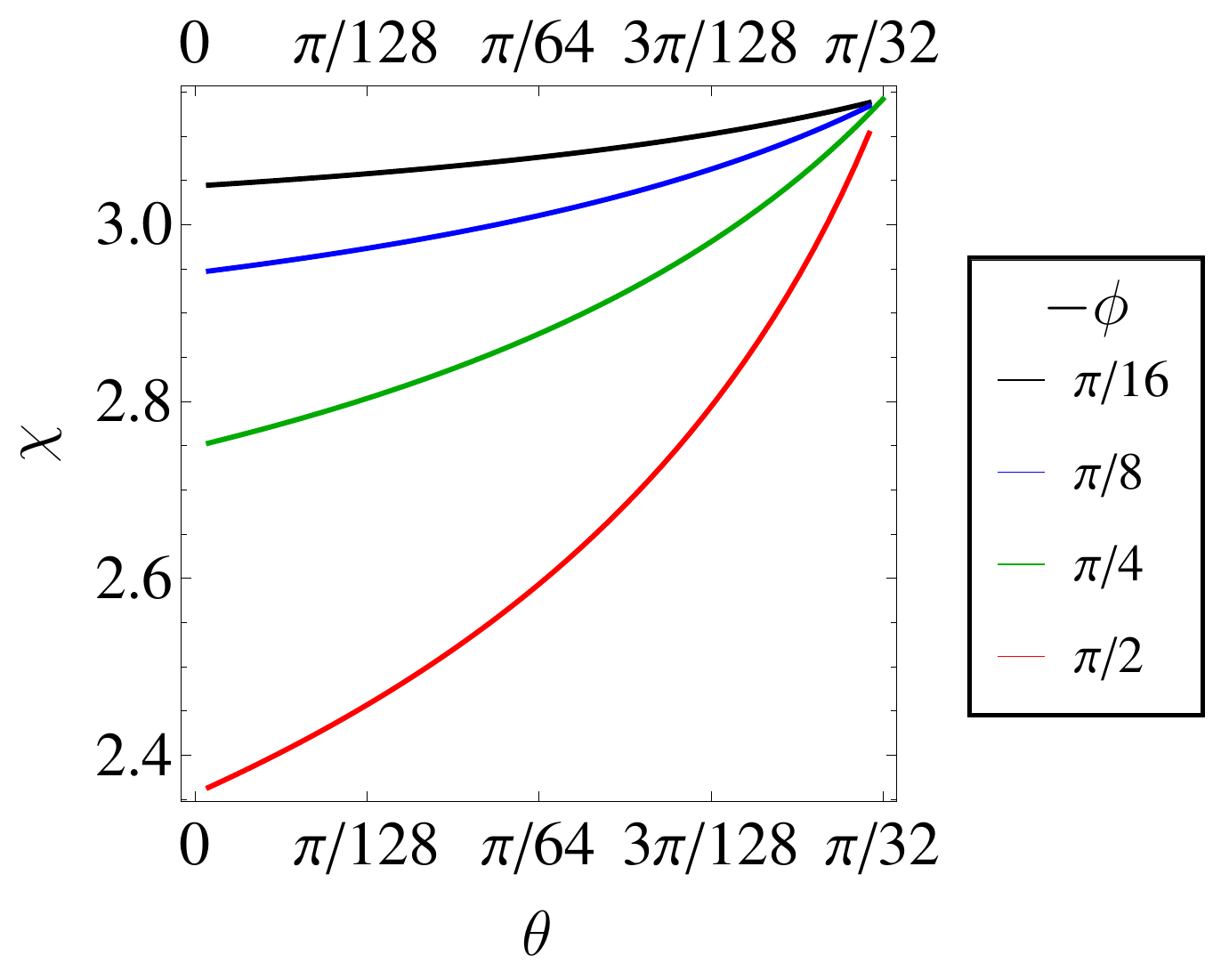}
\caption{As Fig.\ \ref{Fig:Plots_AlphaChiGRS}, but for negative values of $\phi$.}
\label{Fig:Plots_AlphaChiGRS_NegPhi}
\end{figure}

\begin{figure}[ht]
\includegraphics[width=0.49\columnwidth]{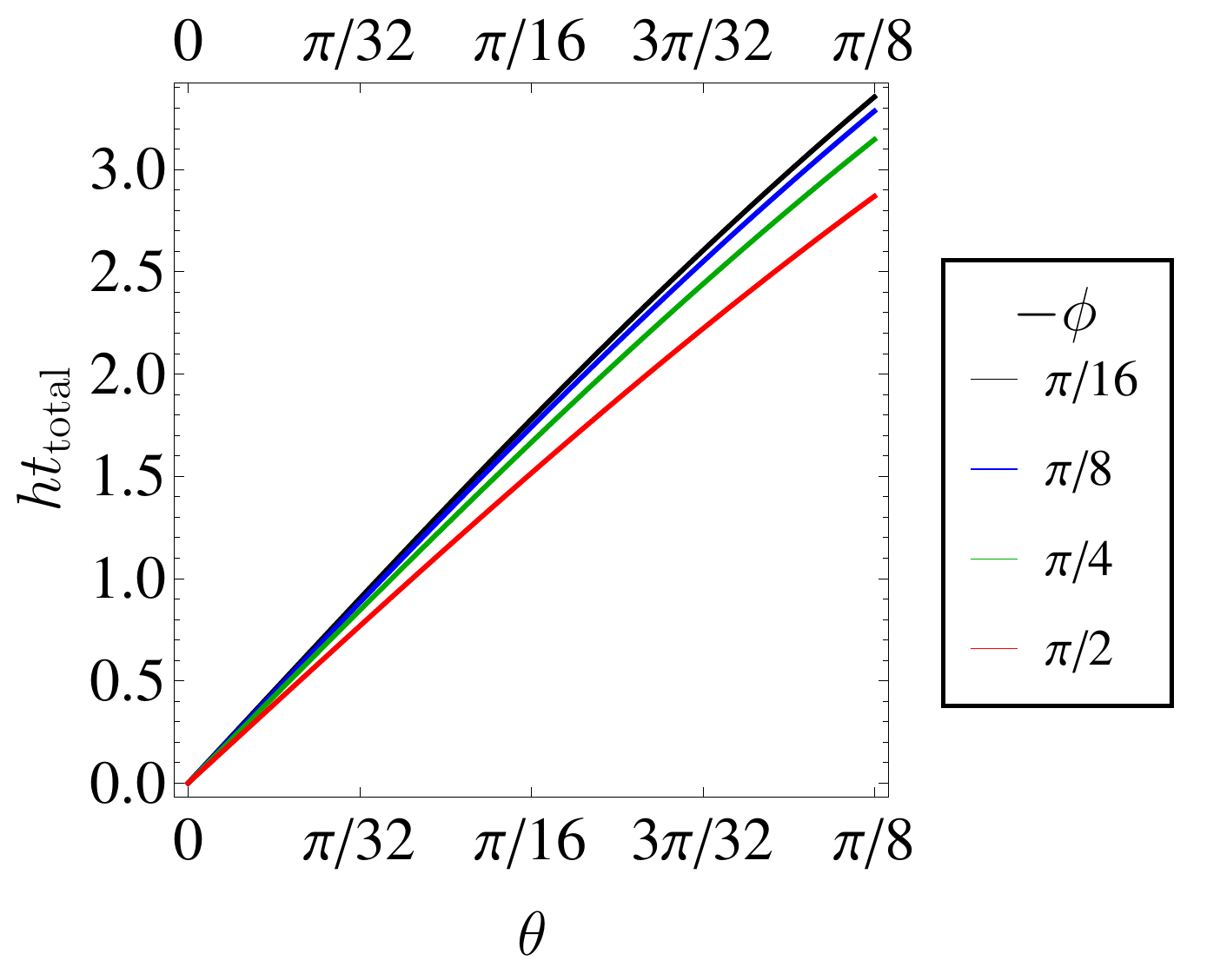}
\includegraphics[width=0.49\columnwidth]{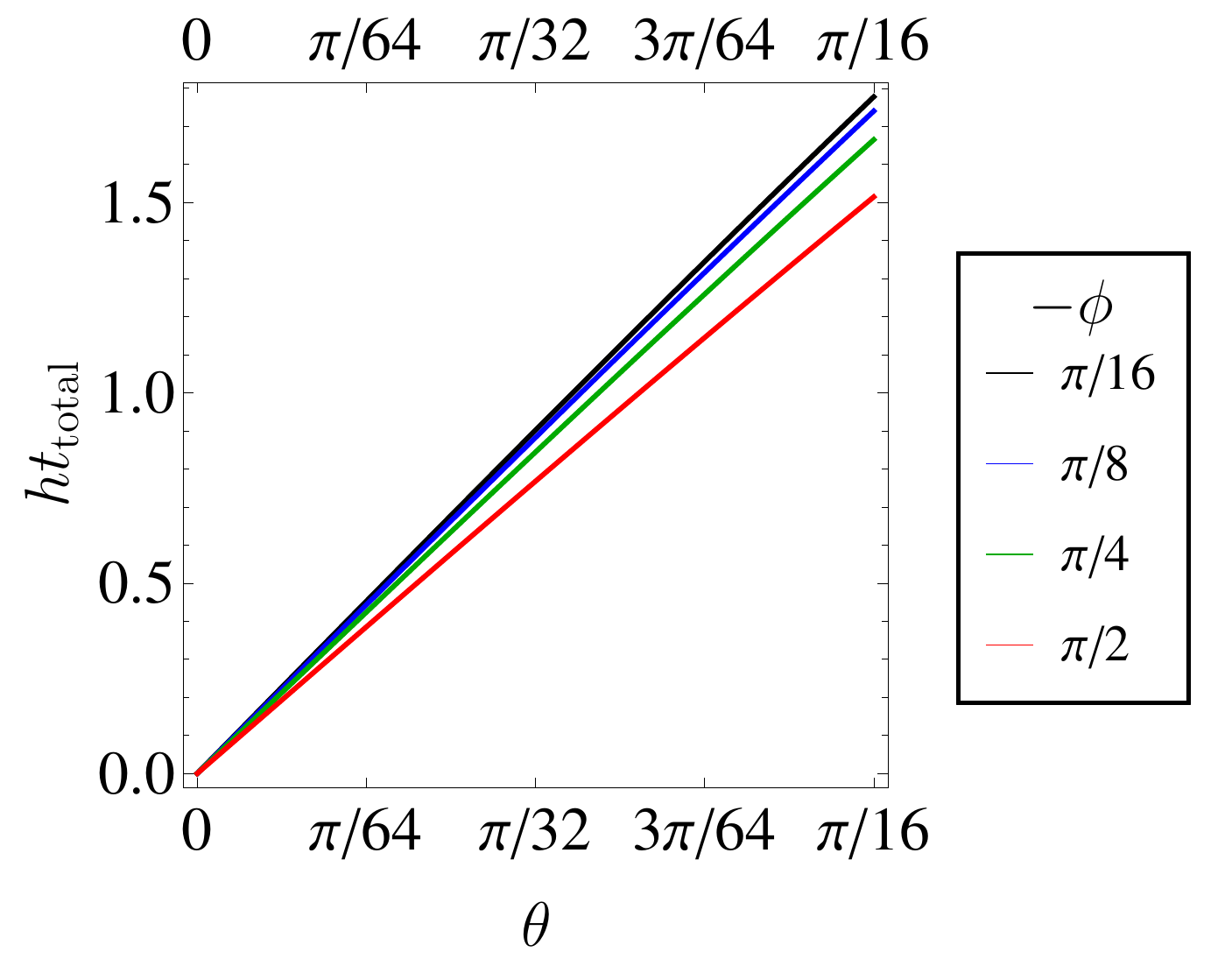}
\includegraphics[width=0.49\columnwidth]{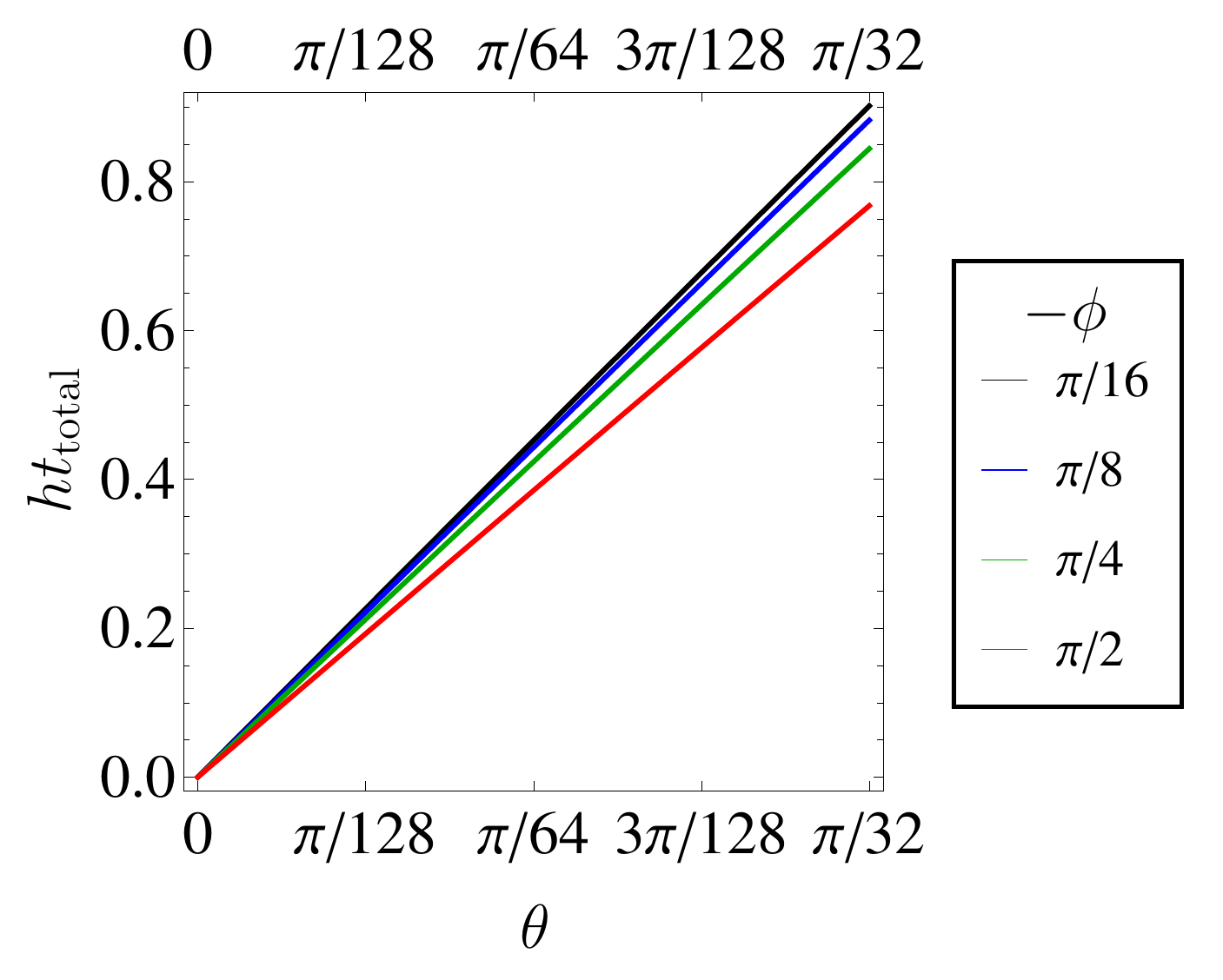}
\caption{As Fig.\ \ref{Fig:Plots_tRTvsGRS}, but for negative values of $\phi$.}
\label{Fig:Plots_tRTvsGRS_NegPhi}
\end{figure}

We will consider solutions to these equations for the same values of $\phi$ as before, starting with the positive values.
In this case, we also choose different values of $\theta'$; here we use $\pi/4$, $\pi/8$, and $\pi/16$.  We provide plots
for each case of $\alpha$ and $\chi$ in Fig.\ \ref{Fig:Plots_AlphaChiGRS} and the total time in Fig.\ \ref{Fig:Plots_tRTvsGRS}.
We see that even this ``generalized Ramon sequence'' is slower than the $\theta$-$2\theta$-$\theta$ sequence for a given
value of $\theta$, regardless of the value of $\theta'$.  This is also the case when we consider negative values of $\phi$;
we plot the values of $\alpha$ and $\chi$ in Fig.\ \ref{Fig:Plots_AlphaChiGRS_NegPhi} and the total time in Fig.\ \ref{Fig:Plots_tRTvsGRS_NegPhi}.
We also see that, similarly to the Ramon sequence, the total execution time of the $z$ rotation is the same for both the
$\theta$-$2\theta$-$\theta$ sequence and the ``generalized Ramon sequence'' when $\theta=\frac{\theta'}{2}$; this is because
the ``generalized Ramon sequence'' reduces to the $\theta$-$2\theta$-$\theta$ sequence in this limit.

\section{Modified $\theta$-$2\theta$-$\theta$ sequence for $x$ rotations} \label{App:GenHZH}
We will discuss here the following sequence for performing $x$ rotations, which is a modified version
of our $\theta$-$2\theta$-$\theta$ sequence for performing $z$ rotations:
\begin{equation}
R(\hat{\vec{x}},\phi)=-R(\theta,\pi)R\left (2\theta-\frac{\pi}{2},\phi\right )R(\theta,\pi). \label{Eq:XRotGH}
\end{equation}
We thus realize a $x$ rotation (with an overall minus sign) by an angle $\phi$ via the following
sequence:

\begin{enumerate}
	\item {Rotate by $\pi$ about an axis at an angle $\theta$ with respect to the $z$ axis.}
	\item {Rotate by $\phi$ about an axis at an angle $2\theta-\frac{\pi}{2}$ with respect to the $z$ axis.}
	\item {Repeat the first rotation.}
\end{enumerate}

The restriction on angles of rotation axes with respect to the $z$ axis requires that $\frac{\pi}{4}<\theta<\frac{\pi}{2}$.

We now give the values of $J$ needed for each type of rotation.  Taking $J_1$ to be the
value of $J$ used for the rotations by $\pi$, then the value of $J$ needed for the rotation
by $\phi$, which we denote by $J_2$, is
\begin{equation}
J_2=\frac{2J_1 h^2}{h^2-J_1^2}.
\end{equation}

\begin{figure}[ht]
\includegraphics[width=0.49\columnwidth]{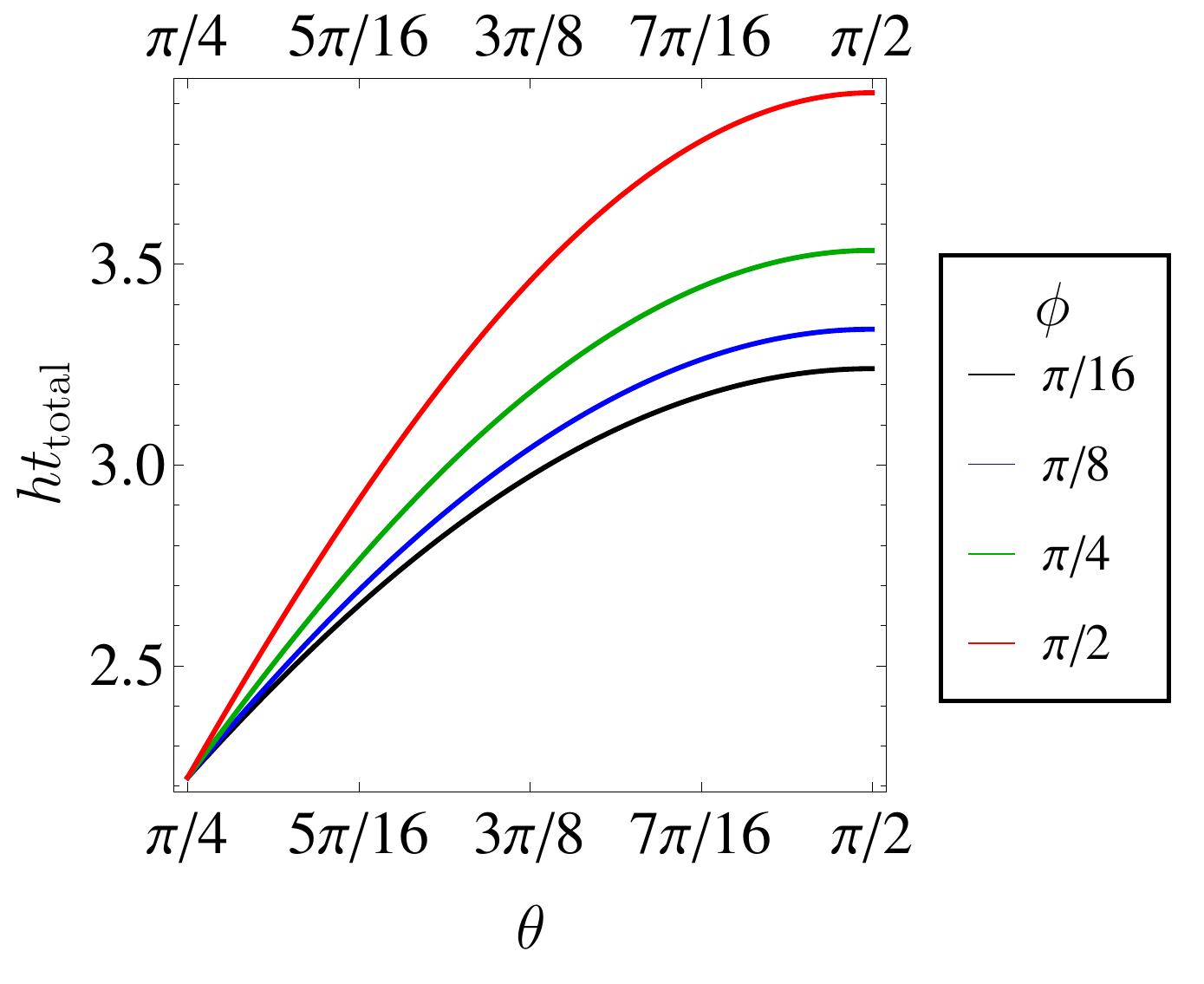}
\includegraphics[width=0.49\columnwidth]{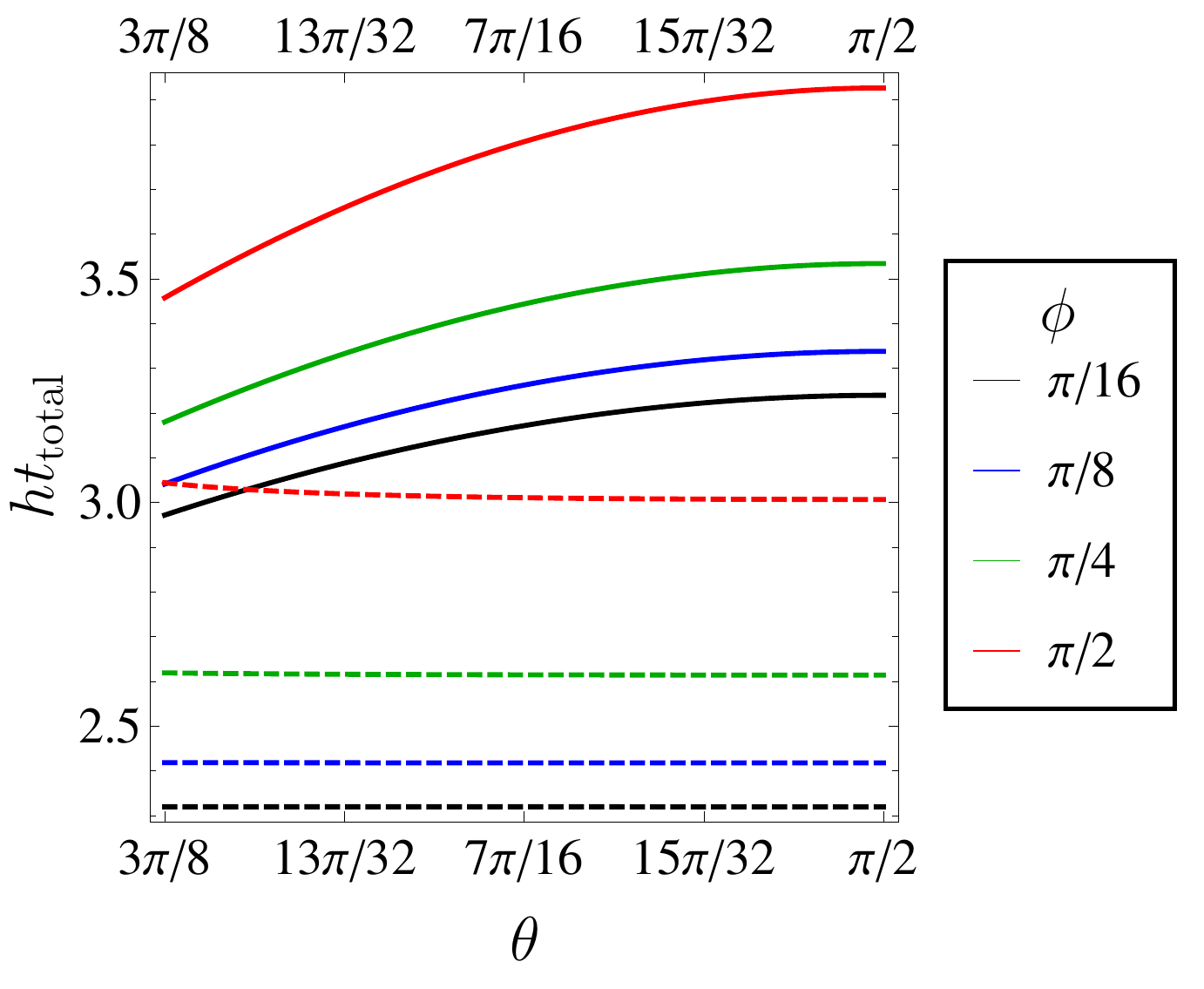}
\includegraphics[width=0.49\columnwidth]{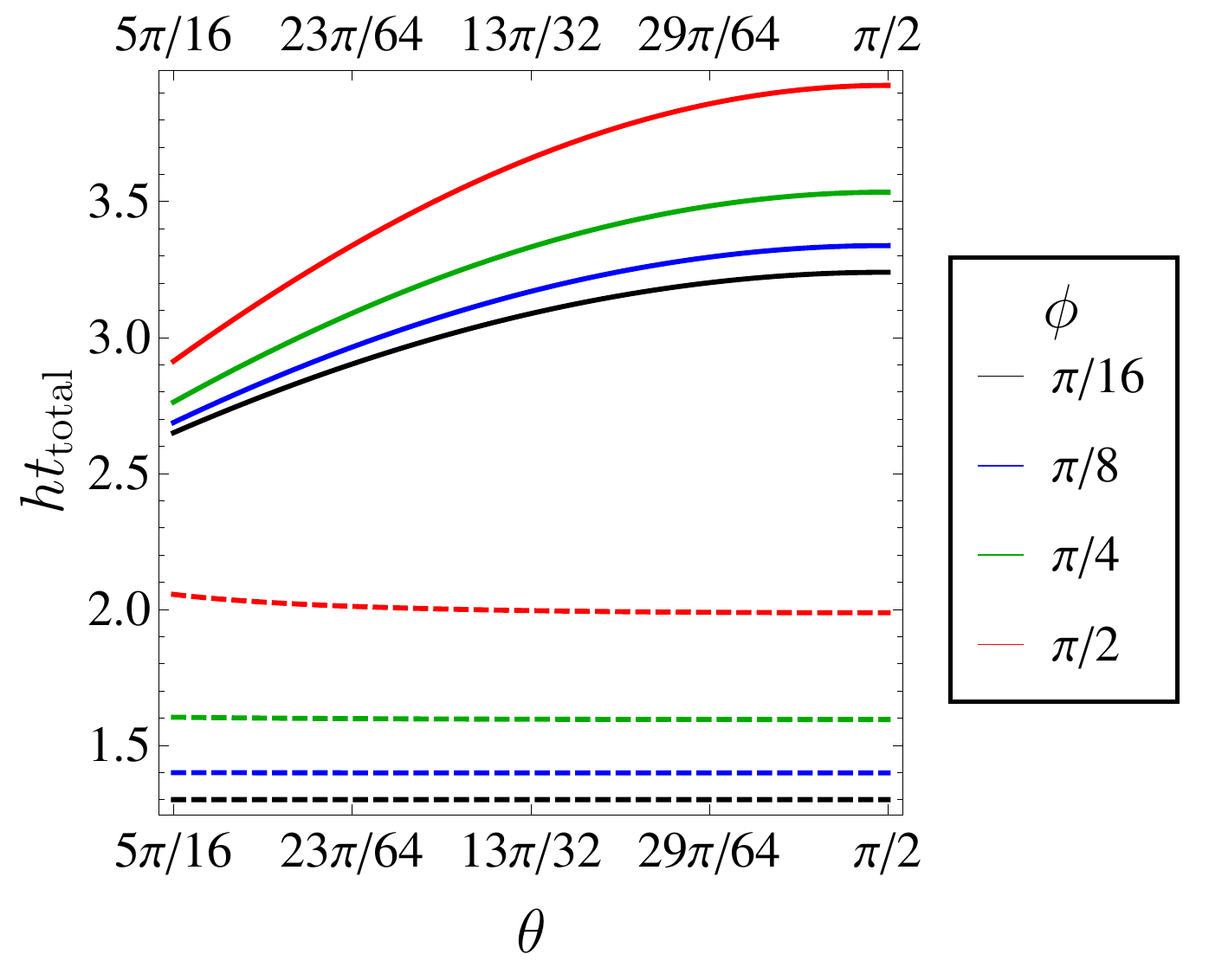}
\includegraphics[width=0.49\columnwidth]{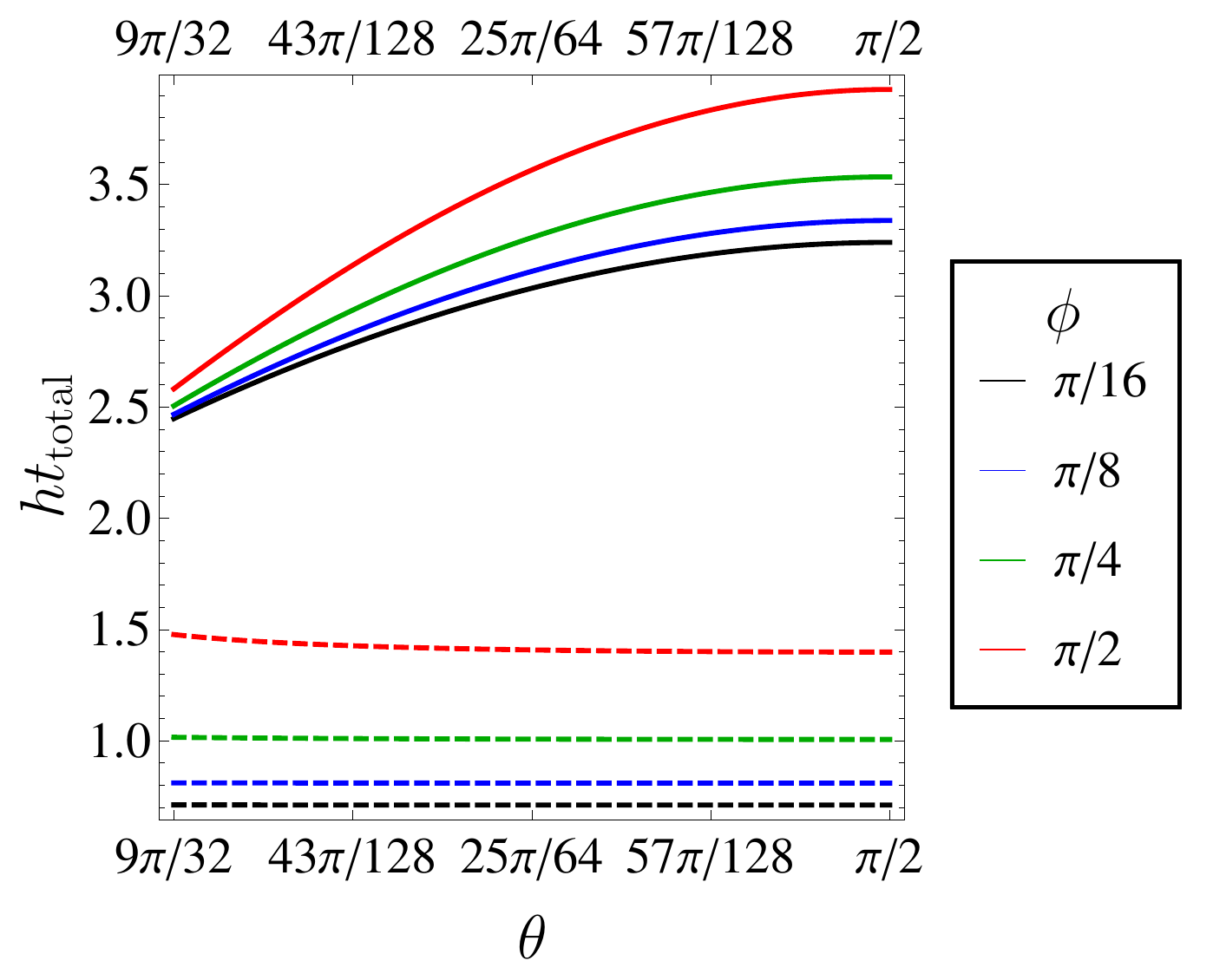}
\caption{(Top left) Plot of the time required to execute the sequence, Eq.\ \eqref{Eq:XRotGH}, for all possible values
of $\theta$, namely, $\pi/4 < \theta < \pi/2$.  The remaining plots compare the times required to execute the sequence,
Eq.\ \eqref{Eq:XRotGH} (solid lines), with the time required to execute the ``modified Ramon sequence'', Eq.\ \eqref{Eq:XRotGGR}
(dashed lines), for $\theta'=\frac{\pi}{4}$ (top right), $\frac{\pi}{8}$ (bottom left), and $\frac{\pi}{16}$ (bottom
right).  All plots shown here are for positive values of $\phi$.}
\label{Fig:tPlotGHvsMGGR_XRot}
\end{figure}

We now compare the timing of this sequence to the modified Ramon sequence.  The total time
required to execute this sequence is
\begin{equation}
ht_{\text{total}}=\pi\sin{\theta}-\tfrac{1}{2}\phi\cos{2\theta}.
\end{equation}
We will now consider four positive values of $\phi$, namely, $\frac{\pi}{2}$, $\frac{\pi}{4}$, $\frac{\pi}{8}$,
and $\frac{\pi}{16}$, as well as the corresponding negative values, and $\theta'=\frac{\pi}{4}$, $\frac{\pi}{8}$,
and $\frac{\pi}{16}$.  We plot the gate times for both sequences, Eqs.\ \eqref{Eq:XRotGH} and \eqref{Eq:XRotGGR},
for positive values of $\phi$ in Fig.\ \ref{Fig:tPlotGHvsMGGR_XRot}. We see that, unlike for the analogous sequences
for performing $z$ rotations, the generalized Ramon-like sequence for performing an $x$ rotation is always faster
than the modified $\theta$-$2\theta$-$\theta$ sequence.

\begin{figure}[ht]
\includegraphics[width=0.49\columnwidth]{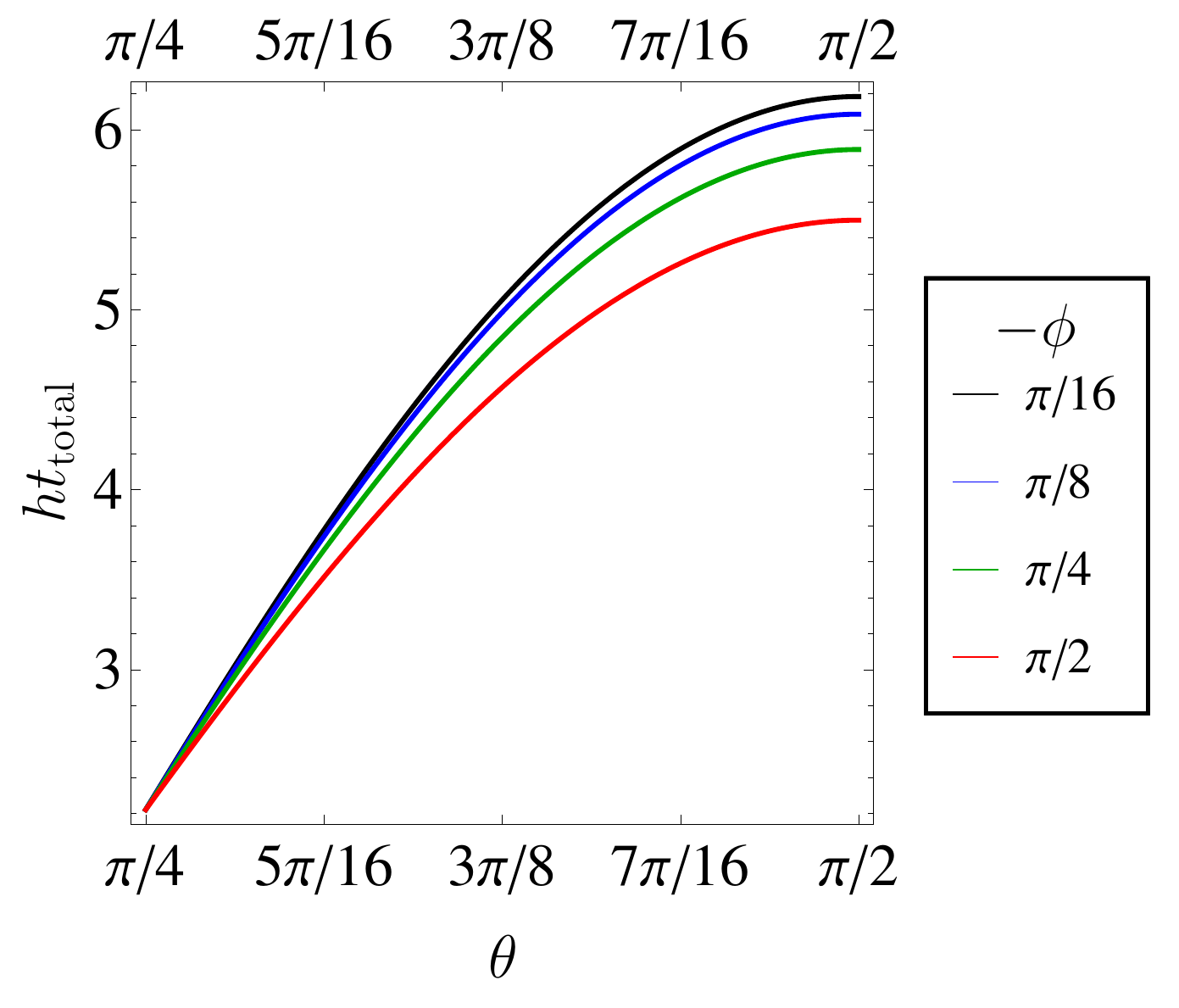}
\includegraphics[width=0.49\columnwidth]{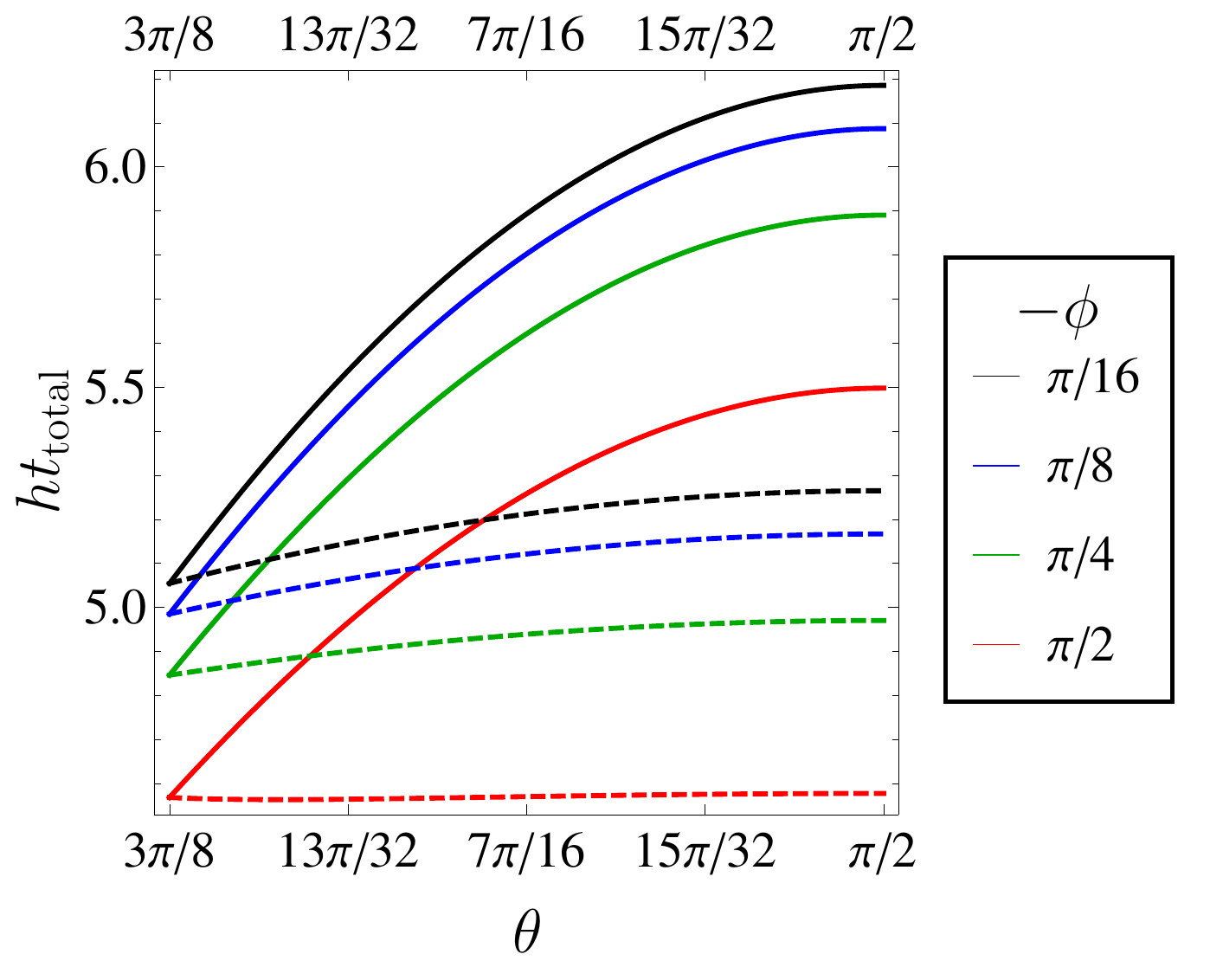}
\includegraphics[width=0.49\columnwidth]{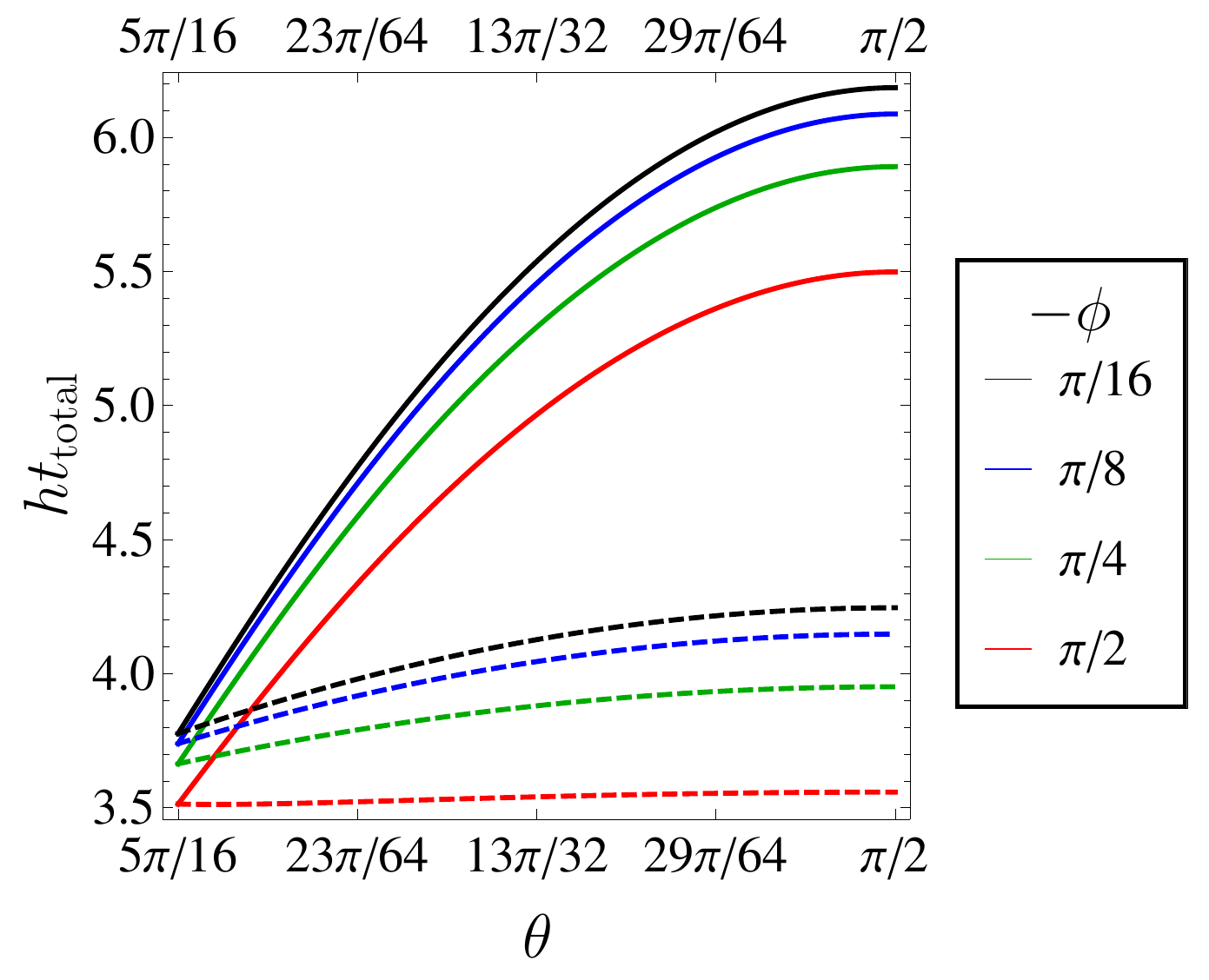}
\includegraphics[width=0.49\columnwidth]{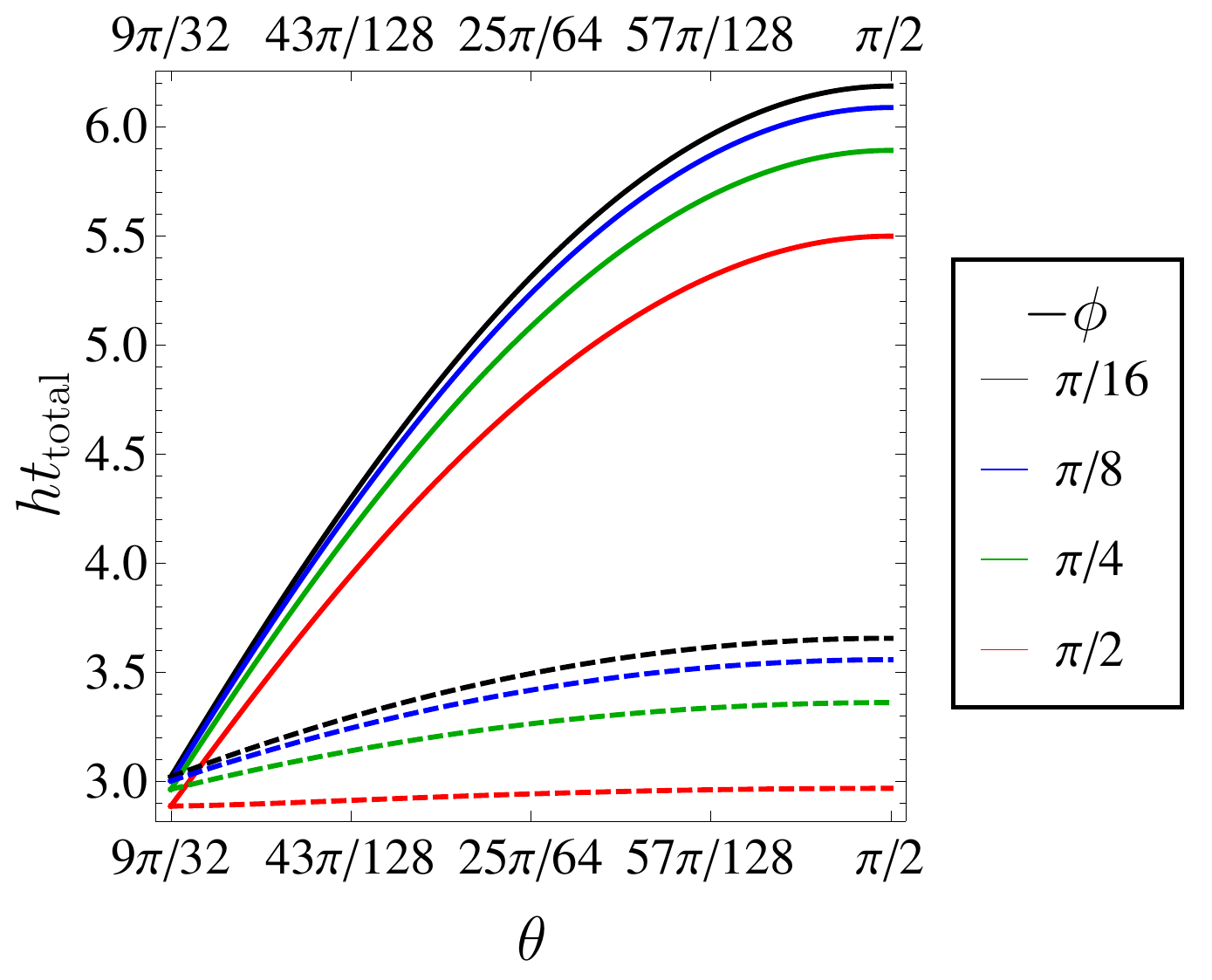}
\caption{As Fig.\ \ref{Fig:tPlotGHvsMGGR_XRot}, but for negative values of $\phi$.}
\label{Fig:tPlotGHvsMGGR_XRot_NegPhi}
\end{figure}

Now let us consider rotations by negative angles.  In this case, because rotations by negative angles are
impossible, we must add $2\pi$ to the rotation by $\phi$ in Eq.\ \eqref{Eq:XRotGH}.  We plot the total times
required to execute the two sequences in Fig.\ \ref{Fig:tPlotGHvsMGGR_XRot_NegPhi}.  In this case, the modified
Ramon sequence is once again the faster sequence.  We thus see that, for both positive and negative rotation
angles, the modified Ramon sequence is faster, in contrast to the analogous sequences for performing $z$ rotations.

\section{Fitting parameters for randomized benchmarking simulations} \label{App:FittingP}
We present here the parameters for the curves that we fit to our randomized
benchmarking simulations.

\begin{table*}[h]
	\centering
	\begin{tabular}{|M{60pt}|M{50pt}|M{62pt}|M{62pt}|M{62pt}|M{62pt}|N}
		
		\hline
		&  \multicolumn{1}{c|}{\multirow{3}{*}{Pulse}} & \multicolumn{2}{c|}{$T_{2}\;(\times 10^{3}\;\mathrm{ns})$}  & \multicolumn{2}{c|}{$n_{0}$} &\\[10pt]		
		\cline{3-6} 
		&  & barrier  &  tilt & barrier & tilt &\\[10pt]

		\hline
		\hline
		\multicolumn{1}{|c|}{\multirow{9}{*}{$h = 23$ MHz, $\sigma_{h}$ = 0 MHz}}
		& UCUO &  733 &  6.81  & 9.35 $\times10^{3} $ & 270 &\\[10pt]
		\cline{2-6}		
		
		& UCO-I &  17.6 &  0.16  & 1.03 $\times10^{3} $ & 12.1 &\\[10pt]
		\cline{2-6}		
		
		& UCO-II &  22.1 & 0.87  & 1.49 $\times10^{3} $ & 63.9 &\\[10pt]
		\cline{2-6}		
		
		& CUO &  $15.9 \times 10^{3}$ & 215  & 1.12 $ \times 10^{7}$ & 1.39 $\times10^{3} $ &\\[10pt]
		\cline{2-6}		
		
		& CO-II &  $7.85\times 10^{3}$ &  73.4  & $1.64\times 10^{6}$ & 0.817 $\times10^{3} $ &\\[10pt]
		\cline{2-6}

		\hline
		\hline
		\multicolumn{1}{|c|}{\multirow{9}{*}{$h = 40$ MHz, $\sigma_{h}$ = 11.5 MHz}}
		& UCUO  & 0.0522  & 0.0451 & 4.10 & 3.59 &\\[10pt]
		\cline{2-6}

		& UCO-I &  0.183 &  0.0500  & 21.6 & 5.80 &\\[10pt]
		\cline{2-6}

		& UCO-II &  0.0815 &  0.0676  & 14.6 & 9.00 &\\[10pt]
		\cline{2-6}		
		
		& CUO &  0.618 &  0.471  & 7.16 & 5.74 &\\[10pt]
		\cline{2-6}		
		
		& CO-II &  3.01 &  1.81  & 62.0 & 44.2 &\\[10pt]
		\cline{2-6}

	   \hline
	   \hline
	   \multicolumn{1}{|c|}{\multirow{9}{*}{ $h = 40$ MHz, $\sigma_{h}$ = 23 MHz}}
	   & UCUO  & 0.0223  & 0.0222 & 2.00 & 2.00 &\\[10pt]
	   \cline{2-6}

	   & UCO-I &  0.0584 &  0.0278  & 7.16 & 3.33 &\\[10pt]
	   \cline{2-6}

	   & UCO-II &  0.0306 &  0.0290  & 4.71 & 4.11 &\\[10pt]
	   \cline{2-6}		
	   
	   & CUO &  0.112 &  0.114  & 1.94 & 1.94 &\\[10pt]
	   \cline{2-6}		
	   
	   & CO-II &  0.280 &  0.286  & 7.49 & 6.44 &\\[10pt]
	   \cline{2-6}		
		
	  \hline		
		
	\end{tabular}
	\caption{Fitting parameters for quasistatic noise.}\label{Tab:FitP_Quasistatic}
\end{table*}

\begin{table*}[h]
	\centering
	\begin{tabular}{|M{50pt}|M{50pt}|M{65pt}|M{65pt}|M{65pt}|M{65pt}|N}
		
		\hline
		&  \multicolumn{1}{c|}{\multirow{3}{*}{Pulse}} & \multicolumn{2}{c|}{$T_{2}\;(\times 10^{3}\;\mathrm{ns})$}  & \multicolumn{2}{c|}{$n_{0}$} &\\[10pt]		
		\cline{3-6} 
		 &  & barrier  &  tilt & barrier & tilt &\\[10pt]

		\hline
		\hline
	\multicolumn{1}{|c|}{\multirow{9}{*}{$\delta h = 0$}}	& UCUO &  91.5 &  1.15  & 2.22 $\times 10^{3}$ & 42.3 &\\[10pt]
		\cline{2-6}		
		
		& UCO-I &  3.00 &  0.0420  & 0.20 $\times 10^{3}$ & 2.92 &\\[10pt]
		\cline{2-6}		
				
		 & UCO-II &  14.8 & 0.124  & 1.17 $\times 10^{3}$ & 9.17 &\\[10pt]
		\cline{2-6}		
			
		& CUO &  175 &  1.92  & 764 & 11.3 &\\[10pt]
		\cline{2-6}		
			
		& CO-II &  39.5 &  0.296  & 415 & 3.68 &\\[10pt]
		\cline{2-6}

		\hline
		\hline
	\multicolumn{1}{|c|}{\multirow{9}{*}{$\delta h \neq 0$}}	& UCUO  & 1.74  & 0.372 & 109 & 24.0 &\\[10pt]
		\cline{2-6}

		& UCO-I &  1.17 &  0.0221  & 135 & 2.70 &\\[10pt]
		\cline{2-6}

	 & UCO-II &  2.38 &  0.0628  & 298 & 8.46 &\\[10pt]
		\cline{2-6}		
		
		& CUO &  45.1 &  0.947  & 439 & 9.98 &\\[10pt]
		\cline{2-6}		
		
		& CO-II &  18.2 &  0.157  & 344 & 3.39 &\\[10pt]
		\cline{2-6}		
		
		\hline

	\end{tabular}
	\caption{Fitting parameters for $1/f^\alpha$ noise with $\alpha_{h}=2.6$ and $\alpha_{J}=0.7$.}\label{Tab:FitP_OneOverF}
\end{table*}

\end{document}